\documentclass[11pt]{article} 

\usepackage{amsmath} 
\DeclareMathOperator*{\argmax}{arg\,max}
\usepackage{graphicx} 
\usepackage{subcaption}
\usepackage{booktabs} 
\usepackage{caption}  

\usepackage{geometry} 
\usepackage{hyperref} 
\usepackage[numbers, sort&compress]{natbib}
\hypersetup{colorlinks=true, linkcolor=blue, citecolor=blue, urlcolor=blue}
\usepackage{longtable}
\usepackage{pdflscape}
\usepackage{svg}

\usepackage{setspace}
\usepackage{titling}
\usepackage{authblk}
 
\pretitle{\begin{center}\fontsize{17}{20}\bfseries\selectfont}
\posttitle{\par\end{center}\vskip 0.8em}

\usepackage{amsfonts} 
\usepackage[utf8]{inputenc}
\usepackage[T1]{fontenc}
\usepackage{newtxtext, newtxmath}
\usepackage{float}
\usepackage{listings} 
\usepackage{xcolor}   
\usepackage{multirow}
\usepackage{algorithm}
\usepackage{algpseudocode}
\usepackage{amsmath}
\usepackage{xcolor}
\usepackage{booktabs} %
\usepackage{makecell}
\usepackage{colortbl}
\usepackage{parskip}
\newcommand{\graycomment}[1]{\hfill \textcolor{gray}
{$\triangleright$\ #1}}
\title{FePySR: A Neural Feature Extraction Framework for Efficient and Scalable Symbolic Regression}
\author[1]{Zhiming Yu}
\author[1,*]{Wangtao Lu}
\author[2,*]{Xin Lai}
\affil[1]{School of Mathematical Sciences, Zhejiang University, Hangzhou, Zhejiang, China}
\affil[2]{Faculty of Medicine and Health Technology, Tampere University, Tampere, Finland}

\begin{document}
\date{}
\maketitle 
\insert\footins{\noindent\footnotesize{$^*$Corresponding authors: \texttt{wangtaolu@zju.edu.cn} (WL); \texttt{xin.lai@tuni.fi} (XL)}}
\begin{abstract}
\small
A fundamental challenge in symbolic regression (SR) is efficiently recovering complex mathematical expressions from observational data. Although this problem is NP-hard, many expressions of practical interest decompose naturally into combinations of nonlinear feature modules, concentrating structural complexity into a small number of reusable components. Here, we introduce FePySR, a two-stage framework that reduces the SR search space by extracting valid features prior to equation search. FePySR first employs a heterogeneous neural network to constrain observational data to a set of candidate expressions, then performs structural optimization within this refined expression space using PySR. Across five standard benchmarks, FePySR outperforms state-of-the-art methods by achieving higher equation recovery rates. On a set of 75 highly complex synthesized equations, FePySR recovers 36 equations, while producing substantially smaller mean squared errors on the remaining unrecovered cases, with reduced computation time compared to PySR. FePySR's first stage also maintains consistent performance under varying numbers of selected top features and increasing levels of noise in the observational data. Applied to ordinary differential equations governing biological systems, FePySR successfully identifies governing equations in 24 out of 100 tests where PySR recovers none. Taken together, FePySR is a generalizable framework that can enhance the SR solvers, enabling the efficient and reliable recovery of symbolic expressions across scientific domains.
\\[10pt]
\textbf{Keywords:} feature engineering, feature discovery, combinatorial search reduction, expression trees, neural network.
\\[10pt]
\textbf{Data and Code Availability:} https://github.com/laixn/FePySR.
\end{abstract}
\newpage
\section{Introduction}
\label{sec:intro} 

Symbolic regression (SR) is a fundamental challenge in machine learning. It aims to discover concise, physically interpretable mathematical models directly from observational data to empower automated scientific discovery. However, SR is computationally challenging due to its proven NP-hard complexity, which comes from its dual computational tasks. First, the feature selection task aims to identify valid features from high-dimensional inputs \cite{virgolin2022symbolic}. Second, the structural optimization task aims to identify optimal mathematical formulas within a vast space of basic operators and primitives \cite{song2024prove}. Both tasks require exhaustive combinatorial search, rendering the overall problem computationally intractable as input dimensionality and equation complexity grow.

Current research approaches can be categorized into three types for addressing the combinatorial explosion of the search space resulting from the NP-hard nature of SR. The first approach involves building more powerful searchers dedicated to efficiently exploring the unconstrained space of equations. These approaches originate from genetic programming (GP) and drive the search process by simulating genetic evolutionary mechanisms, such as crossover and mutation \cite{Koza1994, Schmidt2009}. Subsequent studies have further optimized individual selections \cite{LaCava2016} and offspring generation strategies \cite{Virgolin2017}. Recently, frameworks such as PySR have greatly improved search efficiency through engineering optimization and distributed computing \cite{Cranmer2023PySR}. LASR uses a concept library that is queried by large language models (LLMs) to accelerate SR \cite{Grayeli2024LASR}. 

Besides genetic evolutionary algorithms, deep learning provides a new paradigm for combinatorial exploration. For example, deep symbolic regression (DSR) reformulates the generation of equations as a Markov decision process. It employs risk-seeking policy gradients and enhanced exploration strategies to guide neural networks through non-convex heuristic searches \cite{Petersen2021a, Landajuela2021_exploration}. In such methods, the network typically outputs the probability distribution of symbols in an autoregressive manner \cite{Mundhenk2021, Landajuela2022}, while dynamically incorporating the intrinsic tree structure of symbolic expressions based on current contexts \cite{Petersen2021_priors}. To overcome the limitations of single-method approaches, hybrid frameworks integrate diverse mechanisms. For instance, neural-guided genetic programming (NGGP) uses neural models to initialize high-quality feature populations for evolutionary searches \cite{Mundhenk2021}, and recent advances couple continuous optimization with discrete structural searches \cite{Pettit2025}. Similarly, reinforcement learning has been adopted to guide evolutionary searches in a hybrid system called \cite{RL-GEP}. Furthermore, transformer-based models have fundamentally changed traditional search paradigms by reframing the problem as a sequence-to-sequence translation task. These models leverage massive pre-training data to directly learn the conditional probability distribution from numerical values to combinations of mathematical features, enabling end-to-end generation from data to formulas \cite{biggio2021neural,lee2019set,li2023transformer,valipour2021symbolicgpt, lample2019deep,kamienny2022end,tian2025interactive}.

The second approach aims to constrain the search space. It transforms an infinite-dimensional discrete search problem into a parameter optimization problem within a finite continuous space by imposing inductive biases. A representative example is SINDy, which achieves the decoupled discovery of dominant features by performing sparse regression over a predefined library of basic functions \cite{brunton2016discovering}. Based on this method, Implicit-SINDy overcame the underlying model's inability to handle implicit dynamics and rational fractions \cite{mangan2016inferring}. To address the extreme noise sensitivity introduced by this implicit extension, SINDy-PI constructs a parallelized convex optimization framework that achieves robust identification in high-noise environments \cite{kaheman2020sindy}. Furthermore, ADAM-SINDy integrates continuous space optimization techniques to refine the accuracy of the representation of time-varying parameters within the system \cite{viknesh2024adam}. Another notable work is the equation learner (EQL), which replaces the activation functions of shallow neural networks with a custom library of mathematical operators. When combined with sparse regularization, EQL realizes end-to-end formula discovery within a predefined continuous hyperspace \cite{sahoo2018learning, kim2021integration}.

The third approach advocates simplifying the search space. Instead of directly optimizing the searcher, this paradigm aims to reduce the problem's complexity at its source. Pioneering work in this area is AI Feynman, which uses neural networks as detectors to recursively identify hidden structures in observational data, such as translational symmetry and separability \cite{udrescu2020aifeynman}. The successor AI Feynman 2.0 analyzes the gradient properties during neural network training to reveal modular structures within the computational graph of formulas \cite{udrescu2020aifeynman2}. This strategy decomposes high-dimensional problems into multiple low-dimensional subspaces, which is highly effective in mitigating the NP-hard challenges induced by the curse of dimensionality. Integrating AI Feynman’s philosophy into hybrid frameworks like uDSR \cite{Landajuela2022} fully validates the potential of combining dimensionality reduction with advanced search methods for SR.

AI Feynman and its variants demonstrate that reducing the feature space can simplify the SR problem. Inspired by this, we investigate whether a complementary strategy can be designed to address the alternate root cause of SR's NP-hardness: the combinatorial explosion of operators. This challenge is central to feature engineering and representation learning in SR. Historically, methods addressing this have relied on explicit module discovery, such as automatically defined functions \cite{koza1992genetic} or recursive library learning, such as InceptionSR \cite{gu2025inceptionsr}. While effective for tasks with high structural reusability, these methods couple feature discovery with GP, resulting in substantial computational overhead and code inflation \cite{virgolin2018symbolic}. This coupling also renders the stability and quality of the discovered operators dependent on evolutionary search paths and the structural properties of the target equations \cite{virgolin2017scalable}.

\begin{figure}[h]
    \centering
    \includegraphics[scale=0.18]{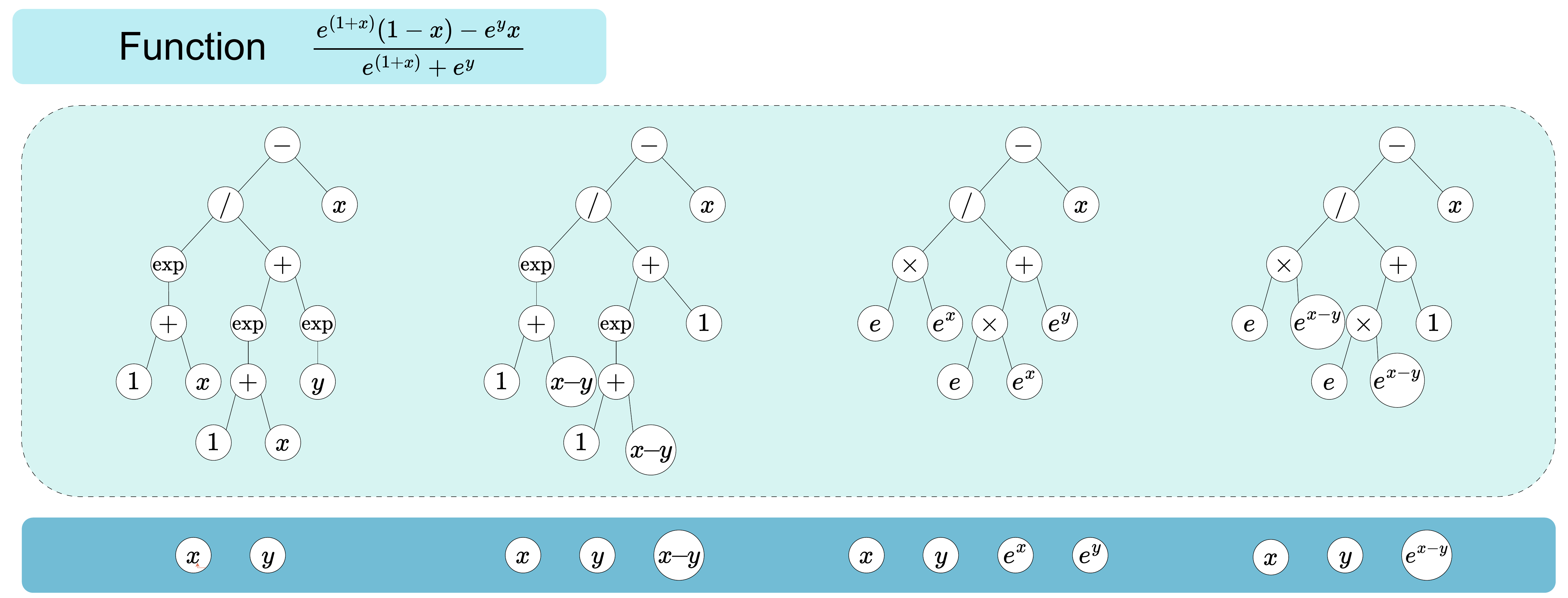}
    \captionsetup{font=small,labelfont=bf} 
    \caption{\textbf{Illustration of expression tree simplification via effective feature extraction}. This figure compares the structural complexity of the expression tree—measured by the total number of nodes (symbols)—required to represent the target symbolic expression $\frac{e^{1+x}(1-x)-e^{y}x}{e^{1+x}+e^{y}}$ under the progressive introduction of different effective features. When relying exclusively on the raw input variables $x$ and $y$ (leftmost), the representation necessitates a 14-node tree structure. Introducing the simple feature $x-y$ reduces the tree complexity to 13 nodes. Further incorporating the nonlinear features $e^x$ and $e^y$, which are highly structurally aligned with the target expression, decreases the node count to 11. Finally, supplying the highly relevant composite feature $e^{x-y}$ (rightmost) yields the most compact representation, requiring only 9 nodes.
    }
    \label{fig:my_figure} 
\end{figure}

We assume that many structurally complex symbolic expressions can be decomposed into combinations of simple features. For example, the expression search tree for the equation $\frac{e^{1+x}(1-x)-e^{y}x}{e^{1+x}+e^{y}}$ has high topological complexity and depth (Figure \ref{fig:my_figure}). However, identifying valid features (e.g., $x-y$, $e^x$, $e^y$, and $e^{x-y}$) from observational data shortens the tree into a shallower, simpler structure. We define valid features as data-derived features that reduce the depth of the target symbolic expression tree, thereby allowing complex topologies to be substituted with streamlined representations. Extracting valid features frees the following search algorithm from reconstructing an equation over a vast vocabulary of atomic operators (e.g., $(\cdot)^2, \sin(\cdot), \cos(\cdot), \exp(\cdot), +, \times$) that requires high computational resources. Instead, the search is conducted within an optimized feature space that allows for the efficient recovery of equations' structures.

Here, we demonstrate that FePySR outperforms existing methods in symbolic expression recovery across several benchmarks. When the complexity of the target equation makes exact recovery impossible, integrating relevant features results in symbolic expressions with lower mean squared error (MSE) than state-of-the-art (SOTA) models. Furthermore, the feature extraction step provides heuristic guidance, even when the extracted features are not exact constituents of the target equation. Features that share structural similarity with valid features steer the search toward promising regions of the expression space. This establishes favorable initial conditions for SR solvers to convergence to the exact solution. Taken together, this feature extraction framework offers the complementary benefits of improved approximation accuracy and heuristic guidance for SR. Its integration with existing SR solvers, such as PySR, establishes a generalizable paradigm for complex SR tasks.

\newpage

\section{Method}

We propose FePySR, a two-stage SR framework. The framework first employs a feature mapping network (FMN) to extract highly expressive nonlinear features from observational data. This streamlines the search space for downstream SR. Next, the data and the identified valid features are concatenated to form a feature representation. This enriched representation is then fed into the PySR solver, which efficiently reconstructs the target symbolic expression within an optimized search space (Figure \ref{FePySR}).

\subsection{FMN's architecture}

In inspired by EQL \cite{sahoo2018learning,kim2021integration}, we develop the proposed FMN that significantly extends the EQL architecture. Conventional EQL networks employ a multi-layer feedforward topology wherein successive layers are mapped via fully connected linear transformations ($h_i = W_i y_{i-1}$). Each layer's activation function $f(\cdot)$ is composed of a predefined library of mathematical primitives (e.g., $id, (\cdot)^2, \sin(\cdot), \times$), thereby projecting the linear combinations into symbolic outputs $y_i = f(h_i)$. In this paradigm, information propagates forward through dense linear mappings, with the goal of directly inducing a complete SR via weight sparsification. However, this architecture faces significant challenges in practice. Shallow networks often have limited nonlinear expressivity due to restricted operator combinations. Deep networks, on the other hand, are highly susceptible to severe gradient instabilities, such as exploding gradients, during training. Consequently, learning efficacy degrades, and the resulting symbolic expressions often become overly convoluted and lose physical meaning.

\begin{figure}[t]
    \centering
    \includegraphics[scale=0.45]{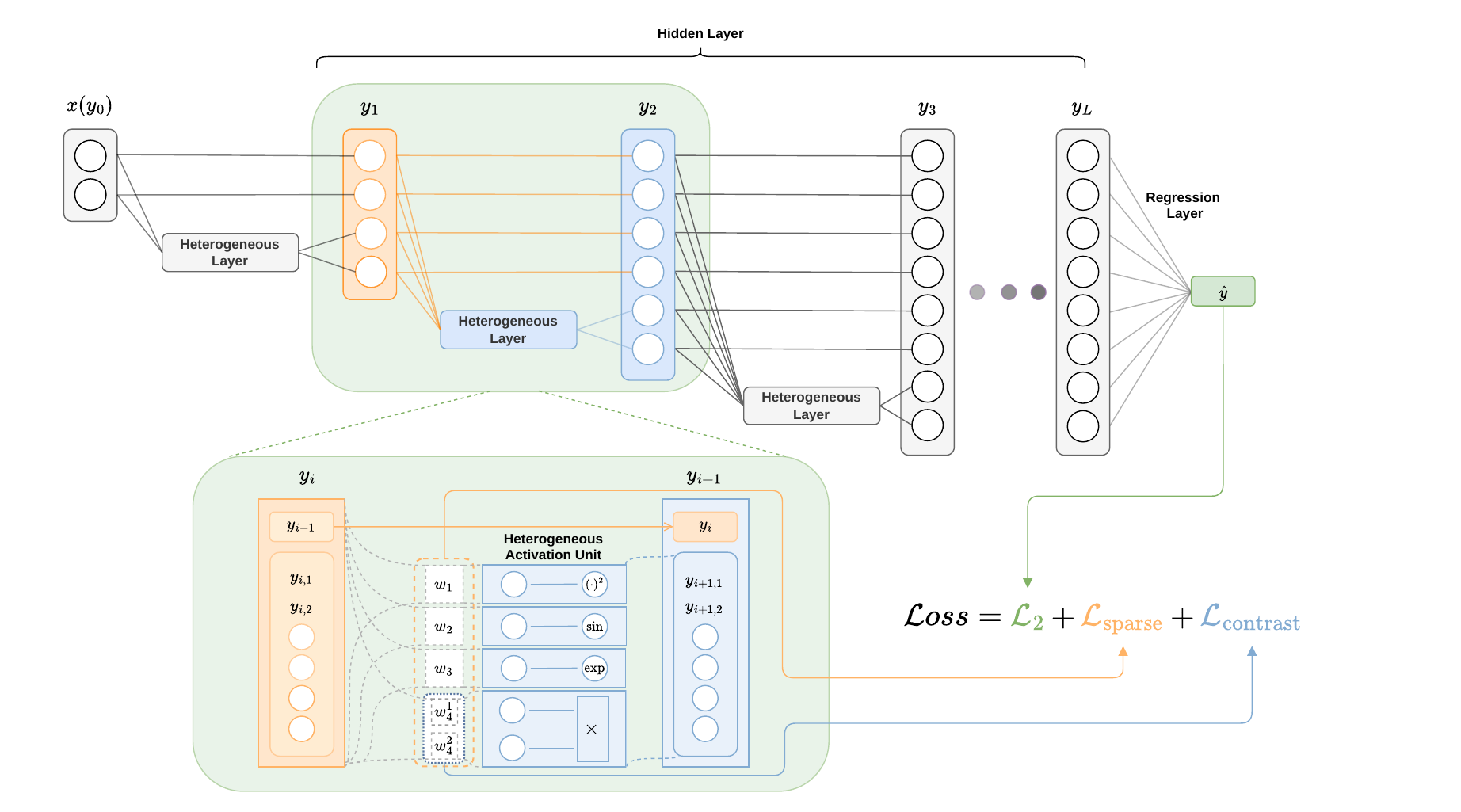}
    \captionsetup{font=small,labelfont=bf} 
    \caption{\textbf{Overall architecture of FMN}. 
    The architecture comprises an input layer, multiple heterogeneous hidden layers, and a regression layer. A heterogeneous activation unit (HAU) uses a specific mathematical primitive (e.g., $(\cdot)^2, \sin(\cdot), \cos(\cdot), \exp(\cdot), +, \times$) as its activation function and learns optimal weights to approximate the target equation's structural components. For simplicity, only four HAUs are presented. A heterogeneous layer is constructed by multiple HAUs arranged in parallel. Training is governed by a loss function including a joint objective integrating a standard squared loss ($\mathcal{L}_2$), a sparsity penalty ($\mathcal{L}_{\text{sparse}}$), and a contrastive loss ($\mathcal{L}_{\text{contrast}}$).}
    \label{FePySR}
\end{figure}

To overcome the limitations inherent to EQL, we modify the training objective and architectural topology. Instead of forcing the network to produce a single, monolithic symbolic expression, we leverage its nonlinear approximation capabilities to learn and reconstruct relevant features that encapsulate the data's intrinsic dynamics. To achieve this goal, we decompose the standard EQL layer into multiple parallel heterogeneous activation units (HAUs) and reorganize the information flow into a dense connectivity scheme similar to fully connected neural networks (Figure \ref{FePySR}). In this architecture, the output of each layer is the concatenation of the results of the current layer's HAU and all the original and derivative features from the preceding layers. This design transforms the process of learning a sparse path for a single formula into the continuous construction of a repository of relevant features through forward propagation. The FMN’s detailed workflow is elucidated below.

\begin{itemize}
  \item \textbf{HAU:} Unlike standard neural networks, which use a uniform activation function (e.g., ReLU), each HAU is designed to perform a specific mathematical operation chosen from a predefined library of candidates. Depending on the required arity, the computation is categorized into unary operators (e.g., $(\cdot)^2, \sin(\cdot)$) and binary operators (e.g., $+, \times$). Accordingly, it is expressed as:
  $$ y_{ij}=f_{ij}(W_{ij}y_{i-1}) \quad \text{or} \quad y_{ij}=f_{ij}(W_{ij_1}y_{i-1},W_{ij_2}y_{i-1}) $$
    where $y_{i-1}$ denotes the output from the $(i-1)$-th layer, and $W_{ij}, W_{ij_1}, W_{ij_2}$ are the trainable weight matrices for the unary and binary HAUs. In contrast to standard transformations ($Wx+b$), the HAU omits the bias term $b$. Here, $f_{ij}$ and $y_{ij}$ denote the specific mathematical operator and the resulting output of the $j$-th HAU in the $i$-th layer, respectively.

  \item \textbf{Heterogeneous Layer:} A heterogeneous layer is constructed by a parallel array of HAUs. To formulate the output of a hidden layer, we concatenate the outputs of the HAUs with the layer's input. The output of the $i$-th layer is expressed as:
  $$y_{i}=[y_{i-1},y_{i,1},\cdots,y_{i,j}]$$

  \item \textbf{Regression Layer:} The output layer of the network adopts a linear, fully connected structure. For an architecture comprising $L$ hidden layers, the final prediction is computed as:
  $$ \hat{y}=y_{L+1}=W_{L+1}y_{L} $$
\end{itemize}

In FMN, each HAU contains a primitive including $\{(\cdot)^2, \sin(\cdot), \cos(\cdot), \exp(\cdot), +, \times\}$. It can be expanded to include nonlinearities such as $\log(\cdot), |(\cdot)|$, and $\sqrt{(\cdot)}$ as dictated by task requirements. Furthermore, inspired by EQL, multiple HAUs are deployed within the same heterogeneous layer. This design reduces the model's sensitivity to stochasticity in parameter initialization, promoting efficient optimization and mitigating the network's tendency to converge to local optima. More details about the model’s configurations and hyperparameters are provided in Appendix \ref{appendixA}.

\subsubsection{FePySR’s Loss Function}
The loss function of FePySR has three terms accounting for a squared error, a sparsity penalty, and a contrastive loss, respectively. This loss function steers the model toward building a robust, highly relevant feature library while simultaneously maximizing approximation accuracy.  The loss function $\mathcal{L}$ is defined as:
$$
\mathcal{L} = \mathcal{L}_{2} + \mathcal{L}_{\text{sparse}} + \mathcal{L}_{\text{contrast}},
$$
where $\mathcal{L}_{2}$ quantifies the errors of FePySR’s predictions against the ground-truth equation’s solutions (i.e., the observational data). $\mathcal{L}_{\text{sparse}}$ is a term that enforces structural parsimony and enhances interpretability by pruning irrelevant features. $\mathcal{L}_{\text{contrast}}$ is a term to prevent the functional collapse of symmetric binary operators during the training. The individual terms are described in detail as follows:
\begin{itemize}
  \item \textbf{Squared Loss ($\mathcal{L}_2$):} This term measures the squared errors between the model’s prediction $\hat{y}$ and the target equation’s solutions $y$, formalized as:
  $$ \mathcal{L}_2 = (\hat{y} - y)^2 $$
  Minimizing this term during training can maximize the model’s prediction accuracy.

  \item \textbf{Sparsity Regularization ($\mathcal{L}_{\text{sparse}}$):} To conduct feature selection during training, we impose an $L_1$ sparsity penalty on the model weights, thereby suppressing the selection of irrelevant features. Unlike conventional regularization schemes, which require global network sparsity, this term is designed to reduce the weight of irrelevant features while maintaining valid ones. Specifically, the sparsity penalty of the weight matrix of the $j$-th HAU in the $i$-th layer as $W_{ij}$ is computed as:
  $$ \mathcal{L}_{\text{sparse}} = \lambda_1 \sum_{i,j} |W_{ij}|, $$
  where $\lambda_1=0.08$ is a regularization coefficient. This ensures that valid features drive the model training while suppressing the influence of irrelevant ones, thereby strengthening the reliability of feature selection without compromising the model’s prediction accuracy.

  \item \textbf{Contrastive Loss ($\mathcal{L}_{\text{contrast}}$):} It is known that that under unconstrained model training, the weights for the two branches of symmetric binary primitives (e.g., $+$, $\times$) often converge, leading to the failure of learning distinct features \cite{bansal2018can}. This pathological weight convergence causes severe functional collapse. For instance, an addition operator could degenerate into a simple scalar multiplier ($x + x = 2x$), and a multiplication operator could collapse into a squared function ($x \times x = x^2$). Such phenomenon limits the operator's degrees of freedom, impairing the model's ability to encompass a diverse, generalized nonlinear feature space. To avoid this, we introduce a contrastive penalty that minimizes the cosine similarity between the input branches, thereby preserving the specificity of binary operators. The corresponding loss is defined as:
  $$ \mathcal{L}_{\text{contrast}} = \lambda_2  \cos(W_1, W_2) = \lambda_2  \frac{W_1^\top W_2}{\|W_1\| \|W_2\|}, $$
   where $W_1$ and $W_2$ represent the weight vectors of the two input branches, and $\lambda_2=0.001$ governs the penalty magnitude. This effectively protects the model from operator degeneration, maintaining a rich and diverse set of features.
\end{itemize}

\subsubsection{Stabilizing Model Training}
A limitation of the EQL architecture is numerical instability, which intensifies with network depth. This instability arises from substituting standard activation functions with mathematical primitives that exhibit domain singularities, such as division, logarithms, and square roots. To overcome this challenge, we use a standard gradient descent optimizer coupled with unit-wise gradient normalization during the backpropagation phase. FMN contains a repository of primitives, such as $\sin(\cdot), \exp(\cdot), \log(\cdot)$, that can induce different gradient scales during training. Consequently, primitives that generate large gradients dominate training, causing operators with smaller gradients to stagnate. Gradient normalization is therefore introduced during training to improve numerical stability, neutralizing gradient explosions and oscillations triggered by division or logarithmic operators near their singularities. This normalization strategy also ensures that all primitives receive comparable update opportunities within a fixed training period (e.g., $100$ epochs), thereby accelerating an unbiased feature selection process.

Although gradient normalization improves training robustness, scaling to deeper networks introduces an additional challenge: the superposition effect of nested exponential operations (($\exp(\cdot)$). During forward propagation, cascading activations through consecutive exponential units trigger exponential growth, leading to numerical overflow in deeper layers. In particular, a two-layer network remains numerically stable, whereas a three-layer topology overflows within a few training epochs and a four-layer variant encounters numerical exceptions during the initial forward computation. To address this, a mask-based structural constraint is introduced for exponential HAUs. This mechanism sets the corresponding weights to zero when the input to an exponential HAU originates directly from the output of a preceding exponential operation, truncating the propagation of numerical overflow through consecutive exponential units. Importantly, this strategy does not eliminate exponential operations, but it removes the exponential superposition structures. This constraint eliminates numerical overflow in deep layers to improve the stability of model training.

\subsubsection{Model Training and Feature Extraction}
\begin{figure}[t]
    \centering
    \includegraphics[scale=0.5]{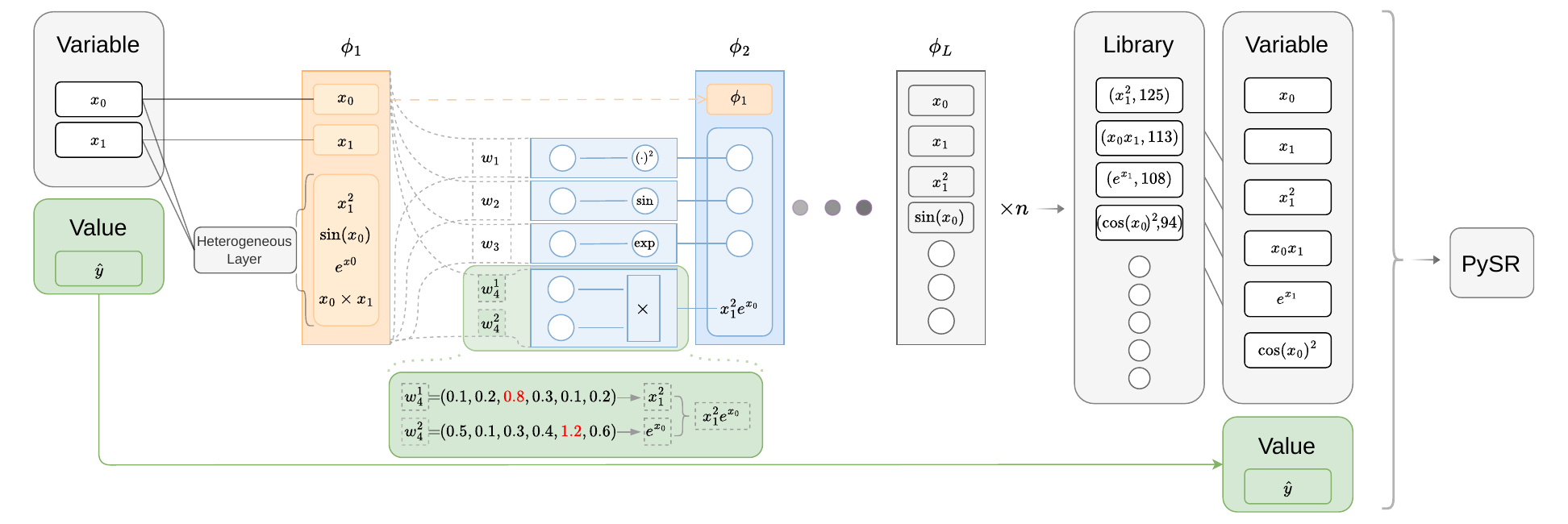}
    \captionsetup{font=small,labelfont=bf} 
    \caption{\textbf{Illustration of the feature extraction process in FMN}. Within heterogeneous layers, the feature corresponding to the term with the maximum absolute weight is first extracted. Subsequently, predefined formulas are applied to the extracted features to obtain new features, constructing a feature library. Finally, the feature library together with the target equation’s observational data are fed into PySR for SR.
    }
    \label{FePySR_tar}
\end{figure}
Given the observational data (i.e., a target equation’s solution), we optimize the FMN by minimizing the loss function defined as:
$$W^* = \arg \min_{W} \mathcal{L}=\arg \min_{W}( \mathcal{L}_{2} + \mathcal{L}_{\text{sparse}} + \mathcal{L}_{\text{contrast}}  )$$
After training (e.g., $100$ epochs), the model weights $W^*$ are fixed. Starting with the initial heterogeneous layer, we derive a candidate feature for each HAU by applying its assigned mathematical primitive to the input feature with the highest absolute weight. As shown in Figure \ref{FePySR_tar}, for the $j$-th HAU in the $i$-th layer, the feature extracted is defined as follows:
\begin{align*}
&h^* = \argmax_{h \in \{1,2,\dots,n_{l}\}} |W_{i,j}^{h}|, \quad  W_{i,j} = [W_{i,j}^{1}, W_{i,j}^{2}, \dots, W_{i,j}^{n_{l}}] \\
&\phi_{i,j}   = f_{i,j}(\Phi_{i-1}^{h^*}) ,  \quad\quad\quad\; \Phi_{i-1}=[\Phi_{i-2},\phi_{i-1,1},\phi_{i-1,2},\cdots,\phi_{i-1,n}]
\end{align*}
where $W_{i,j}$ denotes the weight vector of the $j$-th HAU in the $i$-th layer, $h^*$ is the index of the dominant weight, $\Phi_{i-1}$ denotes the cumulative feature representation up to layer $i-1$, and $f_{i,j}$ represents the activation function of the HAU. Following the dense-connection paradigm, the cumulative feature set for the $i$-th layer is constructed by concatenating the features accumulated up to the $(i-1)$-th layer with the features extracted in the $i$-th layer:
$$\Phi_i=[\Phi_{i-1},\phi_{i,1},\phi_{i,2},\cdots,\phi_{i,n}]$$
Iterating this protocol through to the last layer yields a candidate feature set for a single training run.

We build a feature library using candidate features from multiple independent training runs. Then, we use the SymPy library to simplify all extracted features algebraically into their most parsimonious canonical forms . This allows us to accurately count the number of times structurally identical features occur. 

We then rank the unique features by their cumulative frequencies, identify the top-ranked features, and evaluate their numerical outputs over the input domain. Finally, we concatenate these extracted features with the observational data to construct an enriched representation, thereby completing the first-stage feature extraction (Algorithm \ref{alg:fmn_extraction}). In the second stage, we deploy PySR to perform structural search on this FMN-derived enriched representation, ultimately identifying the target symbolic expression (see Appendix \ref{FePySR-algorithm}).

\begin{algorithm}
\caption{FMN }
\label{alg:fmn_extraction}
\begin{algorithmic}[1]
    \Require Original dataset $\mathcal{D} = \{x, y\}$, Network depth $L$, Epochs $E$
    \Ensure Set of generated features $\Phi$
    \State Initialize network weights $w$                       \graycomment{Phase 1: Tranining FMN}
    \For{$e = 1$ \textbf{to} $E$}
        \State $\mathcal{L} \gets \mathcal{L}_2 + \mathcal{L}_{sparse} + \mathcal{L}_{contrast}$
        \State Update $W \gets W - \eta \frac{\nabla_W \mathcal{L}}{\|\nabla_W \mathcal{L}\|_2 + \epsilon}$  \graycomment{Minimize combined loss}
    \EndFor
    \State Fix optimal parameters $W$
    \State \relax
    \State Initialize feature pool $\Phi_0 \gets \emptyset$   \graycomment{Phase 2: Layer-wise Feature Generation}
    \For{layer $i = 1$ \textbf{to} $L$}
        \State $\mathcal{S}_i \gets \emptyset$ \graycomment{Temp storage for layer $i$}
        \For{unit $j = 1$ \textbf{to} $N_{units}$}
            \State Get weights $W_{i,j} = [W_{i,j}^1, \dots, W_{i,j}^{n_l}]$
            \State $h^* \gets \arg\max_{k} |W_{i,j}^k|$ \graycomment{Select most influential input index}
            \State $\phi_{input} \gets \Phi_{i-1}^{h^*}$ \graycomment{Retrieve feature at index $h^*$}
            \State $\phi_{i,j} \gets f_{i,j}(\phi_{input})$ \graycomment{Apply activation function}
            \State $\mathcal{S}_i \gets \mathcal{S}_i \cup \{ \phi_{i,j} \}$
        \EndFor
        \State $\Phi_i \gets [\Phi_{i-1}, \mathcal{S}_i]$ \graycomment{Residual connection: stack features}
    \EndFor
    
    \State \Return $\Phi(\Phi_L)$
\end{algorithmic}
\end{algorithm}

\newpage

\section{Results}
To comprehensively evaluate the performance of FePySR, we first conduct a systematic benchmarking against SOTA models using five SR datasets including Nguyen \cite{nguyen2014semantically}, Livermore \cite{Mundhenk2021}, Jin \cite{li2025mmsr}, Constant \cite{li2025mmsr}, and R \cite{krawiec2013approximating}. The results demonstrate that FePySR accurately recovers the majority of the equations in the datasets (Sections 3.1). 

To further evaluate FePySR, an LLM-generated benchmark containing 75 complex equations is introduced (Section 3.2). Performance is assessed along two primary metrics. (I) \textbf{Symbolic Recovery Accuracy:} the recovery rate of FePySR is compared against two baseline models, PySR and DSO. (II) \textbf{Approximation Fidelity:} the minimum mean squared error (MSE) achieved by FePySR is compared against that of PySR under equivalent computational settings. FePySR outperforms both baselines on both metrics. Further analysis demonstrates the effectiveness of FePySR in identifying relevant features for SR and its robustness to noise.

Finally, to assess the practical applicability of FePySR, we apply it to ordinary differential equations (ODEs) governing the dynamics of biological regulatory networks \cite{tyson2010functional} (Section 3.3). FePySR recovers the underlying mathematical structures responsible for bistability in gene expression directly from observational data, achieving a success rate of {24\%} where PySR recovers none. This result demonstrates the capacity of FePySR to identify governing equations in complex, real-world biological systems.

\subsection{Standard Benchmark Evaluations}
We compare FePySR wtih SOTA models using five SR datasets such as Nguyen (and its variants), Livermore, Jin, Constant, and R. We evaluate their performance using symbolic recovery accuracy, defined as the proportion of trails in which the discovered expression is mathematically equivalent to the ground-truth equation (e.g., $(x+1)^2$ is considered semantically identical to $x^2+2x+1$). To mitigate the variance induced by stochastic optimization, we perform 100 independent trials for each mathematical expression. The results show that, compared to NGGP, DSR, PySR, PQT, and Eureqa using the Nguyen equations, FePySR achieves a 6.4\%-23.7\% improvement in the overall recovery rate (Table \ref{nguyen_benchmarks}).

Interestingly, Nguyen-12 and Nguyen-2 have similar structures, but the SOTA models perform differently. While almost all SOTA models achieve a 100\% recovery rate on Nguyen-2, they fail completely on Nguyen-12, achieving a 0\% recovery rate. Further analysis of the equations yields two explanations. First, Nguyen-12 is a bivariate equation, which has a larger search space than the univariate equation Nguyen-2. Second, an examination of the search trajectories shows that the SOTA models converge on the nested, factorized form of Nguyen-2 (i.e., $x(x(x(x+1)+1)+1)$), which has only 13 symbolic tokens. In contrast, the factorized form of Nguyen-12 is $x \cdot x \cdot x \cdot (x-1) + 0.5 \cdot y \cdot y - y$, which entails 17 tokens. This elevated structural complexity results in a bottleneck for symbolic searching by the SOTA models. In contrast, FePySR’s feature extraction stage successfully identifies $x^2$ and $y^2$ as valid nonlinear features. This allows Nguyen-12 to be condensed into $x^2(x^2-x) + 0.5y^2 - y$, reducing the symbolic tokens to 11. Consequently, although the input features increase during the symbolic search with PySR, the reduction in the depth of the symbol search tree fundamentally simplifies the search space. This ultimately enables the exact recovery of Nguyen-12.

\begin{table*}[!ht]
\small
\centering
\captionsetup{font=small,labelfont=bf}
\caption{\textbf{Recovery rate comparison between FePySR and other models on the Nguyen dataset.} We test each equation 100 times for FePySR and PySR. The other models’ statistics are taken from their papers.}
\label{nguyen_benchmarks}
\begin{tabular}{lccccccc}
\toprule
\textbf{Benchmark} & \textbf{Expression}&\textbf{FePySR}& \textbf{NGGP} & \textbf{DSR} & \textbf{PySR} & \textbf{PQT} & \textbf{Eureqa} \\
\midrule
Nguyen-1 & $x^3 \!+\! x^2 \!+\! x$ &100\%& 100\% & 100\% & 100\% & 100\% & 100\%  \\
Nguyen-2 & $x^4 \!+\! x^3 \!+\! x^2 \!+\! x$ &100\%& 100\% & 100\% & 100\% & 99\% & 100\%  \\
Nguyen-3 & $x^5 \!+\! x^4 \!+\! x^3 \!+\! x^2 \!+\! x$ &78\%& 100\% & 100\% & 52\% & 86\% & {95\%}  \\
Nguyen-4 & $x^6 \!+\! x^5 \!+\! x^4 \!+\! x^3 \!+\! x^2 \!+\! x$ &100\% & 100\% & 100\% & 62\% & 93\% & 70\%  \\
Nguyen-5 & $\sin(x^2)\cos(x) \!-\! 1$&100\%&100\% & 72\% & 92\% & 73\% & 73\%  \\
Nguyen-6 & $\sin(x) \!+\! \sin(x\!+\!x^2)$&100\%&100\% & 100\% & 100\% & 98\% & 100\%  \\
Nguyen-7 & $\log(x\!+\!1) \!+\! \log(x^2\!+\!1)$&100\%&97\% & 35\% & 19\% & 41\% & 85\%  \\
Nguyen-8 & $\sqrt{x}$&100\%&100\% & 96\% & 100\% & 21\% & 0\%  \\
Nguyen-9 & $\sin(x) \!+\! \sin(y^2)$ &100\%& 100\% & 100\% & 100\% & 100\% & 100\%  \\
Nguyen-10 & $2\sin(x)\cos(y)$ &100\%& 100\% & 100\% & 100\% & 91\% & {64\%} \\
Nguyen-11 & $x^y$ &100\%& 100\% & 100\% & 99\% & 100\% & 100\%  \\
Nguyen-12 & $x^4 \!-\! x^3 \!+\! \frac{1}{2}y^2 \!-\! y$ &95\%& 0\% & 0\% & 0\% & 0\% & 0\%  \\
\midrule
\multicolumn{2}{l}{Average}&\textbf{97.8\%}& 91.4\% & 83.6\% &  77.0\%  & 75.2\% & 73.9\%  \\
\bottomrule
\end{tabular}
\end{table*}

Furthermore, we compare FePySR with NGGP \cite{Mundhenk2021}, PySR \cite{Cranmer2023PySR}, GEGL \cite{tyson2010functional}, and DSR \cite{Petersen2021a} using the Nguyen, R, and Livermore datasets. It is worth noting that both NGGP and GEGL are advanced models with hybrid recurrent neural networks and GP. Our results show that, on average, FePySR outperforms these models by increasing the recovery rate by between 6.19\% and 50.70\% across all three datasets (Table \ref{recovery_rate_comparison_final}). Moreover, FePySR outperforms PySR on four other datasets and increases the recovery rate by an average of 19.18\% (Table \ref{recovery_rate_comparison_final}).

\begin{table}[h!]
    \small
    \centering
    \captionsetup{font=small,labelfont=bf}
    \caption{\textbf{Recovery rate comparison between FePySR and other models on the Nguyen, Livermore, and R datasets.}. We test each equation 100 times for FePySR and PySR. The other models’ statistics are taken from their papers. Recovery rates for individual equations are reported in Table \ref{tab:nguyen_benchmarks} of Appendix \ref{appendixB1}.}
    \label{recovery_rate_comparison_final}
    \begin{tabular}{lccc|c}
        & \multicolumn{4}{c}{Recovery rate (\%)} \\
           & Nguyen & Livermore& R & ALL \\
        \cmidrule(lr){2-5}
        \textbf{FePySR} & \textbf{97.75} & \textbf{77.45} & \textbf{41.33} & \textbf{81.11} \\
        NGGP \cite{Mundhenk2021} & 92.33 & 71.09& 33.33& 74.92   \\
        PySR \cite{Cranmer2023PySR} &77.00   &67.36 &34.67 &70.46  \\
        GEGL \cite{ahn2020guiding} & 86.00  & 56.36 & 33.33 & 64.11 \\
        DSR \cite{Petersen2021a} & 45.19 & 83.58 & 0.00 & 30.41 \\
        \bottomrule
    \end{tabular}
\end{table}

\begin{table}[h!]
    \small
    \centering
    \captionsetup{font=small,labelfont=bf}
    \caption{\textbf{Recovery rate comparison between FePySR and PySR on the Nguyen’, Jin, Constant datasets.} We test each equation 100 times for both methods. Recovery rates for individual equations are reported in Table \ref{tab:nguyen_benchmarks} of Appendix \ref{appendixB1}.}

    \label{recovery_rate_comparison_final2}
    \begin{tabular}{lccc|c}
        & \multicolumn{4}{c}{Recovery rate (\%)} \\
          &Nguyen'  & Jin & Costant & ALL \\
        \cmidrule(lr){2-5}
        \textbf{FePySR} & \textbf{89.67} & \textbf{87.00} & \textbf{91.38} & \textbf{89.55} \\
        PySR \cite{Cranmer2023PySR} &76.33  &76.17  &88.50  &81.15 \\
        \bottomrule
    \end{tabular}
\end{table}

Overall, FePySR achieves high equation recovery rates across standard benchmarks, showing top performance on multiple challenging test cases. This suggests that constructing a refined feature space prior to symbolic search is an effective strategy for reducing structural complexity and improving recovery of complex mathematical expressions.Comprehensive independent results, including evaluations on additional datasets, are detailed in Appendix \ref{appendixB1}.

\subsection{Model Evaluation on Synthesized Benchmarks}
We present a synthetic dataset consisting of complex expressions generated by LLMs. The dataset spans a broad spectrum of equations to emulate the intricate modeling challenges in physics and mathematics. The dataset includes equations ranging from univariate to quadvariate. Importantly, it goes beyond standard polynomial forms by incorporating a wide range of nonlinear dynamics governed by trigonometric, exponential, logarithmic, and hyperbolic functions. 

To rigorously test the capacity of FePySR, we increase the structural complexity of the target equations. We achieve this by incorporating deep operator nesting (e.g., $\sin(\cos(x) + \tan(yz))$), highly convoluted rational fractions (e.g., $\frac{x^2 - y^2}{\sin^2(\pi x)-\cos^2(\pi y)}$), and discontinuous piecewise functions. In total, we generate 75 complex equations. We divide these equations into two groups based on recovery outcomes: recoverable (Table \ref{tab:Recover_function_features_multirow_9}) and unrecoverable (Table \ref{more_complex_data}). For the 36 recoverable equations, we use the recovery rate to compare FePySR with PySR and DSO. Due to their ultra-complex structure, the unrecoverable 39 equations fail all models. Thus, we compare their minimum MSE over 100 runs. By comparing the minimum MSE of FePySR and PySR, we quantitatively assess their proficiency in navigating highly non-convex search spaces and converging on optimal functional approximations.

\subsubsection{Recovery Rate Analysis}
Here, we present the comparison results for the 36 recoverable equations using FePySR, PySR, and DSO (Table \ref{tab:Recover_function_features_multirow_9}). With an average recovery rate of 85.78\%, FePySR significantly outperforms baseline models, including PySR (20.24\%) and DSO (36.11\%) (Table \ref{comparison_vertical}). Notably, FePySR's superior performance is largely due to high recovery rates on some equations. In these equations, FePySR effectively learns essential nonlinear features that correspond to the intrinsic structure of the target equations. This feature extraction function reduces the time for the subsequent feature searching and also enables near 100\% recovery rates in some cases where PySR and DSO fail (Table \ref{full_comparison_condensed}).

\begingroup 
\footnotesize 
\setlength{\tabcolsep}{3pt} 

\begin{longtable}{cccccccc}
    \captionsetup{font=small,labelfont=bf}
    \caption{\textbf{Comparison of FePySR, PySR, and DSO on recoverable equations.} We test each equation 100 times.} \label{comparison_vertical} \\
    \toprule
    \footnotesize 
    {\textbf{Expression}} & \textbf{FePySR}  & \textbf{PySR} & \textbf{DSO} &{\textbf{Expression}} & \textbf{FePySR}  & \textbf{PySR} & \textbf{DSO} \\
    \midrule
    \endfirsthead 

    \toprule
    \footnotesize 
    {\textbf{Expression}}& \textbf{FePySR}  & \textbf{PySR} & \textbf{DSO}&{\textbf{Expression}} & \textbf{FePySR}  & \textbf{PySR} & \textbf{DSO} \\
    \midrule
    \endhead 

    \bottomrule
    \endlastfoot 

    {$x^6 \!+\! x^5 \!+\! x^4 \!+\! x^3 \!+\! x^2 \!+\! x$}&   100\% & 76.8\% & 100\% & {$x^6 \!+\!2 x^5 \!+\!3 x^4 \!+\!4x^3 \!+\! x^2 \!+\! x$}& 98\%& 0\% & 100\% \\
    \cline{2-4}\cline{6-8} 
    {$x^6 \!+\!0.5 x^5 \!+\!0.7 x^4 \!+\!3x^3 \!+\! 5x^2 \!+\! x$}&  14\% & 0\% & 100\% & $\sum_{i=1}^{9}x^{i}$ &  100\% & 0\% & 100\% \\
    \cline{2-4} \cline{6-8}
    {$\sin(x^2)\!\cos(x)\!+\!  \sin(x) \!+\! \sin(x^4 \!\!+\! x^2)$}&  98\% & 57\% & 0\% &{$\frac{xy}{\exp(x)\!+\!\sin(y)^2}$}& 100\% & 0\% & 0\% \\
    \cline{2-4}  \cline{6-8}              
    {$2x^4\!+\!2y^4\!\!-\!\!6x^2y^2$}&   100\% & 32\% &100\% &{$x^5\!+\!3x^3y^2\!\!-\!\!x^2y^3\!+\!y^5$}&  49\% & 0\% & 100\% \\
    \cline{2-4}\cline{6-8} 
    {$e^{x^2\!+\!\sin(y)}$}&  100\% & 100\% & 100\%  &{$x^3\!+\!x^2\!+\!x\!+\!\sin(x)\!+\!\sin(y^2)$}&   100\% & 100\%  &100\% \\
    \cline{2-4}\cline{6-8} 
    {$x^4y \!-\! x^3 \!+\! 0.5y^2\cos(x)\!-\!x$}&  100\%  & 87\% & 0\% &{$x^4y \!-\! x^3 \!+\! 0.5y^2\sin(x)\!-\!x$}& 100\%  & 0\% & 0\% \\
    \cline{2-4}   \cline{6-8}               
    { $ xytanh(x\!+\!y)$}&  99\%  & 0\% &  0\% & {$\cos(y^2) \sin(x) \!-\! 1 \!+\! \sqrt{x^2 \!+\! y^2 \!+\! 1}$}&    94\% & 0\%  & 0\% \\
    \cline{2-4}\cline{6-8} 
    {$\cos(xy) \cos(y) \!-\! 1 \!+\! \sqrt{x^2 \!+\! y^2 \!+\! 1}$}& 39\%  & 0\% & 0\%  &{$\frac{\exp^{(1\!+\!x)}(1 \!-\! x) \!-\! \exp^{y}x}{\exp^{(1\!+\! x)} \!+\!\exp^{y}}$}&  99\% & 9\% &0\%  \\
    \cline{2-4}\cline{6-8} 
    {$\frac{e^{x}\sin(y)\!-\!e^{y}\cos(x)}{xy}$}&  7\%  & 0\% & 0\%  &{$\frac{e^{x}\cos(y)\!-\!e^{y}\sin(x)}{xy}$}&  99\%  & 0\% & 0\%  \\
    \cline{2-4}\cline{6-8} 
    {$e^{\!-\!0.5(\sin^{2}(x) \!+\! \cos^{2}(y))}\cos(xy)$}& 24\%  & 18\% & 0\%  &{$\sin\left(x \!+\! e^{\!-\!y^{2}}\right) \!-\! \cos\left(y \!-\! e^{\!-\!x^{2}}\right)$}&  39\%  & 12\% &  0\% \\
    \cline{2-4}\cline{6-8} 
    {$x^4 \!+\! y^2z^2 \!-\! x^2z^2 \!+\! y^4$}& 100\%  & 0\% & 100\%  &{$x^5 \!-\! y^4z \!+\! z^3x^2 \!-\! xyz \!-\! x \!+\! y$}&  100\%  & 0\% & 100\%  \\
    \cline{2-4}\cline{6-8} 
    {$e^{sin(x)}y^3\!+\!\cos(e^{z})$}&  100\%  & 100\% & 0\%  &{$2(x^2\!+\!y^2)\sin(z)\!+\!x\exp(x\!+\!y)$}&  100\%  & 0\% &0\%  \\
    \cline{2-4}    \cline{6-8} 
    {$0.5\sin(x\!+\!y)z^4\!+\!x^2y^3z$}& 99\%  & 80\% & 0\% &{$\sqrt{z^4\!+\!1}\cos(y\!+\!e^x)$}&   81\% & 7\%  &  0\% \\
    \cline{2-4}\cline{6-8} 
    {$x^4y \!+\! y^2z^3\sin(z) \!+\!xz^4$}&   82\%  & 0\% & 0\%  &{$x^3y^5\!+\!\cos(z^2)*y\!-\!e^{xz}$}&  100\%  & 0\% & 0\%  \\
    \cline{2-4}\cline{6-8} 
    {$z^4\cos(x)e^{\sin(y)}\!+\!xe^{y\!+\!z}$}&  88\%  & 34\% & 0\%  &{$ 0.5\sin(x)z^4\!+\!x^2y^3z$}&  55\%  & 0\% &  0\% \\
    \cline{2-4}\cline{6-8} 
    {$0.9(x^2\!+\!y^2)\sin(3z\!+\!2x)\!+\!xe^{x\!-\!y}$}& 85\%  &0\%  & 0\%  &{$(z^2\!+\!1)/(e^x\!+\!\cos(y)^2)$}&  100\%  & 14\% & 0\% \\
    \cline{2-4}      \cline{6-8} 
    {$(z^2\!+\!1)/(e^y\!+\!\sin(x)^2)$}&  96\%  & 2\% & 0\% &{$x^3y \!-\! y^3z \!+\! z^3h \!-\! h^3x \!+\! xyzh$}&   62\% &0\%  & 100\%  \\
    \cline{2-4}\cline{6-8} 
    {$x^4 \!+\! y^4 \!-\! z^4 \!-\! h^4 \!+\! x^2z^2$}&  81\%  & 0\% & 100\%  &{$x^5 \!-\! y^4 \!+\! z^3 \!-\! h^2 \!+\! xyzh$}&   100\% & 0\% &  100\% \\
    \cline{6-8}
    Average&    &  &   & &   \textbf{85.78}\% & 20.24\% &  36.11\% \\
\end{longtable}
\endgroup

\subsubsection{Sensitivity to Feature Quantities}

To investigate FePySR's sensitivity to the extracted feature space, we vary the number of input features from four to nine and use them for equation recovery. The results show that increasing the number of input features does not necessarily result in faster convergence of the loss trajectories (Figure \ref{loss_convergence}) or fewer training epochs required for successful recovery (Figure \ref{box}). This can be explained by the varying number of valid features identified by FePySR (i.e., features that equal constituent terms in an equation) in the input features. For example, Recover-15 exhibits faster loss convergence and shorter SR runtime as the number of input features increases to six. This improvement is attributable to the additional valid feature introduced by the expanded feature set, which reduces the structural complexity of the downstream search. However, if the number of the identified valid features remains unchanged, then increasing the number of input features increase recovery time with an exception equation (Table \ref{tab:full_comparison}). 

\begin{figure}[htbp]
    \centering
    \captionsetup{font=small,labelfont=bf} 
    \begin{subfigure}[b]{0.53\textwidth}
        \includegraphics[width=\textwidth]{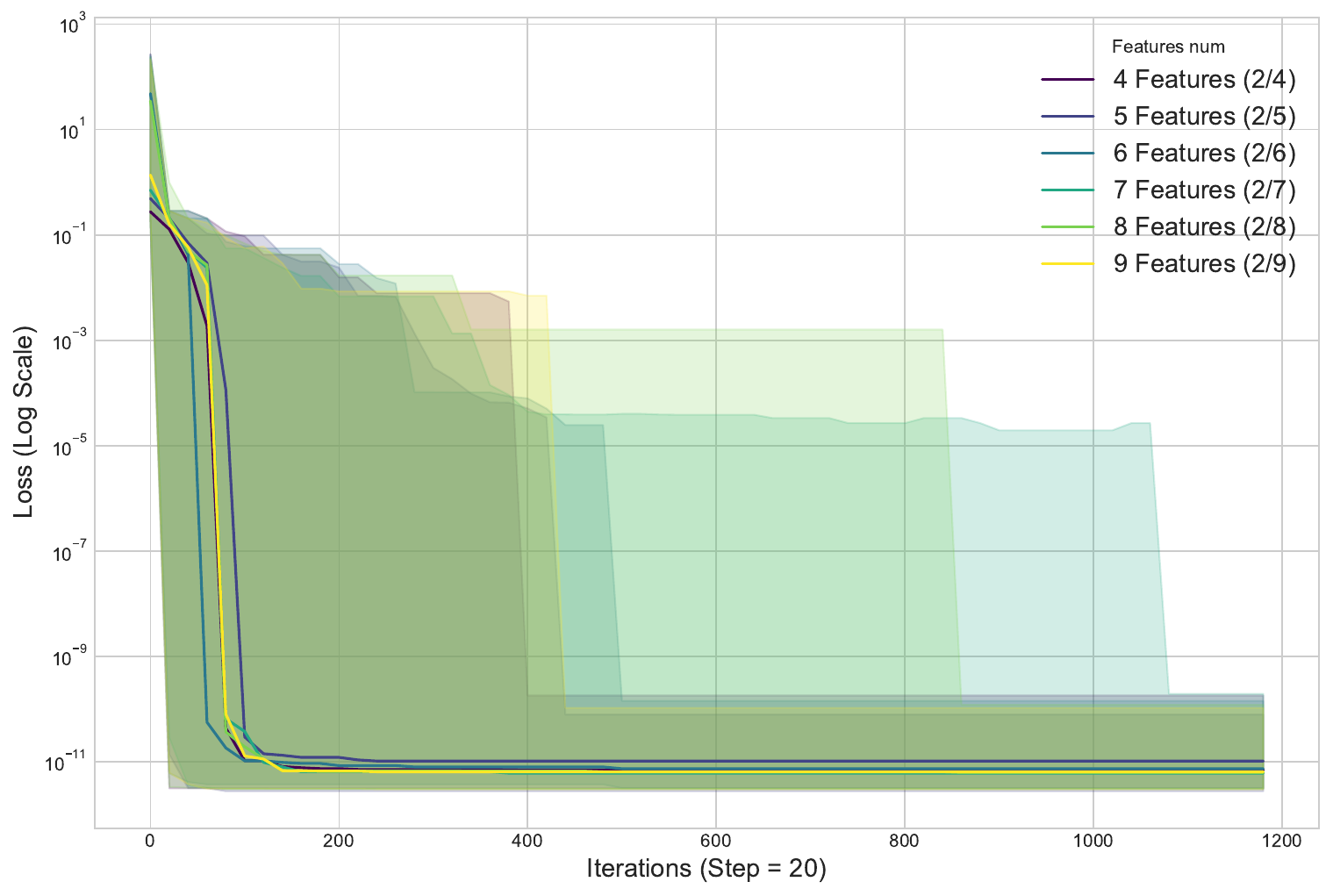}
        \caption{}
        \label{fig:box1}
    \end{subfigure}
    \hspace{0.5cm}
    \begin{subfigure}[b]{0.378\textwidth}
        \includegraphics[width=\textwidth]{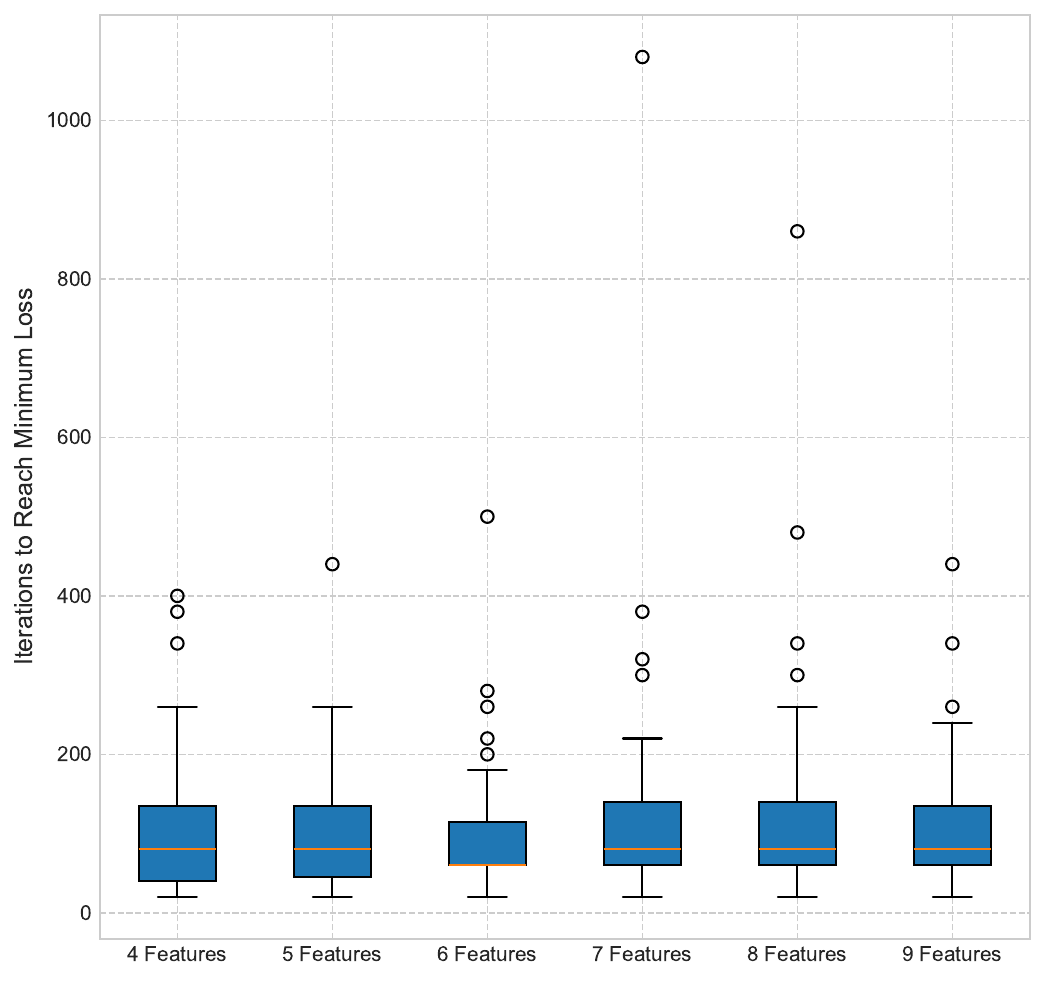}
        \caption{}
        \label{fig:box2}
    \end{subfigure}

    \caption{\textbf{FePySR performance on the Recover-10 equation across different feature counts.} (a) Model loss over training epochs for feature counts ranging from four to nine selected for SR. The x-axis represents the training epoch, with the value in parentheses indicating the recording interval. (b) Distributions of training epochs required to successfully recover the target equation across different feature counts.The x-axis represents the number of selected top-n features. The y-axis represents the number of epochs required for the model to converge to the minimum MSE. Boxes represent the interquartile range (IQR; 25th, 50th, and 75th percentiles), orange lines denote the median, and circles indicate outliers.}
    \label{fig:boxplot_matrix_4x3}
\end{figure}

Particularly, the equation $x^3+x^2+x+\sin(x)+\sin(y^2)$ (Recover-10) defies this expectation. With five input features, FePySR achieves a 98\% recovery rate and an average runtime of 39.8 seconds. Increasing the number of input features to six, despite the absence of additional valid features, increases the model’s success rate to 100\% and reduces the average runtime to 30.9 seconds. The runtime is even shorter than the model that uses four input features. This is further corroborated by the model’s faster loss trajectory convergence (Figure \ref{fig:box1}) and the smaller number of epochs required to recover the equation when the input feature is six (Figure \ref{fig:box2}). To better understand this phenomenon, we analyze PySR’s selection process to identify the target equation, which is the second stage of FePySR. We find that, when the model introduces the sixth input feature $\sin(y)$, it is frequently selected by PySR for SR. Although $\sin(y)$ is an invalid feature and it cannot be used to reduce the depth of the expression tree of the Recover-10 equation, its structure is similar to the valid feature $\sin(y^2)$, suggesting that it has a guidance function for SR at the second stage. Specifically, PySR uses a Pareto front that evaluates the MSE and complexity of candidate equations over iterations to reproduce the input data (i.e., the target equation’s solutions). Since $\sin(y)$ is a partial approximation of the value of $\sin(y^2)$, candidate equations that include $\sin(y)$ (e.g., $x^3+x^2+x+\sin(x)+\sin(y)$) yield significantly smaller MSEs than those that contain other structurally irrelevant features (e.g., $e^x$ or $\log(y)$). As PySR uses GP to process candidate equations, those producing smaller MSEs are more likely to be retained during the evolution (i.e., iteration). This guides PySR to search for a space containing relevant features (e.g., $\sin(\cdot)$). Once $\sin(\cdot)$ is found within PySR’s selection process, subsequent PySR operations (e.g., random substitution and exchanging of elements in candidate equations) can drive the search process to efficiently identify the target equation containing $\sin(y^2)$.

Taken together, structurally similar but invalid features can effectively guide the equation search process in PySR toward those with valid features. This shows that identifying relevant features in FePySR’s first stage can improve the model’s performance.

\subsubsection{Noise Sensitivity Evaluation}
To evaluate the generalization capability and robustness of FePySR, we design a series of noise-sensitivity experiments. Following the experimental protocol introduced in DSR \cite{Petersen2021a}, we add independent Gaussian noise $\mathcal{N}(0, \sigma)$ into the dependent variable $y$ (i.e., the equation’s solutions). The standard deviation $\sigma$ is scaled proportionally to the root mean square (RMS) of the training targets $y$, formulated as $\sigma = \alpha \cdot \text{RMS}(y)$, where $\alpha$ is a coefficient that determines noise’s scale factor.

To systematically assess the impact of noise on FMN’s performance on feature extract (i.e., FePySR’s first stage), we evaluate each dataset across multiple noise levels. Specifically, we set seven noise scale factors: $\alpha \in \{0,\allowbreak 0.02,\allowbreak 0.04,\allowbreak 0.06,\allowbreak 0.08,\allowbreak 0.1,\allowbreak 0.2\}$. For each noise level, we run FMN 256 times utilizing 16 CPU cores for parallel computing. In each run, we find the top 10 most frequently extracted features and identify the valid ones by comparing them against the ground-truth equation. We use three metrics to quantify the robustness of FMN: EFR, which represents the model’s efficiency in identifying valid features; DCG-1, which represents the ranking of the identified valid features; and DCG-2, which represents the confidence of the identified valid features.

\begin{itemize}
    \item \textbf{EFR}: This metric measures the frequency of valid features among the top 10 results, reflecting FMN's ability to filter out irrelevant, noisy terms. A higher EFR value indicates FMN's stronger capability to identify valid features.
    \item \textbf{DCG-1}: Inspired by the widely used discounted cumulative gain (DCG) metric in information retrieval \cite{jarvelin2002cumulated}, this metric evaluates the ranking of valid features. It operates under the assumption that all valid features are equally important, regardless of their extraction frequency. Thus, a higher rank yields a higher score.
    \item \textbf{DCG-2}: This metric extends DCG-1 by incorporating frequency weighting. It posits that the greater the frequency of a valid feature, the greater the model's confidence. Therefore, valid features that are frequently found and highly ranked receive greater weight.
\end{itemize}

The metrics are defined as follows:

\begin{align*}
&\text{EFR}  = \frac{\sum_{i=1}^{10} \text{Freq}(i) \cdot \mathbb{I}(i \in \text{Valid})}{\sum_{j=1}^{10} \text{Freq}(j)}, \\
&\text{DCG-1} = \sum_{i=1}^{10} \frac{\mathbb{I}(i \in \text{Valid})}{\log_2(i + 1)}, \\
&\text{DCG-2} = \sum_{i=1}^{10} \frac{\text{Freq}(i) \cdot \mathbb{I}(i \in \text{Valid})}{\log_2(i + 1)},
\end{align*}

where $\text{Freq}(i)$ denotes the frequency of the feature ranked at position $i$, $\text{Valid}$ represents the set of valid features, and $\mathbb{I}(\cdot)$ is the indicator function, which equals $1$ if the feature belongs to the valid feature set and $0$ otherwise. Detailed calculations of the metrics under each noise level are provided in Appendix \ref{appendixB2.3}.

\begin{figure}[t]
    \centering
    \includegraphics[scale=0.34]{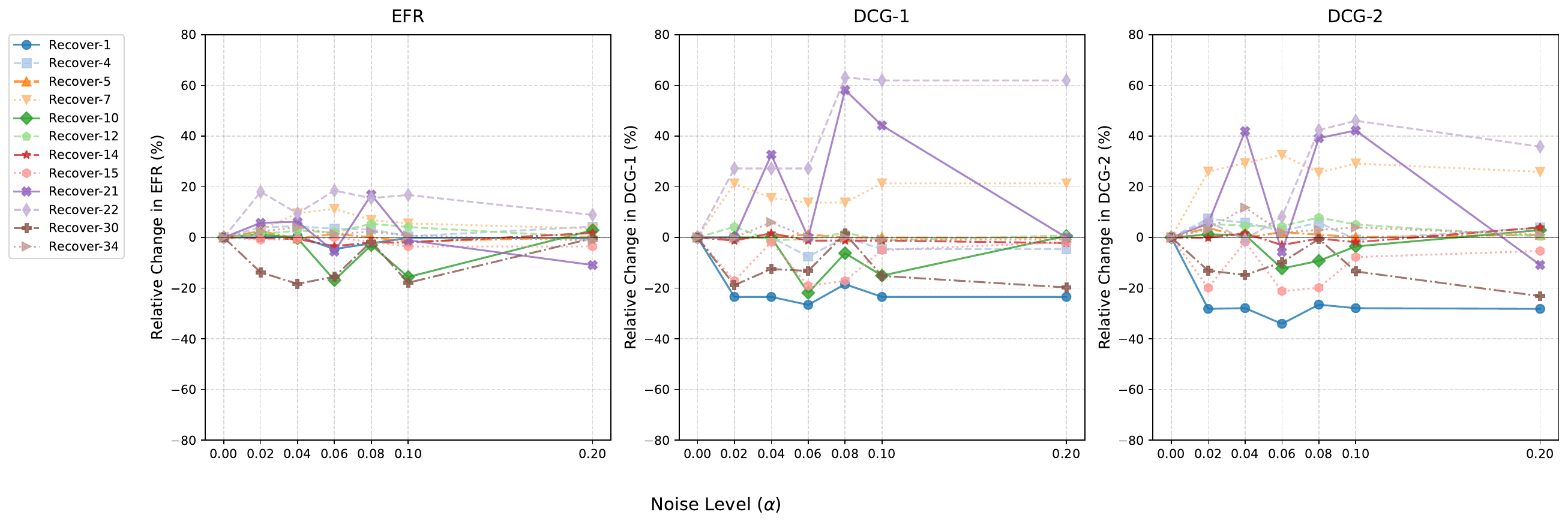}
    \captionsetup{font=small,labelfont=bf} 
    \caption{\textbf{FMN’s sensitivity to noises.} The figure presents the normalized values of three evaluation metrics, namely EFR, DCG-1, and DCG-2, across noise levels ranging from 0.02 to 0.20 applied to the observational data. All values are normalized relative to the zero-noise baseline, where values above unity indicate improved feature extraction performance and values below unity indicate degraded performance. The raw data for drawing this plot found in Appendix \ref{appendixB2.3}. Tables \ref{tab:EFR}-\ref{tab:DCG-2}}.
    \label{Combined_Noise_Sensitivity}
\end{figure}

The three metrics demonstrate that the FMN maintains stable feature selection capability under increasing noise conditions for most equations. Most show deviations of approximately 5\% or less across all metrics and noise levels (Figure \ref{Combined_Noise_Sensitivity}). The consistency across EFR, DCG-1, and DCG-2 indicates that the FMN's feature selection mechanism is robust to input perturbations and irrelevant noisy inputs do not degrade its performance. Among the individual equations, Recover-22 exhibits consistent positive deviations across all three metrics as noise increases. Gains range from approximately 9\% in EFR to 62\% in DCG-1 and 36\% in DCG-2 at the highest noise level. This pattern of convergence across metrics suggests that noise may systematically strengthen the relative prominence of valid features in this equation. Recover-1 exhibits an opposite pattern, demonstrating rapid and early degradation across DCG-1 and DCG-2. Declines of approximately 23.5\% and 28\%, respectively, emerge at the lowest non-zero noise level and persist throughout the entire noise range. Meanwhile, its EFR remains relatively stable. The divergence between the EFR and DCG metrics indicates that, although the quantity of selected features remains intact, their ranking quality and associated confidence deteriorate substantially, even under minimal noise. Recover-21 exhibits the most inconsistent behavior across all three metrics, combining a pronounced negative EFR trend with highly volatile DCG-2 values that rise sharply at intermediate noise levels before collapsing at the highest noise level, and a similarly unstable DCG-1 trajectory. This instability across metrics suggests that Recover-21's feature selection process is structurally sensitive to noise, affecting both selection quality and confidence.

FMN’s robustness stems from the statistical aggregation effect. As we use multiple random initializations and independent training, the noise perturbations captured by the FMN behave in a stochastic and nonsystematic manner. Unlike underlying analytical structures, random noise lacks persistent mathematical patterns. Therefore, it is unable to systematically bias operator-level selections across diverse training trajectories. This mechanism operates like a filter, separating random numerical fluctuations from signals of symbolic expressions. It also explains why the performance of certain equations degrades under high noise conditions. For example, the filter reduces the extraction of valid features with weak signals in Recover-1 and Recover-30. This degrades their index values as the noise level increases. However, these structural advantages at the first stage do not fully translate into the final recovery rate of the target expressions. We find that even under mild noise conditions, the overall success rate of reconstructing the target equation drops to zero (Table \ref{noise_results_extended}). This indicates that the noise robustness of FePySR is primarily limited by its second stage, which employs PySR for symbolic regression. It is well established that SR methods encounter difficulty in recovering symbolic expressions from noisy data. Several approaches have been proposed to address this limitation \cite{reinbold2021robust,sun2025noise}. Because the second stage minimizes pointwise errors on individual noisy datasets, the search process is susceptible to overfitting random fluctuations, thereby diverting the evolutionary search trajectory away from the true symbolic expression. Taken together, the results suggest functional decoupling in noise robustness between the first-stage feature extraction module and the second-stage SR module of FePySR. 

\subsubsection{Model Fitting on Unrecovered Equations}
\label{complex_text}
In theory, with unlimited computational resources, our framework can use SR methods like PySR and DSO to recover mathematical equations of any complexity. In real-world scenarios, however, the utility of any SR method is limited by finite computational resources, such as search time and the total number of equations to be evaluated. It is important to note that the differences in the core algorithms and search strategies between FePySR and DSO make it highly challenging to conduct a fair comparison under identical resource constraints (e.g., fixed runtime or evaluation counts), as any standard metric may introduce an inherent bias toward a specific approach. To avoid this unquantifiable bias and ensure a rigorous evaluation, we focus on a comparative analysis with PySR.

\begin{figure}[H]
    \centering
    \includegraphics[scale=0.19]{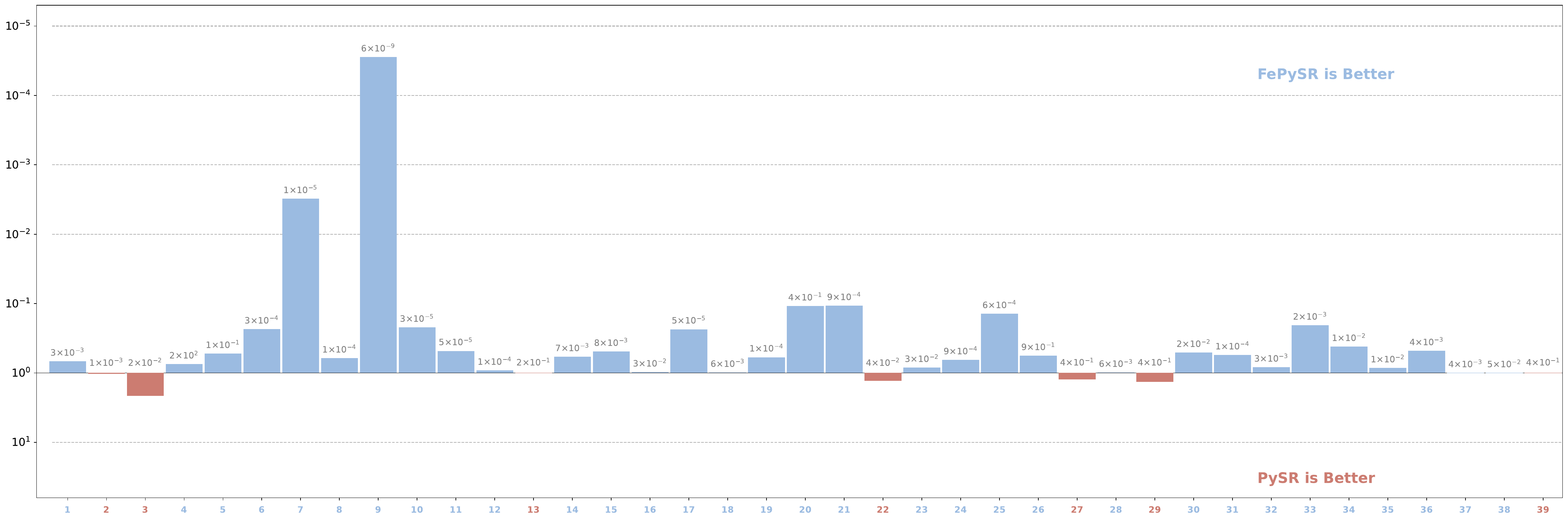}
    \captionsetup{font=small,labelfont=bf} 
    \caption{\textbf{MSE comparison between FePySR and PySR for unrecoverable equations}. The y-axis represents the error ratio (FePySR/PySR) and the x-axis indicates the expression index. Blue bars denote cases where FePySR achieves a lower MSE than PySR, and orange bars denote the reverse. Numerical values above each bar indicate the MSE of FePySR.}
    \label{FePySR_Unrecover_data}
\end{figure}

To evaluate both methods, we performed $50$ tests for each equation using $39$ high-complexity equations synthesized by LLMs (Table \ref{more_complex_data}). We then compare the minimum MSE achieved across all tests (Figure \ref{FePySR_Unrecover_data}). The results show that FePySR achieves a lower minimum MSE in 87.2\% of the tests, whereas PySR outperforms FePySR in only 12.8\% of the instances. A closer inspection of the error ratios reveals FePySR’s advantage in approximating the equations’ solutions. Of the 32 cases in which FePySR is superior, 13 have an error ratio of $<0.5$ than PySR. Three of these cases have an improvement in tenfold (error ratio $<0.1$) with the minimum error ratio of $2.79\times10^{-5}$ (Unrecover-9). Conversely, of the seven cases in which PySR is superior, only one has a relative performance margin greater than twofold (error ratio $> 2.14$; Unrecover-3). Taken together, these findings suggest that, although FePySR cannot discover the exact equations with high complexities, it is more effective than PySR at approximating the equations’ solutions. 

\subsection{Case Study: Modeling Biochemical Dynamics}
Tyson et al. proposes an ODE model to study the dynamics of biological networks \cite{tyson2010functional}. This formalism characterizes a regulatory system between two proteins (represented by $X_1, X_2$) whose expression dynamics are driven by an external stimulus $S$. The proteins’ temporal dynamics are dictated by the nonlinear balance between an activation term (the rate equation $A$) and an inhibition term (the rate equation $B$). These rate equations use exponential function, where the interaction strengths are governed by an activation coefficient $\alpha$ and an inhibition coefficient $\beta$, respectively, while a parameter $\sigma$ modulates the steepness of the response function. By adjusting these parameters’ values, the model can encode different network motifs, such as positive and negative feedback loops, which can result in bistability and oscillation in proteins’ expression levels \cite{tyson2010functional}. This allows the model to serve as a theoretical basis for studying the relationship between network topology and gene expression dynamics.

\begin{equation*}
\begin{aligned}
\frac{dX_1}{dt} &= \gamma_1 \frac{A_1(1 - X_1) - B_1X_1}{A_1 + B_1} \\
\frac{dX_2}{dt} &= \gamma_2 \frac{A_2(1 - X_2) - B_2X_2}{A_2 + B_2} \\[6pt]
A_1 &= \exp\{\sigma(S + a_{10} + a_{12}X_2)\} \\
B_1 &= \exp\{\sigma(\beta_{10} + \beta_{12}X_2)\} \\
A_2 &= \exp\{\sigma(a_{20} + a_{21}X_1)\} \\
B_2 &= \exp\{\sigma(\beta_{20} + \beta_{21}X_1)\}
\end{aligned}
\end{equation*}

\begin{figure}[htbp]

  \centering 
  \hspace{0.00\textwidth}
  \captionsetup{font=small,labelfont=bf} 
  \begin{minipage}[c]{0.45\textwidth}
    \begin{equation*}
    \begin{aligned}
    \frac{dX_1}{dt} &= \frac{e^{S-0.15}(1 - X_1) - e^{-0.4+X2}X_1}{e^{S-0.15} + e^{-0.4+X2}} \\
    \frac{dX_2}{dt} &=  \frac{e^{-0.15+0.5X1}(1 - X_2) - X_2}{e^{-0.15+0.5X1} + 1} \\[6pt]
    \end{aligned}
    \end{equation*}
  \end{minipage}
  \begin{minipage}[c]{0.31\textwidth}
    \centering
    \includegraphics[width=\textwidth]{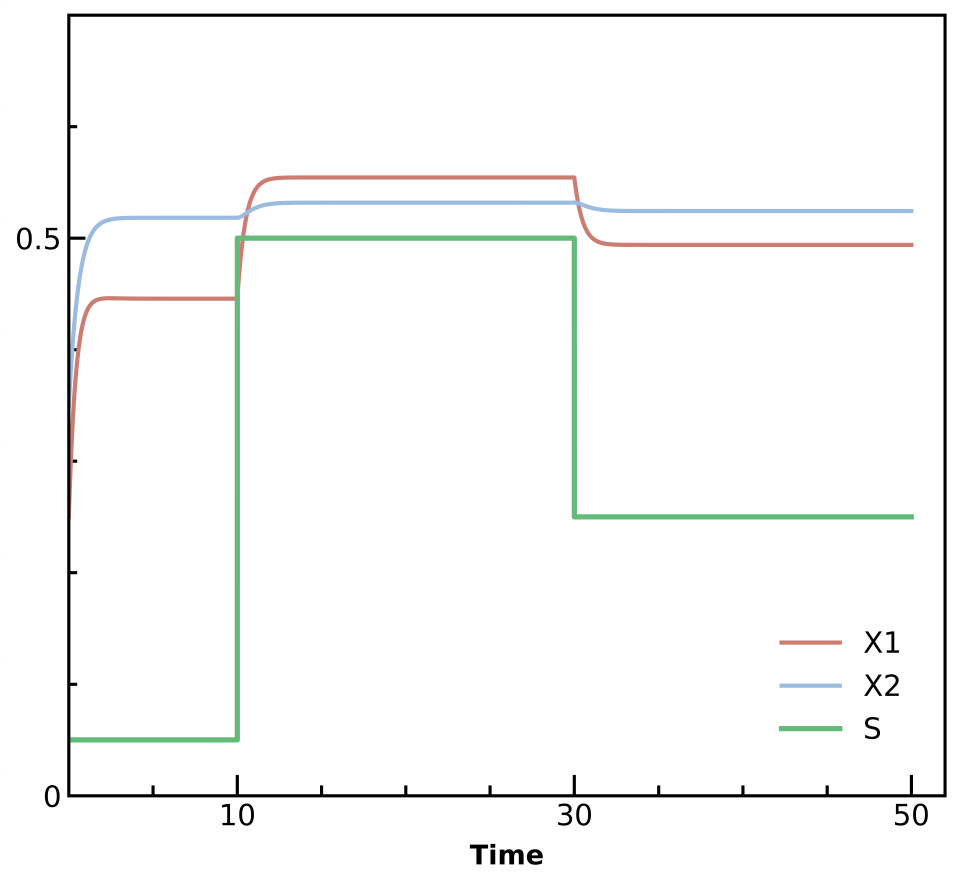}
  \end{minipage}
  \caption{\textbf{Ground-truth equations and corresponding observational data for SR}. (Left) ODEs governing a two-protein regulatory system ($X_1, X_2$) driven by an external stimulus $S$, initialized with specific parameter values. (Right) Time-series dynamics produced by the ODEs. The input signal $S$ drives the temporal evolution of $X_1$ and $X_2$, and the resulting time-series data serves as the observational data for evaluating the equation recovery performance of FePySR relative to PySR.}
  \label{Tyson_pic}
  \hspace{0.04\textwidth}
\end{figure}

Here, we test FePySR’s ability to identify the governing equations from observational data, thereby exploring its efficacy in modeling complex biological systems. To do so, we configure the model to generate data with $\sigma=\gamma_{1}=\gamma_{2}=1$, $\alpha_{10}=-0.15$, $\alpha_{12}=0$,$\alpha_{20}=-0.15$,$\alpha_{21}=0.5$, $\beta_{10}=-0.4$,$\beta_{12}=1$,$\beta_{20}=0$ and $\beta_{21}=0$. The external signal $S$ is partitioned into three piecewise regimes ($0.05, 0.5$, and $0.25$). For each segment, we sample $1000$ data points over $20$ seconds and the model shows protein expression levels caused by a stimulus signal (Figure \ref{Tyson_pic}). The objective is to reconstruct the differential equation governing $\frac{dX_1}{dt}$ as a function of the variables $\{S, X_1, X_2\}$. 

We compare the recovery performance of FePySR and PySR (Table \ref{tab:equation_recovery_comparison}). In $100$ runs under identical settings, PySR completely fails to recover the true structure of the target equation. Its best result is a linear polynomial equation that yields a MSE of $1.23 \times 10^{-10}$. In contrast, FePySR reconstructs the exact nonlinear dynamics in $24$ out of $100$ runs (e.g., $ \frac{dX_1}{dt}=\frac{e^S(1-X_1)-0.779e^{X_2}X1}{e^S + 0.779e^{X_2}}$). Furthermore, FePySR achieves a much smaller MSE of $7.04 \times 10^{-16}$, demonstrating its superiority in recovering complex equations used to model biological systems.

\begin{center} 
\captionsetup{font=small,labelfont=bf, skip=12pt}
\captionof{table}{\textbf{ODE recovery comparison between FePySR and PySR.} For each method, the table reports the best recovered equation with structural similarity to the ground-truth, the recovery rate across 100 independent runs, and the minimum MSE across 100 independent runs.} 
\label{tab:equation_recovery_comparison}
\vspace{1.5ex}
\renewcommand{\arraystretch}{1.5} 
\begin{tabular}{@{}ccc@{}}
\toprule
\textbf{} & {FePySR} & {PySR} \\ \midrule
Best recovered equation & $ \frac{dX_1}{dt}=\frac{e^S(1-X_1)-0.779e^{X_2}X_1}{e^S + 0.779e^{X_2}}$ & $\frac{dX_1}{dt}=-X_1 - 0.249X_2 + 0.249S + 0.562$ \\
Recover rate & $\mathbf{24/100}$ & $0/100$ \\
MSE & $\mathbf{7.04 \times 10^{-16}}$ & $1.23 \times 10^{-10}$ \\ \bottomrule
\end{tabular}
\vspace{1ex}
\small \\
\raggedright
\end{center} 

\clearpage 

\section{Discussion and Conclusion}
Our work is grounded in the theoretical premise that SR is fundamentally an NP-hard problem. SR method development falls into three categories: efficient space exploration, constrained search, and search space simplification. Building on this premise, we introduce FePySR, an AI framework that falls into the space simplification category. It is a complement to dimensionality reduction methods, such as AI Feynman \cite{udrescu2020aifeynman}. FePySR is a two-stage model that combines FMN (a feature extraction neural network) and PySR (a SR method that uses GP to reconstruct mathematical equations). In the first stage, neural networks perform targeted feature engineering, extracting critical latent nonlinear structures from the data and thereby optimizing the search space via dimensionality reduction. Next, the challenging task of structural discovery is reformulated as an evolutionary search over the features discovered in the first stage. Model evaluation shows that FePySR significantly outperforms existing SOTA methods on Nguyen, Livermore, R, Costant, and Jin datasets. For highly complex equations generated by LLMs, FePySR outperforms PySR and DSO by 49\%–65\% for recoverable equations and has smaller MSEs than PySR for most unrecoverable ones. Furthermore, our case study reveals that FePySR can recover complex ODEs of a biological system that is not achievable by PySR. These findings demonstrate FePySR's ability to efficiently recover non-linear equations, even highly complex ones in biological systems.

Though our results demonstrate the efficacy of FePySR in SR, the method has limitations. One such limitation is its limited capacity to handle constants. For example, when given the target equations $\sin(1.5x) $ and $\cos(0.5y)$, the model can identify the base functions $\sin(x)$ and $\cos(y)$, but it struggles to resolve the constants because it lacks a dedicated constant optimization module. In the literature, PySR uses stochastic mutations within its genetic algorithm to iteratively approximate constants. By contrast, DSO and SymbolicGPT employ placeholders for constants in equations and perform gradient-based optimization to determine their exact values. However, adapting these approaches directly to the feature engineering stage in FePySR is challenging. Introducing learnable constants into FMN would fill the feature space with many invalid features, such as $\sin(1.11x)$ and $\sin(1.12x)$, which would obscure the identification of valid features and therefore undermine the model’s training efficiency. To resolve this issue, a possible solution is to use LLMs to maintain a dynamic feature library as demonstrated in LaSR \cite{Grayeli2024LASR}. This approach would leverage LLMs’ reasoning capability to prune invalid features and recommend novel candidates based on feedback from each SR iteration in PySR. Such a dynamic pruning and generation strategy would enhance the precision of feature space and therefore enable the identification of features with constants. In addition, FMN demonstrates robustness in identifying relevant features under noisy observational data. However, the integration of these features in the second stage does not prevent PySR from failing to recover the exact target equation when noise is present. This is consistent with the known limitation that SR methods generally struggle with noisy data, a challenge that has motivated several dedicated mitigation strategies in the literature \cite{reinbold2021robust,sun2025noise}. Future work could investigate the incorporation of noise-robust mechanisms into the second stage of FePySR, such as denoising preprocessing or noise-aware loss formulations, to improve end-to-end recovery performance under realistic data conditions.

In conclusion, the feature space reconstruction strategy proposed in this work offers a new perspective for improving SR methods. Our work demonstrates the potential of combining deep neural networks and robust search algorithms to improve automated symbolic discovery in complex, high-dimensional settings.

\newpage

\section{Data Availability}
FePySR is accessible at https://github.com/laixn/FePySR. 

\section{Author Contributions}
ZY (Data curation [lead], Formal analysis [supporting], Methodology [equal], Software [lead], Visualization [equal], Writing—original draft [equal], Writing—review \& editing [supporting]). WL (Funding acquisition [equal], Project administration [support], Resources [equal], Supervision [supporting], Writing—review \& editing [supporting]). XL (Conceptualization [lead], Formal analysis [lead], Funding acquisition [equal], Investigation [lead], Methodology [equal], Project administration [Lead], Resources [equal], Software [supporting], Supervision [lead], Validation [lead], Visualization [equal], Writing—original draft [equal], Writing—review \& editing [lead]).

\section{Acknowledgements}
We are thankful for Julio R. Banga's insightful discussion and feedback on our paper. XL is funded by the PROFI6 Health Data Science program at Tampere University. WL is partially supported by the National Key Research and Development Program of China (2024YFA1012600, 023YFA1009100), China NSF grant 12174310, and a Key Project of Joint Funds for Regional Innovation and Development (U21A20425), Zhejiang Provincial NSF Distinguished Youth Program (Extended Program LRG25A010001).

\bibliographystyle{unsrtnat}

\newpage

\bibliography{reference}

@book{Koza1994,
  author       = {John R. Koza},
  title        = {Genetic programming 2 - automatic discovery of reusable programs},
  publisher    = {{MIT} Press},
  year         = {1994},
  isbn         = {978-0-262-11189-8},
}

@article{Schmidt2009,
  title={Distilling Free-Form Natural Laws From Experimental Data},
  author={Schmidt, Michael and Lipson, Hod},
  journal={Science},
  volume={324},
  number={5923},
  pages={81--85},
  year={2009},
}

@inproceedings{LaCava2016,
  author       = {William G. La Cava and
                  Lee Spector and
                  Kourosh Danai},
  title        = {Epsilon-Lexicase Selection for Regression},
  booktitle    = {Genetic and Evolutionary Computation Conference},
  pages        = {741--748},
  year         = {2016},
}

@inproceedings{Virgolin2017,
  title={Surrogate Modeling For Genetic Programming by Evolving Model Complexity},
  author={Virgolin, Marco and Alderliesten, Tanja and Bosman, Peter AN},
  booktitle={Genetic Programming Theory and Practice XIV},
  pages={217--236},
  year={2017},
}

@article{Cranmer2023PySR,
  title={Interpretable Machine Learning for Science with {PySR} and {SymbolicRegression.jl}},
  author={Cranmer, Miles},
  journal={arXiv preprint arXiv:2305.01582},
  year={2023}
}

@inproceedings{Grayeli2024LASR,
  author       = {Arya Grayeli and
                  Atharva Sehgal and
                  Omar Costilla{-}Reyes and
                  Miles D. Cranmer and
                  Swarat Chaudhuri},
  title        = {Symbolic Regression with a Learned Concept Library},
  booktitle    = {Advances in Neural Information Processing Systems},
  year         = {2024},
}

@inproceedings{Petersen2021a,
  author       = {Brenden K. Petersen and
                  Mikel Landajuela and
                  T. Nathan Mundhenk and
                  Cl{\'{a}}udio Prata Santiago and
                  Sookyung Kim and
                  Joanne Taery Kim},
  title        = {Deep symbolic regression: Recovering mathematical expressions from
                  data via risk-seeking policy gradients},
  booktitle    = {International Conference on Learning Representations},
  year         = {2021},
}

@inproceedings{Landajuela2021_exploration,
  author    = {Mikel Landajuela and Brenden K. Petersen and Soo K. Kim and Claudio P. Santiago and Ruben Glatt and T. Nathan Mundhenk and Jacob F. Pettit and Daniel M. Faissol},
  title     = {Improving Exploration in Policy Gradient Search: Application to Symbolic Optimization},
  booktitle = {Mathematical Reasoning in General Artificial Intelligence Workshop},
  year      = {2021},
}

@inproceedings{Mundhenk2021,
  author    = {Terrell Mundhenk and Mikel Landajuela and Ruben Glatt and Claudio P. Santiago and Daniel M. Faissol and Brenden K. Petersen},
  title     = {Symbolic Regression via Neural-Guided Genetic Programming Population Seeding},
  booktitle = {Advances in Neural Information Processing Systems},
  volume    = {34},
  pages     = {24912--24923},
  year      = {2021},
}

@inproceedings{Landajuela2022,
  author       = {Mikel Landajuela and
                  Chak Shing Lee and
                  Jiachen Yang and
                  Ruben Glatt and
                  Cl{\'{a}}udio P. Santiago and
                  Ignacio Aravena and
                  Terrell Nathan Mundhenk and
                  Garrett Mulcahy and
                  Brenden K. Petersen},
  title        = {A Unified Framework for Deep Symbolic Regression},
  booktitle    = {Advances in Neural Information Processing Systems},
  year         = {2022},
}

@inproceedings{petersen2021_priors,
  title={Incorporating domain knowledge into neural-guided search via in situ priors and constraints},
  author={Petersen, Brenden K and Santiago, Claudio and Landajuela, Mikel},
  booktitle    = {International Conference on Machine Learning},
  publisher    = {{PMLR}},
  year         = {2021},
}

@inproceedings{Pettit2025,
  author    = {Jacob F. Pettit and Chak Shing Lee and Jiachen Yang and Alex Ho and Daniel M. Faissol and Brenden K. Petersen and Mikel Landajuela},
    title     = {{DisCo-DSO}: Coupling Discrete and Continuous Optimization for Efficient Generative Design in Hybrid Spaces},
    booktitle = {{AAAI} Conference on Artificial Intelligence },
    pages     = {27117--27125},
    year      = {2025}
}

@inproceedings{RL-GEP,
  author       = {Hengzhe Zhang and
                  Aimin Zhou},
  title        = {{RL-GEP:} Symbolic Regression via Gene Expression Programming and Reinforcement Learning},
  booktitle    = {International Joint Conference on Neural Networks},
  pages        = {1--8},
  year         = {2021},
}

@inproceedings{biggio2021neural,
  author       = {Luca Biggio and
                  Tommaso Bendinelli and
                  Alexander Neitz and
                  Aur{\'{e}}lien Lucchi and
                  Giambattista Parascandolo},
  title        = {Neural Symbolic Regression that scales},
  booktitle    = {International Conference on Machine Learning},
  pages        = {936--945},
  year         = {2021},
}

@inproceedings{li2023transformer,
  author       = {Wenqiang Li and
                  Weijun Li and
                  Linjun Sun and
                  Min Wu and
                  Lina Yu and
                  Jingyi Liu and
                  Yanjie Li and
                  Songsong Tian},
  title        = {Transformer-based model for symbolic regression via joint supervised
                  learning},
  booktitle    = {International Conference on Learning Representations},
  year         = {2023},
  bibsource    = {dblp computer science bibliography, https://dblp.org}
}

@article{li2025mmsr,
  author       = {Yanjie Li and
                  Jingyi Liu and
                  Min Wu and
                  Lina Yu and
                  Weijun Li and
                  Xin Ning and
                  Wenqiang Li and
                  Meilan Hao and
                  Yusong Deng and
                  Shu Wei},
  title        = {{MMSR:} Symbolic regression is a multi-modal information fusion task},
  journal      = {Inf. Fusion},
  volume       = {114},
  pages        = {102681},
  year         = {2025},
  bibsource    = {dblp computer science bibliography, https://dblp.org}
}

@article{tian2025interactive,
  title={Interactive Symbolic Regression with Co-Design Mechanism Through Offline Reinforcement Learning},
  author={Tian, Yuan and Zhou, Wenqi and Viscione, Michele and Dong, Hao and Kammer, David S. and Fink, Olga},
  journal={Nature Communications},
  volume={16},
  number={1},
  pages={3930},
  year={2025},
}

@article{valipour2021symbolicgpt,
  title={{SymbolicGPT} : A generative transformer model for symbolic regression},
  author={Valipour, Mojtaba and You, Bowen and Panju, Maysum and Ghodsi, Ali},
  journal={arXiv preprint arXiv:2106.14131},
  year={2021}
}

@inproceedings{lample2019deep,
  author       = {Guillaume Lample and
                  Fran{\c{c}}ois Charton},
  title        = {Deep Learning For Symbolic Mathematics},
  booktitle    = {International Conference on Learning Representations},
  year         = {2020},
  bibsource    = {dblp computer science bibliography, https://dblp.org}
}

@inproceedings{lee2019set,
  author       = {Juho Lee and
                  Yoonho Lee and
                  Jungtaek Kim and
                  Adam R. Kosiorek and
                  Seungjin Choi and
                  Yee Whye Teh},
  title        = {Set Transformer: {A} Framework for Attention-based Permutation-Invariant
                  Neural Networks},
  booktitle    = {International Conference on Machine Learning},
  pages        = {3744--3753},
  year         = {2019},
}

@inproceedings{kamienny2022end,
  author       = {Pierre{-}Alexandre Kamienny and
                  St{\'{e}}phane d'Ascoli and
                  Guillaume Lample and
                  Fran{\c{c}}ois Charton},
  title        = {End-to-end Symbolic Regression with Transformers},
  booktitle    = {Advances in Neural Information Processing Systems},
  year         = {2022},
  bibsource    = {dblp computer science bibliography, https://dblp.org}
}

@article{udrescu2020aifeynman,
  title={AI Feynman: A physics-inspired method for symbolic regression},
  author={Udrescu, Silviu-Marian and Tegmark, Max},
  journal={Science Advances},
  volume={6},
  number={16},
  pages={eaay2631},
  year={2020}
}

@inproceedings{udrescu2020aifeynman2,
  author       = {Silviu{-}Marian Udrescu and
                  Andrew K. Tan and
                  Jiahai Feng and
                  Orisvaldo Neto and
                  Tailin Wu and
                  Max Tegmark},
  title        = {{AI} Feynman 2.0: Pareto-optimal symbolic regression exploiting graph
                  modularity},
  booktitle    = {Advances in Neural Information Processing Systems},
  year         = {2020},
  bibsource    = {dblp computer science bibliography, https://dblp.org}
}

@inproceedings{sahoo2018learning,
  author       = {Subham S. Sahoo and
                  Christoph H. Lampert and
                  Georg Martius},
  title        = {Learning Equations for Extrapolation and Control},
  booktitle    = {International Conference on Machine Learning},
  pages        = {4439--4447},
  year         = {2018},
}

@article{kim2021integration,
  author       = {Samuel Kim and
                  Peter Y. Lu and
                  Srijon Mukherjee and
                  Michael Gilbert and
                  Li Jing and
                  Vladimir Ceperic and
                  Marin Soljacic},
  title        = {Integration of Neural Network-Based Symbolic Regression in Deep Learning
                  for Scientific Discovery},
  journal      = {{IEEE} Trans. Neural Networks Learn. Syst.},
  volume       = {32},
  number       = {9},
  pages        = {4166--4177},
  year         = {2021},
  bibsource    = {dblp computer science bibliography, https://dblp.org}
}

@article{brunton2016discovering,
  title={Discovering Governing Equations from Data by Sparse Identification of Nonlinear Dynamical Systems},
  author={Brunton, Steven L. and Proctor, Joshua L. and Kutz, J. Nathan},
  journal={Proceedings of the National Academy of Sciences},
  volume={113},
  number={15},
  pages={3932--3937},
  year={2016},
}

@article{mangan2016inferring,
  author       = {Niall M. Mangan and
                  Steven L. Brunton and
                  Joshua L. Proctor and
                  J. Nathan Kutz},
  title        = {Inferring Biological Networks by Sparse Identification of Nonlinear
                  Dynamics},
  journal      = {{IEEE} Trans. Mol. Biol. Multi Scale Commun.},
  volume       = {2},
  number       = {1},
  pages        = {52--63},
  year         = {2016},
  bibsource    = {dblp computer science bibliography, https://dblp.org}
}

@article{kaheman2020sindy,
  title={SINDy-PI: a robust algorithm for parallel implicit sparse identification of nonlinear dynamics},
  author={Kaheman, Kadierdan and Kutz, J Nathan and Brunton, Steven L},
  journal={Proceedings of the Royal Society A},
  volume={476},
  number={2242},
  pages={20200279},
  year={2020},
}

@article{viknesh2024adam,
  title={Adam-sindy: An efficient optimization framework for parameterized nonlinear dynamical system identification},
  author={Viknesh, Siva and Tatari, Younes and Christenson, Chase and Arzani, Amirhossein},
  journal={arXiv preprint arXiv:2410.16528},
  year={2024}
}

@inproceedings{krawiec2013approximating,
  author       = {Krzysztof Krawiec and
                  Tomasz Pawlak},
  title        = {Approximating geometric crossover by semantic backpropagation},
  booktitle    = {Genetic and Evolutionary Computation Conference},
  pages        = {941--948},
  year         = {2013},
}

@article{nguyen2014semantically,
  author       = {Nguyen Quang Uy and
                  Nguyen Xuan Hoai and
                  Michael O'Neill and
                  Robert I. McKay and
                  Edgar Galv{\'{a}}n L{\'{o}}pez},
  title        = {Semantically-based crossover in genetic programming: application to real-valued symbolic regression},
  journal      = {Genetic Programming and Evolvable Machines},
  volume       = {12},
  number       = {2},
  pages        = {91--119},
  year         = {2011},
  bibsource    = {dblp computer science bibliography, https://dblp.org}
}

@book{koza1992genetic,
  author       = {John R. Koza},
  title        = {Genetic Programming: On the Programming of Computers by Means of Natural Selection},
  publisher    = {{MIT} Press},
  year         = {1993},
}

@inproceedings{virgolin2018symbolic,
  author       = {Marco Virgolin and
                  Tanja Alderliesten and
                  Arjan Bel and
                  Cees Witteveen and
                  Peter A. N. Bosman},
  title        = {Symbolic Regression and Feature Construction with {GP-GOMEA} Applied to Radiotherapy Dose Reconstruction of Childhood Cancer Survivors},
  booktitle    = {Genetic and Evolutionary Computation Conference},
  pages        = {1395--1402},
  year         = {2018}
}

@inproceedings{virgolin2017scalable,
  author       = {Marco Virgolin and
                  Tanja Alderliesten and
                  Cees Witteveen and
                  Peter A. N. Bosman},
  title        = {Scalable genetic programming by gene-pool optimal mixing and input-space
                  entropy-based building-block learning},
  booktitle    = {Genetic and Evolutionary Computation Conference},
  pages        = {1041--1048},
  year         = {2017},
  bibsource    = {dblp computer science bibliography, https://dblp.org}
}

@inproceedings{gu2025inceptionsr,
  author={Gu, Edward and Alford, Simon and Costilla-Reyes, Omar and Cranmer, Miles and Ellis, Kevin},
  title={{InceptionSR}: Recursive Symbolic Regression for Equation Synthesis},
  booktitle = {{AAAI} Conference on Artificial Intelligence},
  year={2025},
}

@article{virgolin2022symbolic,
  author       = {Marco Virgolin and
                  Solon P. Pissis},
  title        = {Symbolic Regression is {NP}-hard},
  journal      = {Transactions on Machine Learning Research},
  year         = {2022},
  bibsource    = {dblp computer science bibliography, https://dblp.org}
}

@article{song2024prove,
  title={Prove symbolic regression is {NP}-hard by symbol graph},
  author={Song, Jinglu and Lu, Qiang and Tian, Bozhou and Zhang, Jingwen and Luo, Jake and Wang, Zhiguang},
  journal={arXiv preprint arXiv:2404.13820},
  year={2024}
}

@inproceedings{ahn2020guiding,
  author       = {Sungsoo Ahn and
                  Junsu Kim and
                  Hankook Lee and
                  Jinwoo Shin},
  title        = {Guiding Deep Molecular Optimization with Genetic Exploration},
  booktitle    = {Advances in Neural Information Processing Systems},
  year         = {2020},
  bibsource    = {dblp computer science bibliography, https://dblp.org}
}

@article{tyson2010functional,
  title     = {Functional Motifs in Biochemical Reaction Networks},
  author    = {Tyson, John J. and Albert, R{\'e}ka and Goldbeter, Albert and Ruoff, Peter and Sible, Jill C.},
  journal   = {Annual Review of Physical Chemistry},
  volume    = {61},
  pages     = {219--240},
  year      = {2010},
}

@article{jarvelin2002cumulated,
  author       = {Kalervo J{\"{a}}rvelin and
                  Jaana Kek{\"{a}}l{\"{a}}inen},
  title        = {Cumulated gain-based evaluation of {IR} techniques},
  journal      = {ACM Transactions on Information Systems},
  volume       = {20},
  number       = {4},
  pages        = {422--446},
  year         = {2002},
  bibsource    = {dblp computer science bibliography, https://dblp.org}
}

@article{bansal2018can,
  title={Can we gain more from orthogonality regularizations in training deep networks?},
  author={Bansal, Nitin and Chen, Xiaohan and Wang, Zhangyang},
  journal={Advances in Neural Information Processing Systems},
  year={2018}
}

@article{reinbold2021robust,
  title={Robust learning from noisy, incomplete, high-dimensional experimental data via physically constrained symbolic regression},
  author={Reinbold, Patrick AK and Kageorge, Logan M and Schatz, Michael F and Grigoriev, Roman O},
  journal={Nature communications},
  volume={12},
  number={1},
  pages={3219},
  year={2021},
  publisher={Nature Publishing Group UK London}
}

@inproceedings{sun2025noise,
  title={Noise-resilient symbolic regression with dynamic gating reinforcement learning},
  author={Sun, Chenglu and Shen, Shuo and Tao, Wenzhi and Xue, Deyi and Zhou, Zixia},
  booktitle={Proceedings of the AAAI Conference on Artificial Intelligence},
  year={2025}
}

\appendix
\renewcommand{\thetable}{A\arabic{table}}
\renewcommand{\thefigure}{A\arabic{figure}}
\renewcommand{\thealgorithm}{A\arabic{algorithm}}

\setcounter{figure}{0} 
\setcounter{table}{0}  
\newpage
\appendix

\section*{\centering Appendix}
This appendix details the training configuration of the FePySR parameters, along with the training procedures applied to the benchmark datasets and additional expressions evaluated in the experiments.

\section{Implementation of FePySR}
\label{appendixA}
FePySR operates in two sequential stages. In the first stage, FMN extracts nonlinear features from the observational data and concatenates them with the original inputs to form an augmented feature space. In the second stage, this augmented representation is passed to the PySR solver, which reconstructs the target symbolic expression within the reduced search space. The model parameters are shown in Table \ref{FePySR-hyperparameters}.

\begin{table}[htbp]
\centering
\captionsetup{font=small,labelfont=bf} 
\caption{\textbf{FePySR Hyperparameters.} FMN-par contains FMN’s hyperparameters such as batch size, learning rate, and epochs used in the first. The fun-net defines the mathematical function set, defaulting to $\{(\cdot)^2, \sin(\cdot), \cos(\cdot), \exp(\cdot), +, \times\}$. net-depth controls the neural network’s depth and is set to 3 layers for univariate and 4 layers for multivariate functions, respectively. Following configuration, the network is trained repeatedly in parallel, using random weight initializations for each run. num-experiments is the number of training runs, and num-workers is the number of parallel processes. pysr-num specifies the number of target features (default: 4) used for SR. After the first stage, all extracted features are aggregated, deduplicated with frequency counts, and sorted. The top selected features are concatenated with the original data for SR in the second stage. PySR performs symbolic regression on this enriched representation. Its default operators are divided into unary-operators (defaulting to $\{\sin(\cdot), \cos(\cdot), \exp(\cdot), \sqrt{(\cdot)}\}$) and binary-operators (defaulting to $\{ \times, +, -, / \}$). The full pseudo-code of this process is detailed in Algorithm \ref{FePySR-algorithm}.}

\label{FePySR-hyperparameters}
\begin{tabular}{ccc}
\toprule
Category & Parameter& Default Value \\
\midrule
FMN-par     & batch size    &50       \\
        &learning rate             &0.5       \\
        &epoches   &100       \\
        &fun-net        &$\{(\cdot)^2,\sin(\cdot),\cos(\cdot), \exp(\cdot),+, \times\}$       \\
        &net-depth      &4       \\
Parallel-par    &num-experiments      &16       \\
        &num-workers      &8       \\
Data-symbol-par    &pysr-num      &4       \\
PySR-par    &unary-operators      &$\{\sin(\cdot),\cos(\cdot), \exp(\cdot),\sqrt(\cdot) \}$       \\
        &binary-operators      &$\{ \times,+,-,/ \}$       \\
\bottomrule
\end{tabular}
\end{table}

\begin{algorithm}
\caption{FePySR}
\label{FePySR-algorithm}
\begin{algorithmic}[1]
    \Require Original dataset $\mathcal{D} = \{x, y\}$, Hyperparameters $\{\Theta_{FMN}, \Theta_{P}, \Theta_{S}\}$
    \Ensure Symbolic expression $\mathcal{F} $
    
    \State Extract parameters: $\{bs, \eta,E, L\} \gets \Theta_{FMN}$ \graycomment{Initialize FMN}
    \State Extract parallel settings: $\{N_{exp}, N_{w}\} \gets \Theta_{P}$ \graycomment{Initialize Parallelism}
    \State $K \gets \Theta_{S}$ \graycomment{Target number of features}
    
    \State $\Phi \gets \emptyset$
    \For{$i = 1$ \textbf{to} $N_{exp}$} \graycomment{Iterative Feature Extraction}
        \State $\Phi_i  = \{\phi_{i1},\phi_{i2}, \dots \} \gets \text{FMN}(x, y; \Theta_{FMN})$ \graycomment{Extract features (Alg.\ref{alg:fmn_extraction})}
        \State $\Phi \gets \Phi \cup \Phi_i$ \graycomment{Update candidate feature pool}
    \EndFor
    \State $\Phi_{top} \gets \text{Select}(\Phi, K)$ \graycomment{Rank and select $K$ features}
    \State $\mathcal{D}_{aug} \gets \{x, \Phi_{top}, y\}$ \graycomment{Augment original data}
    \State $\mathcal{F}  \gets \text{PySR}(\mathcal{D}_{aug})$ \graycomment{Final symbolic search}
    
    \State \Return $\mathcal{F}$
\end{algorithmic}
\end{algorithm}
\newpage

\section{Model Evaluation and Comparison}

\subsection{Model Evaluation on Standard Datasets}
\label{appendixB1}

\begin{longtable}{ccccc}
\captionsetup{font=small,labelfont=bf}
\caption{\textbf{Comparison of between FePySR and PySR on the Nguyen, Nguyen', Livermore, Jin, Constan, and R datasets.} We test each equation 100 times for both models. $U(a, b, c)$ indicates $c$ points sampled uniformly at random between $a$ and $b$ per input variable; $E(a, b, c)$ denotes $c$ evenly spaced points between $a$ and $b$ per input variable. For all benchmark datasets, the feature extraction stage utilizes the operator set $\{(\cdot)^2, \exp(\cdot), \sin(\cdot), \cos(\cdot), \times, + \}$, whereas the PySR solving stage employs $\{+, -, \times, /, \sin(\cdot), \cos(\cdot), \exp(\cdot), \log(\cdot)\}$.}

 \label{tab:nguyen_benchmarks} \\ 
 \toprule 
  \rowcolor{gray!20} 
 \textbf{Dataset} & \textbf{Equation} & \textbf{Input data} & \textbf{FePySR} & \textbf{PySR} \\ 
 \endfirsthead 
 \toprule 
 \endhead 
 \bottomrule 
 \endfoot  
 \bottomrule 
 \endlastfoot

 Nguyen-1 & $x^3 + x^2 + x$ & U(-1,1,20) & 100\% & 100\% \\ 
 Nguyen-2 & $x^4 + x^3 + x^2 + x$ & U(-1,1,20) & 100\% & 100\% \\ 
 Nguyen-3 & $x^5 + x^4 + x^3 + x^2 + x$ & U(-1,1,20) & 78\% & 52\% \\ 
 Nguyen-4 & $x^6 + x^5 + x^4 + x^3 + x^2 + x$ & U(-1,1,20) & 100\% & 62\% \\ 
 Nguyen-5 & $\sin(x^2)\cos(x) - 1$ & U(-1,1,20) & 100\% & 92\% \\ 
 Nguyen-6 & $\sin(x) + \sin(x + x^2)$ & U(-1,1,20) & 100\% & 100\% \\ 
 Nguyen-7 & $\log(x + 1) + \log(x^2 + 1)$ & U(0,2,20) & 100\% & 19\% \\ 
 Nguyen-8 & $\sqrt{x}$ & U(0,4,20) & 100\% & 100\% \\ 
 Nguyen-9 & $\sin(x) + \sin(y^2)$ & U(0,1,20) & 100\% & 100\% \\ 
 Nguyen-10 & $2\sin(x)\cos(y)$ & U(0,1,20) & 100\% & 100\% \\ 
 Nguyen-11 & $x^y$ & U(0,1,20) & 100\% & 99\% \\ 
 Nguyen-12 & $x^4 - x^3 + \frac{1}{2}y^2 - y$ & U(0,1,20) & 95\% & 0\% \\ 
 \rowcolor{gray!20} 
\bf{Nguyen average} &  &  & 97.75\% & 77\% \\ 
 Nguyen-2' & $4x^4 + 3x^3 + 2x^2 + x$ & U(-1, 1, 20) & 76\% & 84\% \\ 
 Nguyen-5' & $\sin(x^2)\cos(x) - 2$ & U(-1, 1, 20) & 99\% & 96\% \\ 
 Nguyen-8'' & $\sqrt[3]{x^2}$ & U(0, 4, 20) & 99\% & 100\% \\ 
 Nguyen-1$^{c}$ & $3.39x^3 + 2.12x^2 + 1.78x$ & U(-1, 1, 20) & 100\% & 100\% \\ 
 Nguyen-5$^{c}$ & $\sin(x^2)\cos(x) - 0.75$ & U(-1, 1, 20) & 99\% & 76\% \\ 
 Nguyen-7$^{c}$ & $\log(x+1.4) + \log(x^2+1.3)$ & U(0, 2, 20) & 65\% & 2\% \\ 
 \rowcolor{gray!20} 
\bf{Nguyen' average} &  &  & 89.67\% & 76.33\% \\ 
Livermore-1 & $\frac{1}{3} + x + \sin(x^2)$ & U(-3, 3, 100) & 100\% &100\% \\
Livermore-2 & $\sin(x^2)\cos(x) - 2$ & U(-3, 3, 100) & 100\% & 96\% \\
Livermore-3 & $\sin(x^3)\cos(x^2) - 1$ & U(-3, 3, 100) & 38\% & 60\% \\
Livermore-4 & $\log(x+1) + \log(x^2+1) + \log(x)$ & U(-3, 3, 100) & 100\% & 43\% \\
Livermore-5 & $x^4 - x^3 + x^2 - y$ & U(-3, 3, 100) & 100\% & 100\% \\
Livermore-6 & $4x^4 + 3x^3 + 2x^2 + x$ & U(-3, 3, 100) & 100\% & 100\% \\
Livermore-7 & $\frac{(\exp(x) - \exp(-x))}{2}$ & U(-1, 1, 100) & 0\% &0\% \\
Livermore-8 & $\frac{(\exp(x) + \exp(-x))}{2}$ & U(-1, 1, 100) & 0\% &0\% \\
Livermore-9 & $\sum_{i=1}^{9}x^i$ & U(-1, 1, 100) & 19\% &0\% \\
Livermore-10 & $6  \sin(x)\cos(y)$ & U(-3, 3, 100) & 100\% &100\% \\
Livermore-11 & $\frac{x^2 y^2}{(x+y)}$ & U(-3, 3, 100) & 76\% &100\% \\
Livermore-12 & $\frac{x^5}{y^3}$ & U(-3, 3, 100) & 18\% & 3\% \\
Livermore-13 & $x^{\frac{1}{3}}$ & U(-3, 3, 100) & 100\% &100\% \\
Livermore-14 & $x^3 + x^2 + x + \sin(x) + \sin(y^2)$ & U(-1, 1, 100) & 87\% &0\% \\
Livermore-15 & $x^{\frac{1}{5}}$ & U(-3, 3, 100) & 100\% &80\% \\
Livermore-16 & $x^{\frac{2}{3}}$ & U(-3, 3, 100) & 98\% &100\% \\
Livermore-17 & $4\sin(x)\cos(y)$ & U(-3, 3, 100) & 100\% &100\% \\
Livermore-18 & $\sin(x^2)\cos(x) - 5$ & U(-3, 3, 100) & 100\% &100\% \\
Livermore-19 & $x^5 + x^4 + x^2 + x$ & U(-3, 3, 100) & 100\% &100\% \\
Livermore-20 & $e^{-x^2}$ & U(-3, 3, 100) & 100\% &100\% \\
Livermore-21 & $\sum_{i=1}^{8}x^i$ & U(-1, 1, 20) & 68\% &0\% \\
Livermore-22 & $e^{-0.5x^2}$ & U(-3, 3, 100) & 100\% &100\% \\
 \rowcolor{gray!20} 
\bf{Livermore average} &  &  & 77.45\% & 67.36\% \\ 
 Jin-1 & $2.5x^4 - 1.3x^3 + 0.5y^2 - 1.7y$ & U(-3, 3, 100) & 99\% & 45\% \\ 
 Jin-2 & $8.0x^2 + 8.0y^3 - 15.0$ & U(-3, 3, 100) & 100\% & 100\% \\ 
 Jin-3 & $0.2x^3 + 0.5y^3 - 1.2y - 0.5x$ & U(-3, 3, 100) & 23\% & 12\% \\ 
 Jin-4 & $1.5\exp(x) + 5.0\cos(y)$ & U(-3, 3, 100) & 100\% & 100\% \\ 
 Jin-5 & $6.0\sin(x)\cos(y)$ & U(-3, 3, 100) & 100\% & 100\% \\ 
 Jin-6 & $1.35xy + 5.5\sin((x - 1.0)(y - 1.0))$ & U(-3, 3, 100) & 100\% & 100\% \\
  \rowcolor{gray!20} 
\bf{Jin average} &  &  & 87\% & 76.17\% \\  
 Constan-1 & $3.39x^3 + 2.12x^2 + 1.78x$ & U(-4, 4, 100) & 100\% & 100\% \\ 
 Constan-2 & $\sin(x^2) \cos(x) - 0.75$ & U(-4, 4, 100) & 97\% & 98\% \\ 
 Constan-3 & $\sin(1.5x) \cos(0.5y)$ & U(0.1, 4, 100) & 100\% & 100\% \\ 
 Constan-4 & $2.7x^y$ & U(0.3, 4, 100) & 100\% &100 \% \\ 
 Constan-5 & $\sqrt{1.23x}$ & U(0.1, 4, 100) & 100\% & 100\% \\ 
 Constan-6 & $x^{0.423}$ & U(0, 4, 100) & 100\% & 100\% \\ 
 Constan-7 & $2\sin(1.3x) \cos(y)$ & U(-4, 4, 100) & 100\% & 100\% \\ 
 Constan-8 & $\ln(x + 1.4) + \ln(x^2 + 1.3)$ & U(0, 4, 100) & 34\% & 10\% \\ 
  \rowcolor{gray!20} 
\bf{Constan average} &  &  & 91.375\% & 88.5\% \\ 
 R-1 & $\frac{(x+1)^3}{x^2-x+1} $ & E(-5, 5, 100) & 6\% & 8\% \\ 
 R-2 & $\frac{(x^5-3x^3+1)}{x^2+1} $ & E(-4, 4, 100) & 18\% & 6\% \\ 
 R-3 & $ \frac{x^6+x^5}{x^4+x^3+x^2+x}  $ & E(-4, 4, 100) & 100\% & 90\% \\
 \rowcolor{gray!20} 
\bf{R average} &  &  & 41.33\% & 34.67\% \\
 \rowcolor{gray!20} 
\bf{All average}&  &  & 84.07\% & 72.51\% \\
\end{longtable}
\newpage

\subsection{Model Evaluations on Equations Generated by LLMs}
\label{appendixB2}
\captionsetup{font=small,labelfont=bf}
\begin{longtable}{lcc}
    \caption{\textbf{Features extracted by FMN on recoverable equations.} Input data for the equations consist of 500 points sampled from a Gaussian distribution. The network utilizes the operator set $\{(\cdot)^2, \exp(\cdot), \sin(\cdot), \cos(\cdot), \times, + \}$ and adopts 3-layer and 4-layer architectures for univariate and multivariate equations, respectively. Numbers in parentheses indicate feature frequencies within the feature library, while blue features represent valid ones.} 
    
    \label{tab:Recover_function_features_multirow_9} \\
    
    \toprule
    \textbf{Identifier} & \textbf{Equation} & \textbf{Features} \\
    \midrule
    \endfirsthead
    
    \multicolumn{3}{c}{\tablename\ \thetable{} } \\
    \toprule
    \textbf{Identifier} & \textbf{Equation} & \textbf{Features} \\
    \midrule
    \endhead
    
    \bottomrule
    \endlastfoot
    
    \multirow{2}{*}{Recover-1} & \multirow{2}{*}{$x^6 \!+\! x^5 \!+\! x^4 \!+\! x^3 \!+\! x^2 \!+\! x$} & $\cos(x)(127), \textcolor{blue}{x^4(120)}, \textcolor{blue}{x^2(118)}, \sin(x)(108), e^{x}(98)$ \\
     & & $2x^2(44), \sin(x^2)(42), e^{x^2}(32), e^{\sin(x)}(26)$ \\ \cline{3-3}
    
    \multirow{2}{*}{Recover-2} & \multirow{2}{*}{$x^6 \!+\!2x^5 \!+\!3x^4 \!+\!4x^3 \!+\! x^2 \!+\! x$} & $\textcolor{blue}{x^4(124)}, \cos(x)(119), \textcolor{blue}{x^2(109)}, \sin(x)(92), e^{x}(86)$ \\
     & & $\sin(\sin(x))(67), 2x^2(62), e^{x^2}(49), \cos(\sin(x))(38)$ \\ \cline{3-3}
    
    \multirow{2}{*}{Recover-3} & \multirow{2}{*}{$x^6 \!+\!0.5x^5 \!+\!0.7x^4 \!+\!3x^3 \!+\! 5x^2 \!+\! x$} & $\cos(x)(119), \sin(x)(118), \textcolor{blue}{x^4(118)}, \textcolor{blue}{x^2(116)}, e^{x}(100)$ \\
     & & $2x^2(53), e^{x^2}(40), \sin(\sin(x))(24), \cos(\sin(x))(23)$ \\ \cline{3-3}
    
    \multirow{2}{*}{Recover-4} & \multirow{2}{*}{$\sum_{i=1}^{9}x^i$} & $\textcolor{blue}{x^4(192)}, \cos(\cos(x))(124), e^{x}(108), \textcolor{blue}{x^2(95)}, \sin(x)(73)$ \\
     & & $\cos(\cos(x))^2(68), 2\cos(\cos(x))(67), \sin(x^2)(62), 2x^2(57)$ \\ \cline{3-3}
    
    \multirow{2}{*}{Recover-5} & \multirow{2}{*}{$\sin(x^2)\cos(x)\!+\! \sin(x) \!+\! \sin(x^4 \!+\! x^2)$} & $\textcolor{blue}{\sin(x)(131)}, \textcolor{blue}{\cos(x)(125)}, e^{x}(95), \textcolor{blue}{x^2(84)}, \textcolor{blue}{x^4(41)}$ \\
     & & $\sin(x)^2(31), e^{x^2}(27), e^{\sin(x)}(26), \cos(\cos(x))(18)$ \\ \cline{3-3}
    
    \multirow{2}{*}{Recover-6} & \multirow{2}{*}{$\frac{xy}{e^{x}\!+\!\sin(y)^2}$} & $\textcolor{blue}{\sin(y)(105)}, \textcolor{blue}{xy(81)}, y^2(68), \textcolor{blue}{e^{y}(65)}, \cos(y)(64)$ \\
     & & $x + y(56), x^2(56), \cos(x)(54), e^{x}(50)$ \\ \cline{3-3}
    
    \multirow{2}{*}{Recover-7} & \multirow{2}{*}{$2x^4\!+\!2y^4\!-\!6x^2y^2$} & $\sin(y)(137), e^{y}(107), \textcolor{blue}{y^2(106)}, \textcolor{blue}{x^2(102)}, \cos(y)(75)$ \\
     & & $\textcolor{blue}{xy(75)}, \cos(x)(68), 2xy(59), 2y(58)$ \\ \cline{3-3}
    
    \multirow{2}{*}{Recover-8} & \multirow{2}{*}{$x^5\!+\!3x^3y^2\!-\!x^2y^3\!+\!y^5$} & $\textcolor{blue}{x^2(119)}, e^{x}(117), \sin(y)(95), \textcolor{blue}{y^2(83)}, \cos(x)(71)$ \\
     & & $\sin(x)(62), e^{y}(56), \textcolor{blue}{2y(51)}, \textcolor{blue}{2x^2(51)}$ \\ \cline{3-3}
    
    \multirow{2}{*}{Recover-9} & \multirow{2}{*}{$e^{x^2\!+\!\sin(y)}$} & $\textcolor{blue}{x^2(169)}, \textcolor{blue}{e^{x}(96)}, \cos(\cos(x))(93), \sin(x)(84), \cos(x)(76)$ \\
     & & $2x(71), \cos(\cos(x))^2(58), y^2(55), \textcolor{blue}{\sin(y)(49)}$ \\ \cline{3-3}
    
    \multirow{2}{*}{Recover-10} & \multirow{2}{*}{$x^3\!+\!x^2\!+\!x\!+\!\sin(x)\!+\!\sin(y^2)$} & $e^{x}(132), \textcolor{blue}{\sin(x)(103)}, \cos(\cos(x))(100), \textcolor{blue}{x^2(95)}, 2x(69)$ \\
     & & $\sin(y)(64), \cos(x)(61), xy(57), x + y(48)$ \\ \cline{3-3}
    
    \multirow{2}{*}{Recover-11} & \multirow{2}{*}{$x^4y \!-\! x^3 \!+\! 0.5y^2\cos(x)\!-\!x$} & $\textcolor{blue}{y^2(115)}, \sin(x)(99), \textcolor{blue}{x^2(90)}, e^{y}(65), \cos(y)(65)$ \\
     & & $e^{x}(61), \textcolor{blue}{\cos(x)(54)}, \cos(\cos(x))(53), \sin(y)(49)$ \\ \cline{3-3}
    
    \multirow{2}{*}{Recover-12} & \multirow{2}{*}{$x^4y \!-\! x^3 \!+\! 0.5y^2\sin(x)\!-\! x$} & $\textcolor{blue}{x^2(186)}, \textcolor{blue}{\sin(x)(86)}, \cos(\cos(x))(74), e^{y}(60), \cos(x)(58)$ \\
     & & $e^{x}(57), \sin(\cos(x))(57), \textcolor{blue}{y^2(51)}, \cos(y)(44)$ \\ \cline{3-3}
    
    \multirow{2}{*}{Recover-13} & \multirow{2}{*}{$xytanh(x\!+\!y)$} & $\textcolor{blue}{e^{xy}(74)}, \textcolor{blue}{xy(63)}, \textcolor{blue}{x+y(60)}, \cos(\cos(x))(58), \sin(y)(57)$ \\
     & & $\cos(\cos(y))(55), \sin(x)(54), \textcolor{blue}{e^{y}(49)}, \cos(x)(45)$ \\ \cline{3-3}
    
    \multirow{2}{*}{Recover-14} & \multirow{2}{*}{$\cos(y^2) \sin(x) \!-\! 1 \!+\! \sqrt{x^2 \!+\! y^2 \!+\! 1}$} &  $\textcolor{blue}{\sin(x)(97)}, e^x(94), xy(82), \textcolor{blue}{x^2(78)}, \cos(y)(72)$ \\
     & & $x\!+\!y(63),\textcolor{blue}{y^2(57)},\cos(\cos(y))(54),\sin(y)(51),e^{y}(47)$ \\ \cline{3-3}
    
    \multirow{2}{*}{Recover-15} & \multirow{2}{*}{$\cos(xy) \cos(y) \!-\! 1 \!+\! \sqrt{x^2 \!+\! y^2 \!+\! 1}$} & $\textcolor{blue}{xy(83)}, \textcolor{blue}{\cos(y)(69)}, \textcolor{blue}{x + y(68)}, \textcolor{blue}{y^2(63)}, \sin(y)(63)$ \\
     & & $x^2(59), e^{x}(58), \sin(x)(55), \cos(\cos(y))(50)$ \\ \cline{3-3}
    
    \multirow{2}{*}{Recover-16} & \multirow{2}{*}{$\frac{e^{(1\!+\!x)}(1 \!-\! x) \!-\! e^{y}x}{e^{(1\!+\! x)} \!+\!e^{y}}$} & $\textcolor{blue}{e^{x}(84)}, \sin(y)(82), \sin(x)(75), \textcolor{blue}{-x + y(74)}, \textcolor{blue}{e^{y-x}(73)}$ \\
     & & $-x^2 + y(65), xy(63), y^2(62), x - y^2(62)$ \\ \cline{3-3}
    
    \multirow{2}{*}{Recover-17} & \multirow{2}{*}{$\frac{e^{x}\sin(y)\!-\! e^{y}\cos(x)}{xy}$} & $\textcolor{blue}{\sin(y)(94)}, \textcolor{blue}{\cos(x)(88)}, \textcolor{blue}{2xy(83)}, xy^2(82), \textcolor{blue}{xy(78)}$ \\
     & & $\cos(\cos(x))(77), 2x(68), y^2(60), \textcolor{blue}{e^{y}(60)}$ \\ \cline{3-3}
    
    \multirow{2}{*}{Recover-18} & \multirow{2}{*}{$\frac{e^{x}\cos(y)\!-\! e^{y}\sin(x)}{xy}$} & $\textcolor{blue}{\sin(x)(73)}, \textcolor{blue}{\cos(y)(63)}, x^2(62), \textcolor{blue}{e^{x}(58)}, \textcolor{blue}{e^{y}(55)}$ \\
     & & $\sin(y)(55), \textcolor{blue}{2xy(53)}, \textcolor{blue}{xy(47)}, y^2(46)$ \\ \cline{3-3}
    
    \multirow{2}{*}{Recover-19} & \multirow{2}{*}{$e^{\!-\!0.5(\sin^{2}(x) \!+\! \cos^{2}(y))}\cos(xy)$} & $\textcolor{blue}{xy(79)}, \sin(y)(78), \textcolor{blue}{x + y(70)}, \sin(x)(64), e^{x}(58)$ \\
     & & $xy^2(53), y^2(51), \cos(\cos(x))(51), \cos(y)(49)$ \\ \cline{3-3}
    
    \multirow{2}{*}{Recover-20} & \multirow{2}{*}{$\sin\left(x \!+\! e^{\!-\!y^{2}}\right) \!-\! \cos\left(y \!-\! e^{\!-\!x^{2}}\right)$} & $\sin(y)(96), xy(80), x + y(80), \textcolor{blue}{y^2(76)}, \textcolor{blue}{e^{y}(75)}$ \\
     & & $\sin(x)(69), \textcolor{blue}{e^{x}(63)}, \cos(\cos(y))(61), \cos(y)(55)$ \\ \cline{3-3}
    
    \multirow{2}{*}{Recover-21} & \multirow{2}{*}{$x^4 \!+\! y^2z^2 \!-\! x^2z^2 \!+\! y^4$} & $\textcolor{blue}{y^2(162)}, \textcolor{blue}{x^2(96)}, e^{y}(93), \textcolor{blue}{z^2(75)}, \sin(y)(74)$ \\
     & & $\cos(y)(74), 2x(68), e^{x}(59), \sin(z)(56)$ \\ \cline{3-3}
    
    \multirow{2}{*}{Recover-22} & \multirow{2}{*}{$x^5 \!-\! y^4z \!+\! z^3x^2 \!-\! xyz \!-\! x \!+\! y$} & $\textcolor{blue}{x^2(143)}, e^{x}(123), \textcolor{blue}{y^2(97)}, \textcolor{blue}{z^2(94)}, 2x(80)$ \\
     & & $\sin(y)(72), \sin(z)(69), \cos(x)(68), \sin(x)(60)$ \\ \cline{3-3}
    
    \multirow{2}{*}{Recover-23} & \multirow{2}{*}{$e^{\sin(x)}y^3\!+\!\cos(e^{z})$} & $e^{y}(110), \textcolor{blue}{y^2(74)}, \textcolor{blue}{\sin(x)(60)}, xy(52), \cos(x)(47)$ \\
     & & $\cos(\cos(y))(45), \sin(y)(43), \textcolor{blue}{e^{z}(39)}, \cos(\cos(x))(39)$ \\ \cline{3-3}
    
    \multirow{2}{*}{Recover-24} & \multirow{2}{*}{$2(x^2\!+\!y^2)\sin(z)\!+\!xe^{x\!+\!y}$} & $\textcolor{blue}{\sin(z)(96)}, \textcolor{blue}{x^2(83)}, \textcolor{blue}{e^{x}(77)}, \textcolor{blue}{y^2(62)}, z^2(60)$ \\
     & & $\cos(\cos(z))(54), \cos(z)(52), \textcolor{blue}{e^{y}(41)}, \sin(y)(40)$ \\ \cline{3-3}
    
    \multirow{2}{*}{Recover-25} & \multirow{2}{*}{$0.5\sin(x\!+\!y)z^4\!+\!x^2y^3z$} & $\textcolor{blue}{z^2(103)}, \textcolor{blue}{x^2(83)}, \textcolor{blue}{y^2(81)}, \textcolor{blue}{\sin(x)(70)}, \sin(z)(64)$ \\
     & & $\textcolor{blue}{\sin(y)(61)}, \cos(z)(60), e^{z}(57), xy(50)$ \\ \cline{3-3}
    
    \multirow{2}{*}{Recover-26} & \multirow{2}{*}{$\sqrt{z^4\!+\!1}\cos(y\!+\!e^x)$} & $\sin(x)(82), \textcolor{blue}{z^2(71)}, \cos(x)(58), \textcolor{blue}{e^{z}(41)}, \sin(z)(39)$ \\
     & & $e^{y}(37), e^{x}(37), \cos(\cos(x))(37), y^2(36)$ \\ \cline{3-3}
    
    \multirow{2}{*}{Recover-27} & \multirow{2}{*}{$x^4y \!+\! y^2z^3\sin(z) \!+\!xz^4$} & $\textcolor{blue}{x^2(88)}, \textcolor{blue}{y^2(73)}, \textcolor{blue}{z^2(69)}, e^{x}(68), \textcolor{blue}{\sin(z)(63)}$ \\
     & & $\cos(z)(63), \cos(\cos(z))(57), e^{z}(53), \sin(x)(48)$ \\ \cline{3-3}
    
    \multirow{2}{*}{Recover-28} & \multirow{2}{*}{$x^3y^5\!+\!\cos(z^2)y\!-\! e^{xz}$} & $\textcolor{blue}{x^2(107)}, 2y(82), \textcolor{blue}{y^2(74)}, \cos(\cos(x))(72), \textcolor{blue}{e^{y}(71)}$ \\
     & & $\textcolor{blue}{e^{x}(65)}, \sin(x)(55), \sin(y)(49), \cos(x)(46)$ \\ \cline{3-3}
    
    \multirow{2}{*}{Recover-29} & \multirow{2}{*}{$z^4\cos(x)e^{\sin(y)}\!+\!xe^{y\!+\!z}$} & $\textcolor{blue}{e^{z}(111)}, \textcolor{blue}{z^2(92)}, \cos(\cos(z))(77), \sin(z)(74), \cos(z)(71)$ \\
     & & $x^2(65), \sin(x)(61), e^{x}(61), \cos(z)^2(58)$ \\ \cline{3-3}
    
    \multirow{2}{*}{Recover-30} & \multirow{2}{*}{$0.5\sin(x)z^4\!+\!x^2y^3z$} & $\textcolor{blue}{z^2(75)}, \textcolor{blue}{\sin(x)(58)}, \cos(\cos(z))(55), \textcolor{blue}{y^2(54)}, \sin(y)(40)$ \\
     & & $\textcolor{blue}{x^2(38)}, e^{x}(38), e^{yz}(38), \cos(z)(37)$ \\ \cline{3-3}
    
    \multirow{2}{*}{Recover-31} & \multirow{2}{*}{$0.9(x^2\!+\!y^2)\sin(3z\!+\!2x)\!+\!xe^{x\!-\! y}$} & $\textcolor{blue}{e^{x}(153)}, \textcolor{blue}{x^2(137)}, \sin(z)(67), 2x(64), \cos(\cos(z))(64)$ \\
     & & $\textcolor{blue}{y^2(61)}, \cos(z)(48), \sin(y)(45), y + z(36)$ \\ \cline{3-3}
    
    \multirow{2}{*}{Recover-32} & \multirow{2}{*}{$(z^2\!+\!1)/(e^x\!+\!\cos(y)^2)$} & $\textcolor{blue}{e^{x}(89)}, \textcolor{blue}{z^2(84)}, y^2(70), \sin(z)(58), \sin(x)(56)$ \\
     & & $\cos(\cos(z))(51), \cos(z)(49), e^{z}(47), \sin(y)(45)$ \\ \cline{3-3}
    
    \multirow{2}{*}{Recover-33} & \multirow{2}{*}{$(z^2\!+\!1)/(e^y\!+\!\sin(x)^2)$} & $\textcolor{blue}{z^2(74)}, \textcolor{blue}{\sin(x)(67)}, \textcolor{blue}{e^{y}(67)}, e^{z}(65), \cos(\cos(x))(63)$ \\
     & & $x^2(54), \cos(x)(46), \sin(y)(39), y^2(33)$ \\ \cline{3-3}
    
    \multirow{2}{*}{Recover-34} & \multirow{2}{*}{$x^3y \!-\! y^3z \!+\! z^3h \!-\! h^3x \!+\! xyzh$} & $\textcolor{blue}{z^2(71)}, \textcolor{blue}{h^2(71)}, \textcolor{blue}{x^2(70)}, \textcolor{blue}{y^2(67)}, e^{x}(60)$ \\
     & & $\sin(h)(55), \sin(y)(55), \cos(x)(54), \sin(x)(48)$ \\ \cline{3-3}
    
    \multirow{2}{*}{Recover-35} & \multirow{2}{*}{$x^4 \!+\! y^4 \!-\! z^4 \!-\! h^4 \!+\! x^2z^2$} & $\textcolor{blue}{x^2(95)}, \textcolor{blue}{y^2(78)}, \sin(x)(72), \textcolor{blue}{z^2(69)}, \cos(x)(63)$ \\
     & & $e^{x}(60), \textcolor{blue}{h^2(59)}, \sin(y)(50), e^{h}(42)$ \\ \cline{3-3}
    
    \multirow{2}{*}{Recover-36} & \multirow{2}{*}{$x^5 \!-\! y^4 \!+\! z^3 \!-\! h^2 \!+\! xyzh$} & $\textcolor{blue}{x^2(124)}, 2x(112), \textcolor{blue}{y^2(99)}, e^{x}(87), \textcolor{blue}{z^2(81)}$ \\
     & & $\sin(y)(71), \sin(x)(58), \cos(y)(55), \cos(x)(47)$ \\

\end{longtable}

\newpage
\subsubsection{Detailed Exact Recovery Rate Analysis}
\label{appendixB2.1}
\begin{longtable}{cccccccc}
  \captionsetup{font=small,labelfont=bf}
\caption{\textbf{Comparision of FePySR, PySR, and DSO on the recoverable equations.} We test each equation 100 times for all methods. FePySR employs only the top four features for SR.
}

\label{full_comparison_condensed} \\
\toprule
\multirow{2}{*}{\textbf{Identifier}} & \textbf{FePySR} & \textbf{PySR} & \textbf{DSO} & \multirow{2}{*}{\textbf{Identifier}} & \textbf{FePySR} & \textbf{PySR} & \textbf{DSO} \\
                            & \textbf{Runtime } & \textbf{Runtime } & \textbf{Runtime } &                             & \textbf{Runtime } & \textbf{Runtime } & \textbf{Runtime } \\
\midrule
\endfirsthead
\multicolumn{8}{c}{\tablename\ \thetable{}} \\
\toprule
\multirow{2}{*}{\textbf{Identifier}} & \textbf{FePySR} & \textbf{PySR} & \textbf{DSO} & \multirow{2}{*}{\textbf{Identifier}} & \textbf{FePySR} & \textbf{PySR} & \textbf{DSO} \\
                            & \textbf{Runtime } & \textbf{Runtime } & \textbf{Runtime } &                             & \textbf{Runtime } & \textbf{Runtime } & \textbf{Runtime } \\
\midrule
\endhead
\bottomrule
\endlastfoot

\multirow{2}{*}{Recover-1} & 100\% & 76.8\% & 100\% & \multirow{2}{*}{Recover-19} & 24\% & 18\% & 0\% \\
                           & 4.82s & 8.92s & 9.33s &                            & 46.5s & 59.5s & \textbackslash \\
 \cline{2-4}  \cline{6-8}
\multirow{2}{*}{Recover-2} & 98\% & 0\% & 100\% & \multirow{2}{*}{Recover-20} & 39\% & 12\% & 0\% \\
                           & 14.0s & \textbackslash & 8.87s &                            & 77.5s & 135.9s & \textbackslash \\
 \cline{2-4}  \cline{6-8}
\multirow{2}{*}{Recover-3} & 14\% & 0\% & 100\% & \multirow{2}{*}{Recover-21} & 100\% & 0\% & 100\% \\
                           & 47.6s & \textbackslash & 8.84s &                           & 6.74s & \textbackslash & 19.7s \\
 \cline{2-4}  \cline{6-8}
\multirow{2}{*}{Recover-4} & 100\% & 0\% & 100\% & \multirow{2}{*}{Recover-22} & 100\% & 0\% & 100\% \\
                           & 8.51s & \textbackslash & 8.79s &                           & 49.3s & \textbackslash & 17.4s \\
 \cline{2-4}  \cline{6-8}
\multirow{2}{*}{Recover-5} & 98\% & 57\% & 0\% & \multirow{2}{*}{Recover-23} & 100\% & 100\% & 0\% \\
                           & 16.3s & 159s & \textbackslash &                           & 10.5s & 23.7s & \textbackslash \\
 \cline{2-4}  \cline{6-8}
\multirow{2}{*}{Recover-6} & 100\% & 0\% & 0\% & \multirow{2}{*}{Recover-24} & 100\% & 0\% & 0\% \\
                           & 22.0s & \textbackslash & \textbackslash &                            & 8.69s & \textbackslash & \textbackslash \\
 \cline{2-4}  \cline{6-8}
\multirow{2}{*}{Recover-7} & 100\% & 32\% & 100\% & \multirow{2}{*}{Recover-25} & 99\% & 80\% & 0\% \\
                           & 7.21s & 32.5s & 15.1s &                          & 20.9s & 35.1s & \textbackslash \\
 \cline{2-4}  \cline{6-8}
\multirow{2}{*}{Recover-8} & 49\% & 0\% & 100\% & \multirow{2}{*}{Recover-26} & 81\% & 7\% & 0\% \\
                           & 26.1s & \textbackslash & 18.9s &                          & 114s & 79.6s & \textbackslash \\
 \cline{2-4}  \cline{6-8}
\multirow{2}{*}{Recover-9} & 100\% & 100\% & 100\% & \multirow{2}{*}{Recover-27} & 82\% & 0\% & 0\% \\
                           & 5.41s & 24.1s & 251s &                           & 36.0s & \textbackslash & \textbackslash \\
 \cline{2-4}  \cline{6-8}
\multirow{2}{*}{Recover-10} & 100\% & 100\% & 100\% & \multirow{2}{*}{Recover-28} & 100\% & 0\% & 0\% \\
                            & 31.8s & 18.5s & 1256s &                          & 26.1s & \textbackslash & \textbackslash \\
 \cline{2-4}  \cline{6-8}
\multirow{2}{*}{Recover-11} & 100\% & 87\% & 0\% & \multirow{2}{*}{Recover-29} & 88\% & 34\% & 0\% \\
                            & 20.0s & 68.8s & \textbackslash &                           & 92.3s & 27.6s & \textbackslash \\
 \cline{2-4}  \cline{6-8}
\multirow{2}{*}{Recover-12} & 100\% & 0\% & 0\% & \multirow{2}{*}{Recover-30} & 55\% & 0\% & 0\% \\
                            & 15.5s & \textbackslash & \textbackslash &                          & 52.0s & \textbackslash & \textbackslash \\
 \cline{2-4}  \cline{6-8}
\multirow{2}{*}{Recover-13} & 99\% & 0\% & 0\% & \multirow{2}{*}{Recover-31} & 85\% & 0\% & 0\% \\
                            & 244s & \textbackslash & \textbackslash &                         & 82.7s & \textbackslash & \textbackslash \\
 \cline{2-4}  \cline{6-8}
\multirow{2}{*}{Recover-14} & 94\% & 0\% & 0\% & \multirow{2}{*}{Recover-32} & 100\% & 14\% & 0\% \\
                            & 78.8s & \textbackslash & \textbackslash &                         & 76.2s & 164s & \textbackslash \\
 \cline{2-4}  \cline{6-8}
\multirow{2}{*}{Recover-15} & 39\% & 0\% & 0\% & \multirow{2}{*}{Recover-33} & 83\% & 2\% & 0\% \\
                            & 137s & \textbackslash & \textbackslash &                        & 62.4s & 91.6s & \textbackslash \\
 \cline{2-4}  \cline{6-8}
\multirow{2}{*}{Recover-16} & 99\% & 0\% & 0\% & \multirow{2}{*}{Recover-34} & 62\% & 0\% & 100\% \\
                            & 73.2s & \textbackslash & \textbackslash &                        & 63.5s & \textbackslash & 45.5s \\
 \cline{2-4}  \cline{6-8}
\multirow{2}{*}{Recover-17} & 7\% & 0\% & 0\% & \multirow{2}{*}{Recover-35} & 81\% & 0\% & 100\% \\
                            & 205s & \textbackslash & \textbackslash &                       & 57.6s & \textbackslash & 36.7s \\
 \cline{2-4}  \cline{6-8}
\multirow{2}{*}{Recover-18} & 99\% & 9\% & 0\% & \multirow{2}{*}{Recover-36} & 100\% & 0\% & 100\% \\
                            & 51.3s & 299s & \textbackslash &                          & 83.3s & \textbackslash & 35.9s

\end{longtable}
\newpage

\subsubsection{Feature Quantity Sensitivity Analysis}
\label{appendixB2.2}
\begin{longtable}{lcccccc}
  \captionsetup{font=small,labelfont=bf}
\caption{\textbf{Effect of the number of selected top-n features on the recovery rate and runtime of FePySR.} Results are reported for 12 recoverable equations, with the number of input features increasing from four to nine. The notation $a/b$ indicates that $a$ out of $b$ selected features successfully reduce the search space for SR.} 
\label{tab:full_comparison} \\

\toprule
\textbf{Identifier} & \textbf{4-Features} & \textbf{5-Features} & \textbf{6-Features} &\textbf{7-Features} &\textbf{8-Features} &\textbf{9-Features} \\
\midrule
\endfirsthead

\multicolumn{7}{c}{\tablename\ \thetable{}} \\
\toprule
\textbf{Identifier} & \textbf{4-Features} & \textbf{5-Features} & \textbf{6-Features} &\textbf{7-Features} &\textbf{8-Features} &\textbf{9-Features}  \\
\midrule
\endhead

\bottomrule
\endlastfoot


\multirow{3}{*}{Recover-1} 
  &2/4 & 2/5 & 2/6 & 2/7 & 2/8 & 2/9 \\
  & 100\% & 100\% & 100\% & 100\% & 100\% & 100\% \\
  & 4.822s & 5.257s & 5.612s & 6.065s & 6.083s & 6.075s \\
\cline{2-7}

\multirow{3}{*}{Recover-4} 
  &2/4 & 2/5 & 2/6 & 2/7 & 2/8 & 2/9 \\
  & 100\% & 100\% & 99\% & 100\% & 100\% & 100\% \\
  & 8.514s & 8.3099s & 8.649s & 10.18s & 10.37s & 10.123s \\
\cline{2-7}

\multirow{3}{*}{Recover-5} 
  &3/4 & 4/5 & 4/6 &4 /7 & 4/8 & 4/9 \\
  & 100\% & 100\% & 100\% & 100\% & 100\% & 100\% \\
  & 16.25s & 11.73s & 13.476s & 15.87s & 20.181s & 28.853s \\
\cline{2-7}

\multirow{3}{*}{Recover-7}
  &2/4 & 2/5 & 3/6 & 3/7 & 3/8 & 3/9 \\
  & 100\% & 100\% & 100\% & 100\% & 100\% & 100\% \\
  & 7.208s & 7.534s & 7.860s & 8.745s & 9.610s & 9.644s \\
\cline{2-7}

\multirow{3}{*}{Recover-10}
  &2/4 & 2/5 & 2/6 & 2/7 & 2/8 & 2/9 \\
  & 100\% & 98\% & 100\% & 98\% & 98\% & 96\% \\
  & 31.845s & 39.762s & 30.890s &  47.211s & 48.001s &64.177s \\
\cline{2-7}

\multirow{3}{*}{Recover-12}
  &2/4 & 2/5 & 2/6 & 2/7 & 3/8 & 3/9 \\
  & 100\% & 100\% & 99\% & 100\% & 100\% & 100\% \\
  & 15.484s & 18.159s & 23.564s & 12.037s & 13.966s & 13.219s \\
\cline{2-7}

\multirow{3}{*}{Recover-14}
  &2/4 & 2/5 & 2/6 & 3/7 & 3/8 & 3/9 \\
  & 94\% & 93\% & 90\% & 100\% & 100\% & 100\% \\
  & 78.679s & 114.423s & 125.12s & 14.685s & 16.146s & 19.111s \\
\cline{2-7}

\multirow{3}{*}{Recover-15}
  &3/4 & 3/5 & 4/6 & 4/7 & 4/8 & 4/9 \\
  & 39\% & 31\% & 69\% & 73\% & 63\% & 60\% \\
  & 133.630s & 163.948s & 86.313s & 87.653s & 94.026s & 91.506s \\
\cline{2-7}

\multirow{3}{*}{Recover-21}
  &3/4 & 3/5 & 3/6 & 3/7 & 3/8 & 3/9 \\
  & 100\% & 100\% & 99\% & 100\% & 100\% & 100\% \\
  & 6.741s & 6.699s & 6.648s & 6.700s & 6.714s & 7.489s \\
\cline{2-7}

\multirow{3}{*}{Recover-22}
  &3/4 & 3/5 & 3/6 & 3/7 & 3/8 & 3/9 \\
  & 84\% & 89\% & 91\% & 88\% & 84\% & 85\% \\
  & 49.263s & 64.217s & 52.036s & 70.791s & 58.829s & 59.513s \\
\cline{2-7}

\multirow{3}{*}{Recover-30}
  &3/4 & 3/5 & 4/6 & 4/7 & 4/8 & 4/9 \\
  & 55\% & 43\% & 95\% & 97\% & 94\% & 95\% \\
  & 51.993s & 50.541s & 28.483s & 18.764s & 17.34s & 24.02s \\
\cline{2-7}

\multirow{3}{*}{Recover-34}
  &4/4 & 4/5 & 4/6 & 4/7 & 4/8 & 4/9 \\
  & 62\% & 63\% & 57\% & 56\% & 44\% & 37\% \\
  & 63.513s & 68.0413s & 70.343s & 70.414s & 71.270s & 70.471s \\

\end{longtable}

\newpage

\begin{figure}[H]
    \centering
    \captionsetup{font=small,labelfont=bf}
    \captionsetup[subfigure]{labelformat=empty}
    \begin{subfigure}[b]{0.28\textwidth}
        \includegraphics[width=\textwidth]{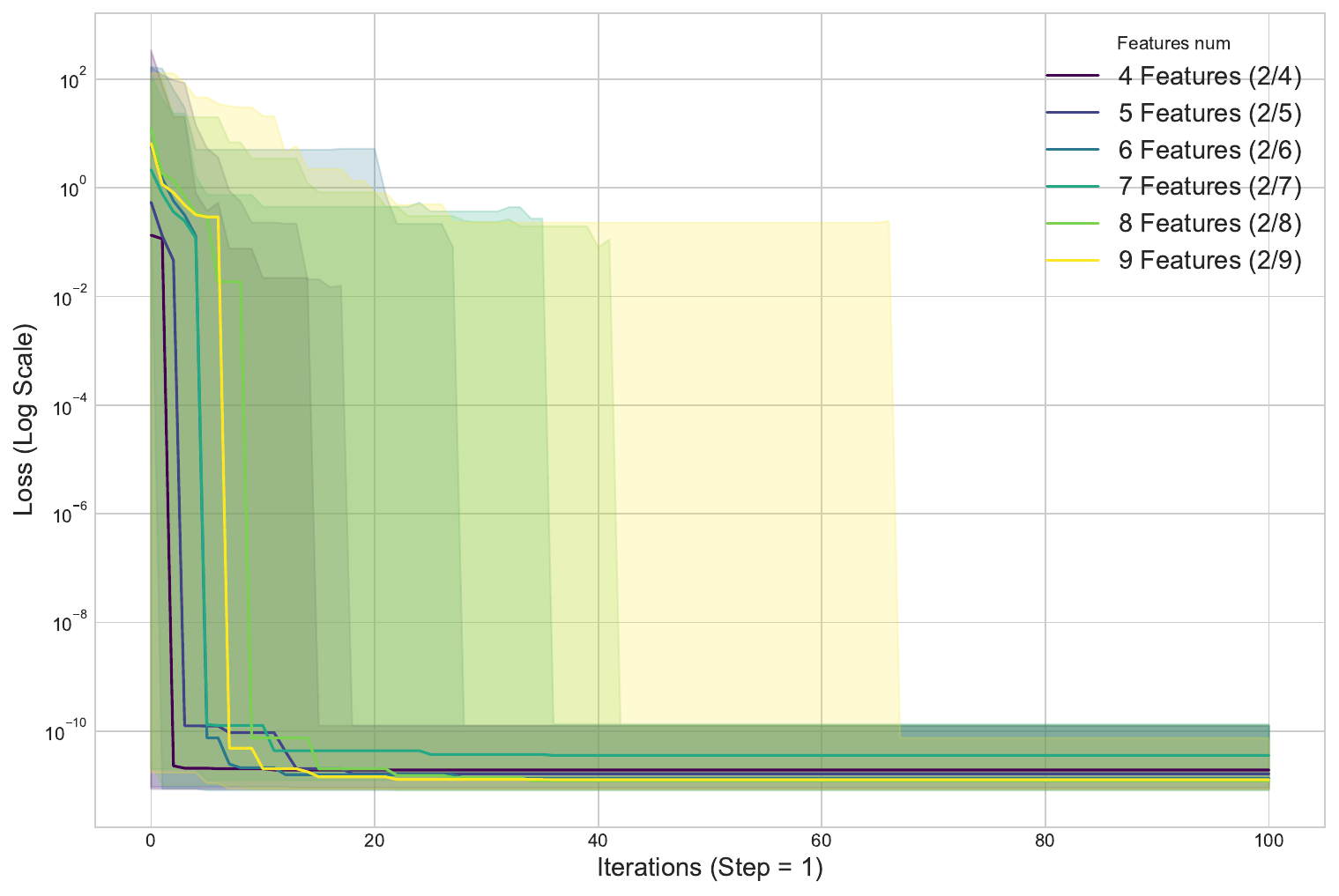}
        \caption{Recover-1}
        \label{fig:loss1}
    \end{subfigure}
    \hfill
    \begin{subfigure}[b]{0.28\textwidth}
        \includegraphics[width=\textwidth]{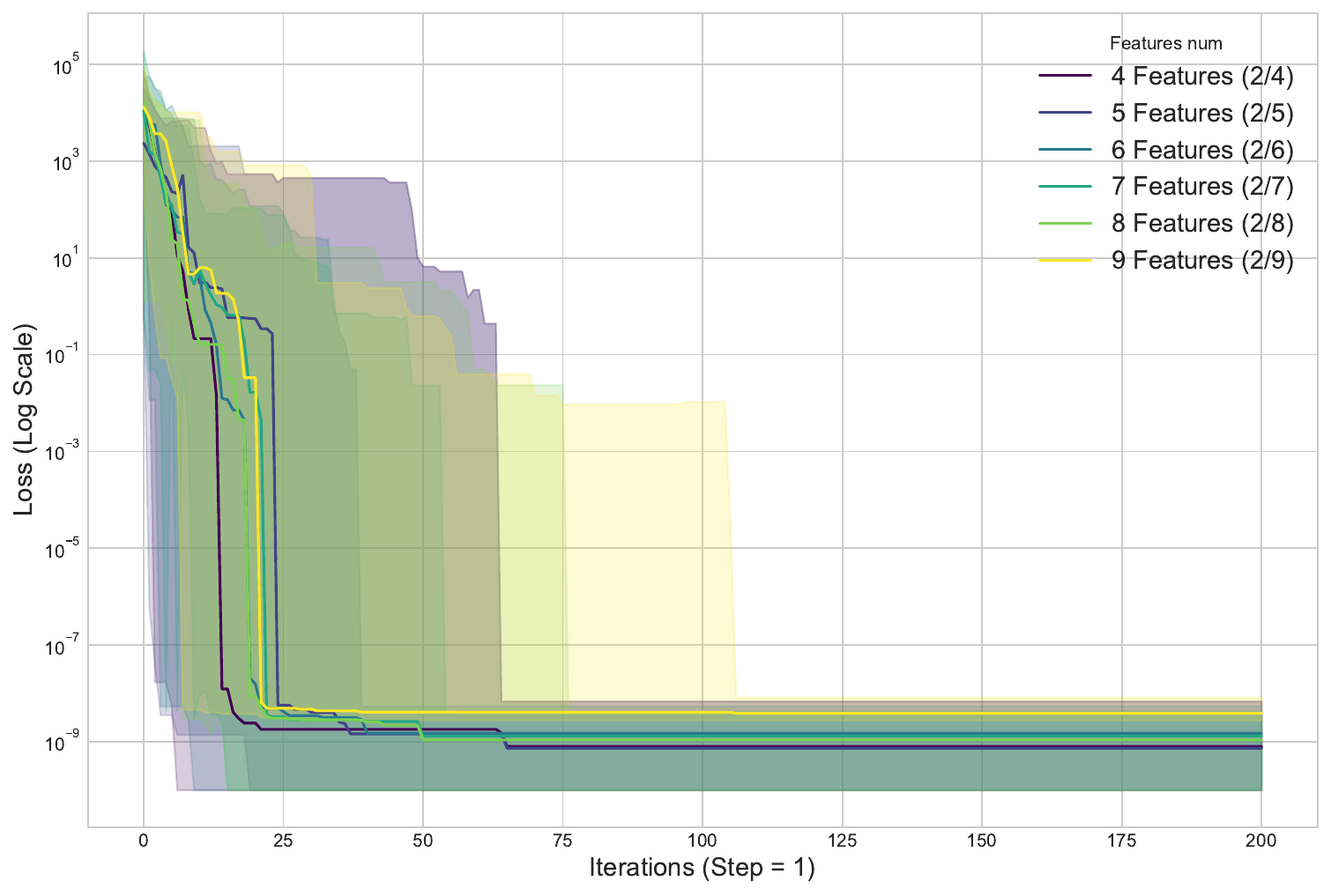}
        \caption{Recover-4}
        \label{fig:loss2}
    \end{subfigure}
    \hfill
    \begin{subfigure}[b]{0.28\textwidth}
        \includegraphics[width=\textwidth]{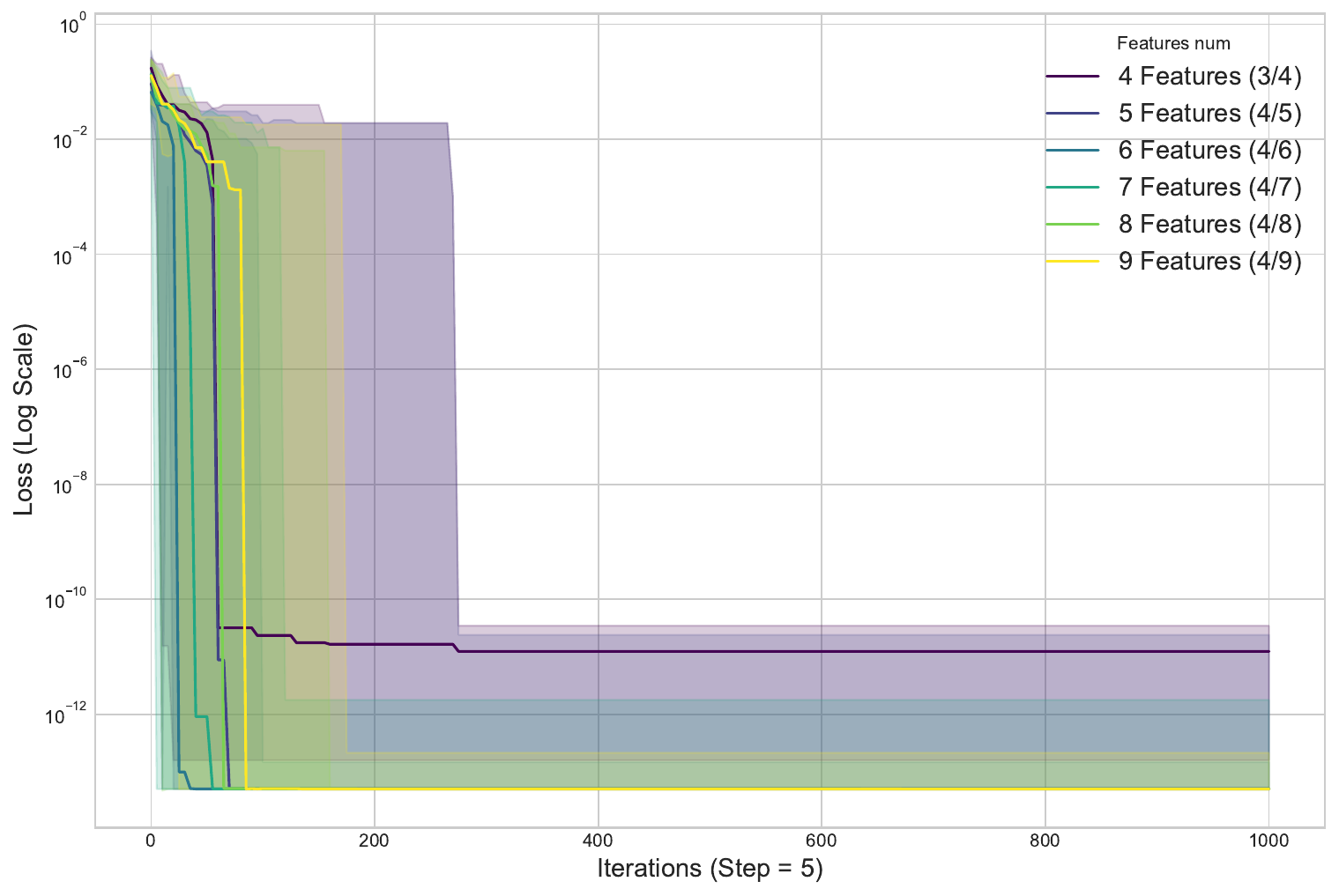}
        \caption{Recover-5}
        \label{fig:loss3}
    \end{subfigure}

    \vspace{1em} 

    \begin{subfigure}[b]{0.28\textwidth}
        \includegraphics[width=\textwidth]{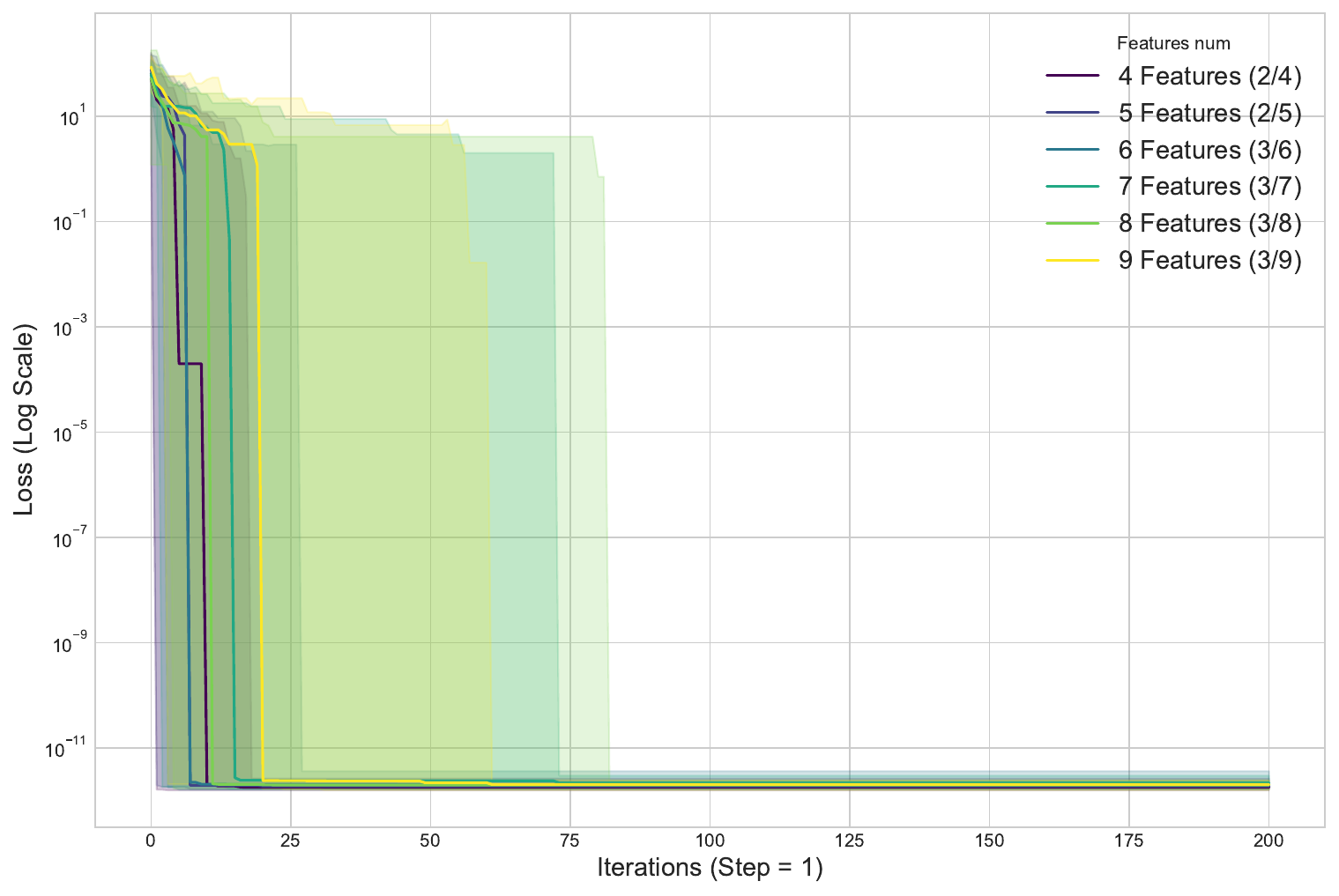}
        \caption{Recover-7}
        \label{fig:loss4}
    \end{subfigure}
    \hfill
    \begin{subfigure}[b]{0.28\textwidth}
        \includegraphics[width=\textwidth]{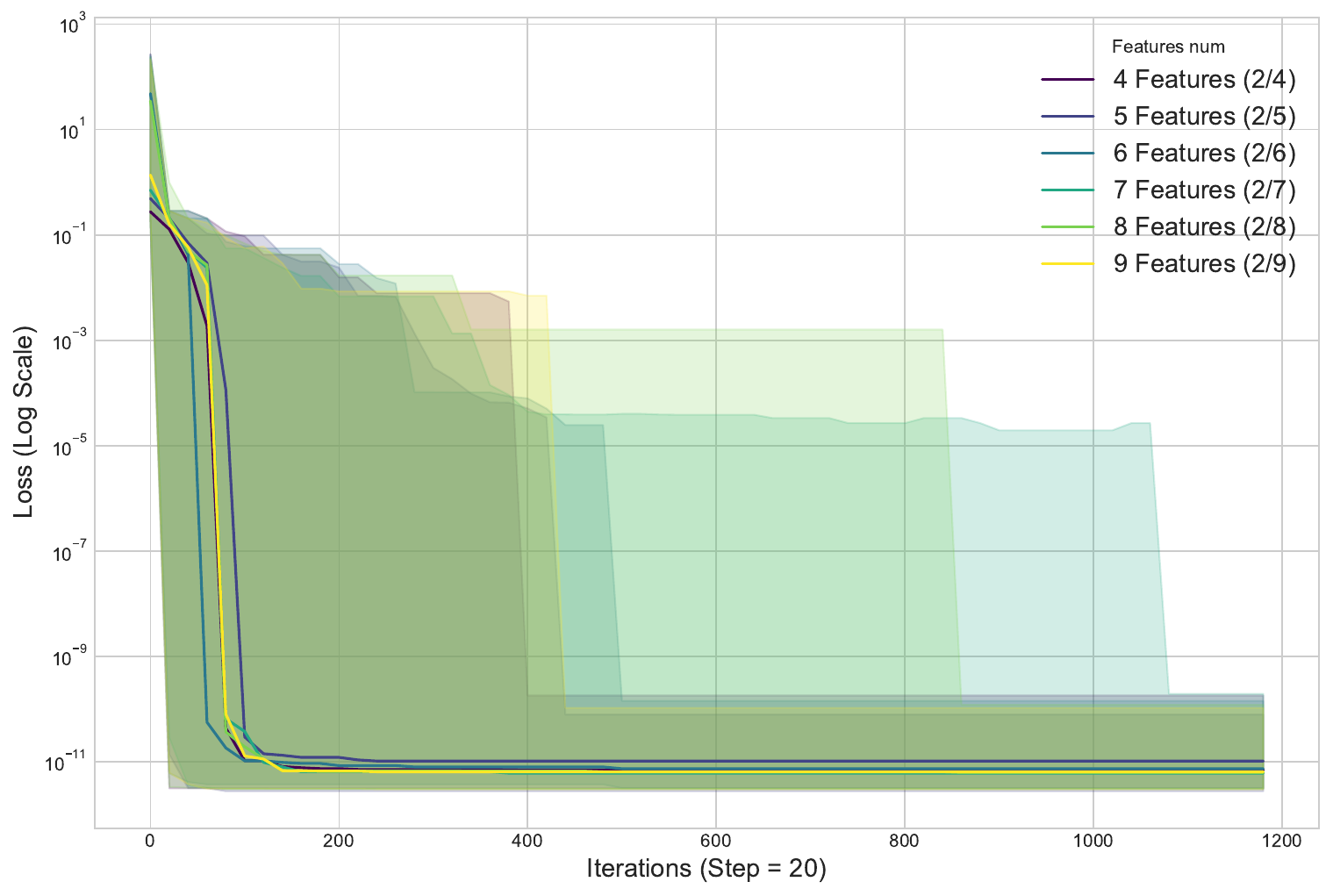}
        \caption{Recover-10}
        \label{fig:loss5}
    \end{subfigure}
    \hfill
    \begin{subfigure}[b]{0.28\textwidth}
        \includegraphics[width=\textwidth]{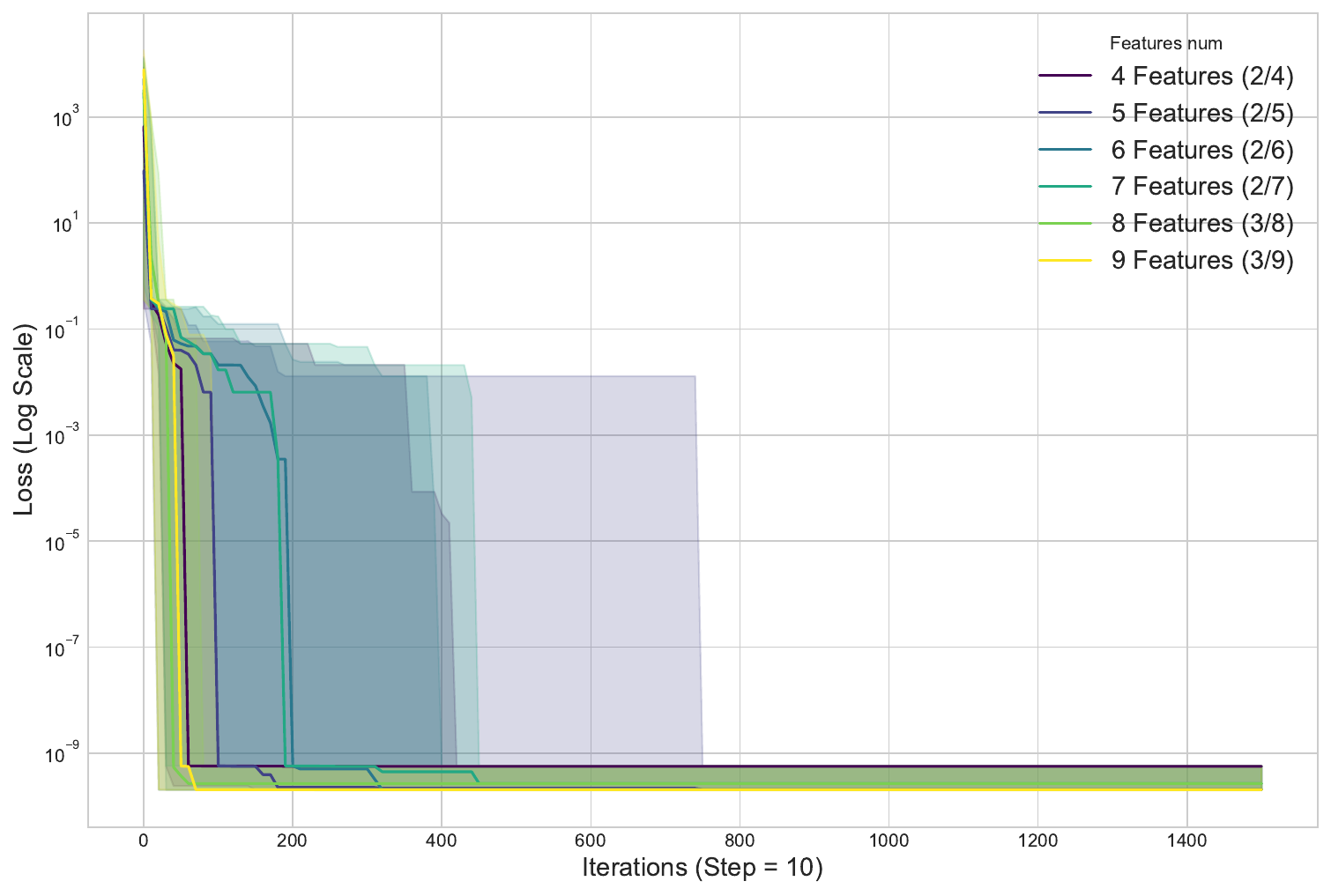}
        \caption{Recover-12}
        \label{fig:loss6}
    \end{subfigure}

    \vspace{1em} 

    \begin{subfigure}[b]{0.28\textwidth}
        \includegraphics[width=\textwidth]{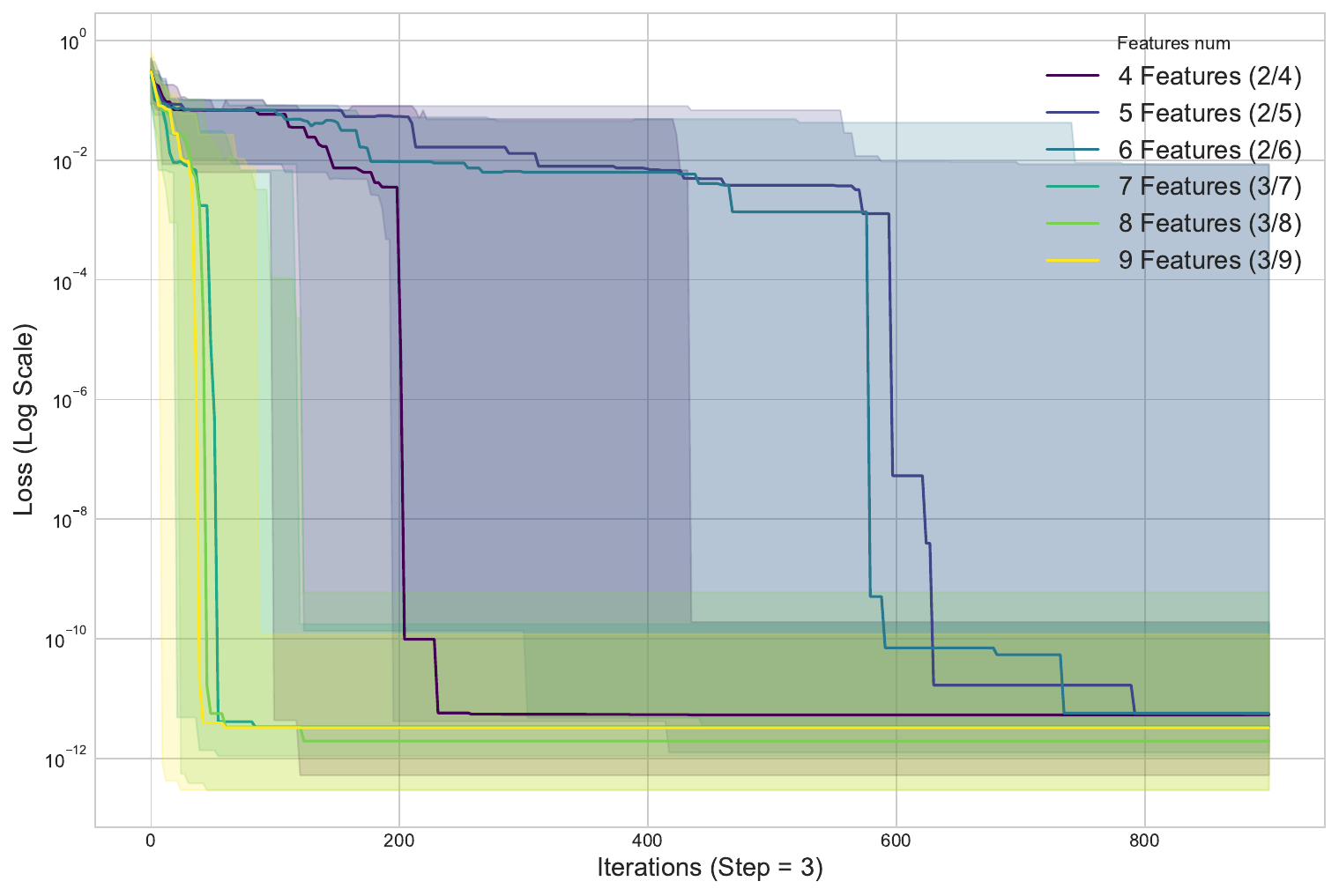}
        \caption{Recover-14}
        \label{fig:loss7}
    \end{subfigure}
    \hfill
    \begin{subfigure}[b]{0.28\textwidth}
        \includegraphics[width=\textwidth]{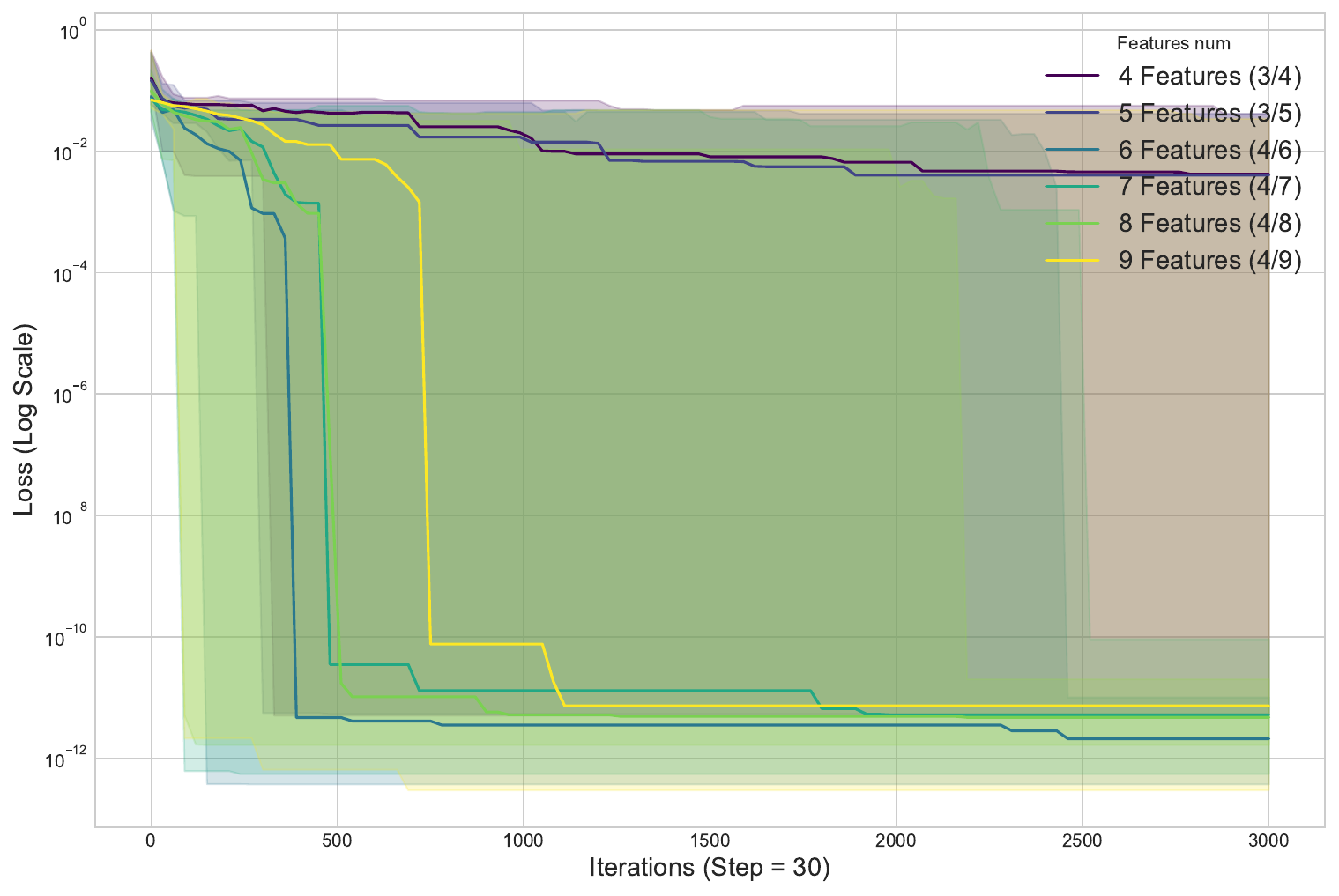}
        \caption{Recover-15}
        \label{fig:loss8}
    \end{subfigure}
    \hfill
    \begin{subfigure}[b]{0.28\textwidth}
        \includegraphics[width=\textwidth]{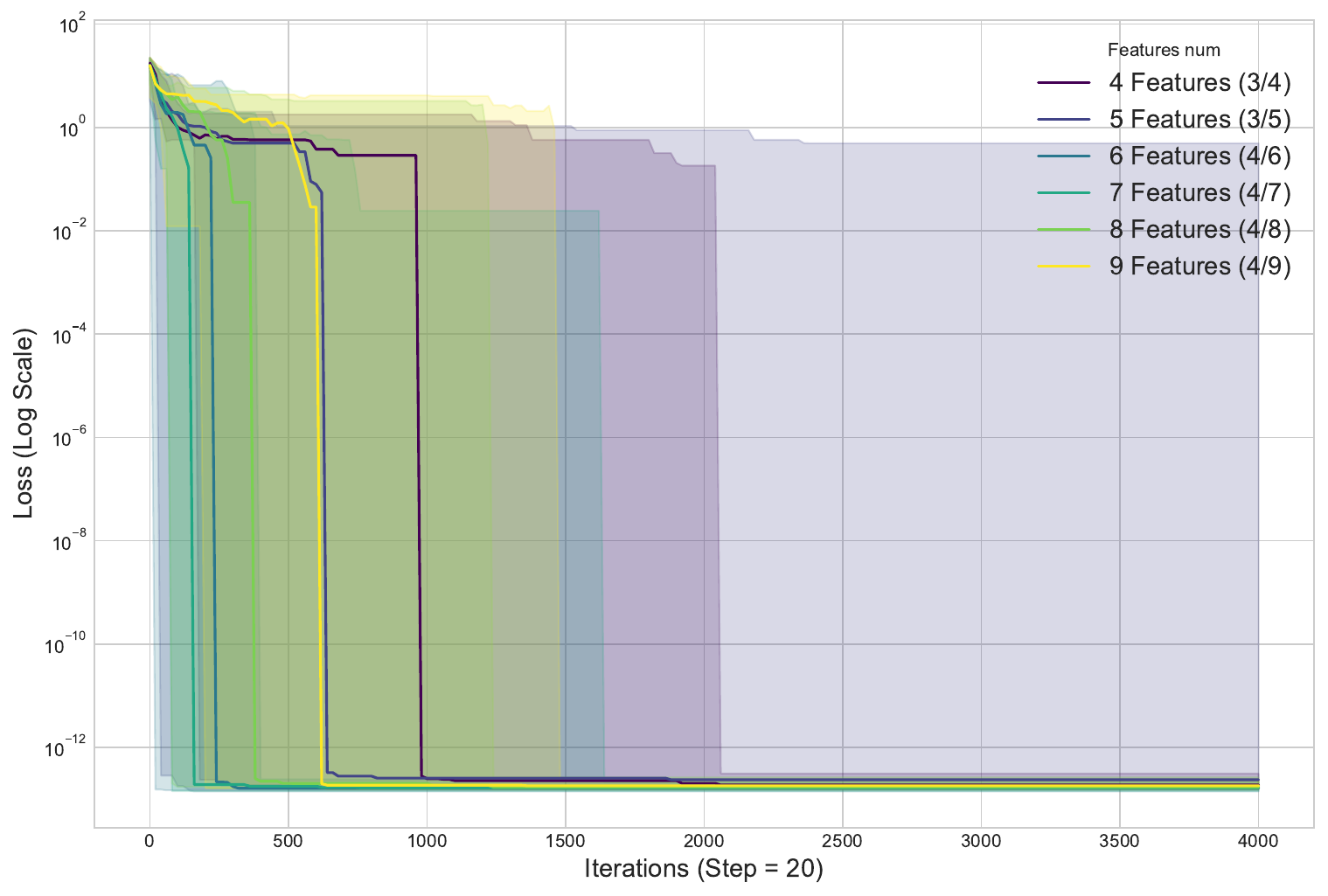}
        \caption{Recover-21}
        \label{fig:loss9}
    \end{subfigure}

    \vspace{1em} 

    \begin{subfigure}[b]{0.28\textwidth}
        \includegraphics[width=\textwidth]{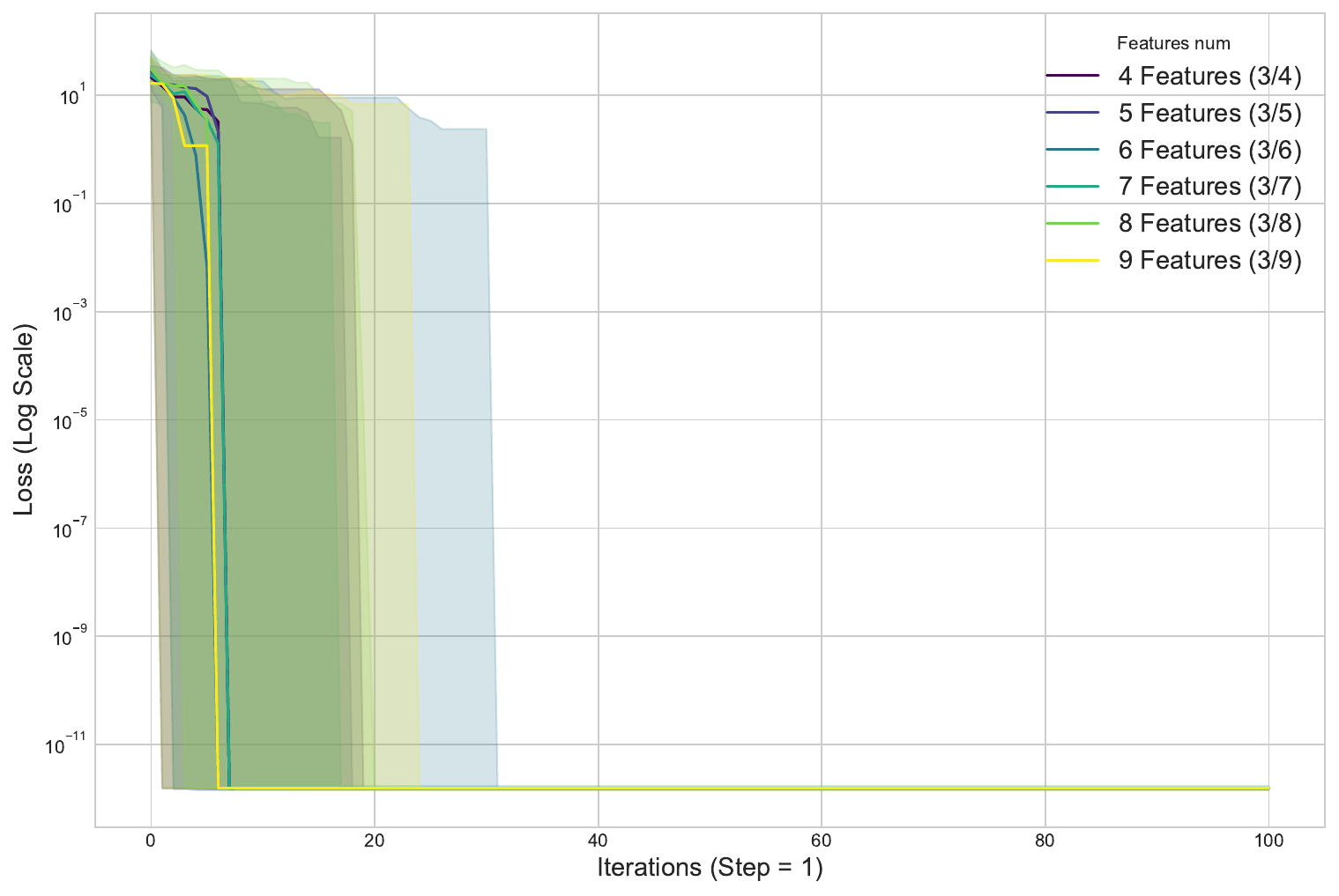}
        \caption{Recover-22}
        \label{fig:loss10}
    \end{subfigure}
    \hfill
    \begin{subfigure}[b]{0.28\textwidth}
        \includegraphics[width=\textwidth]{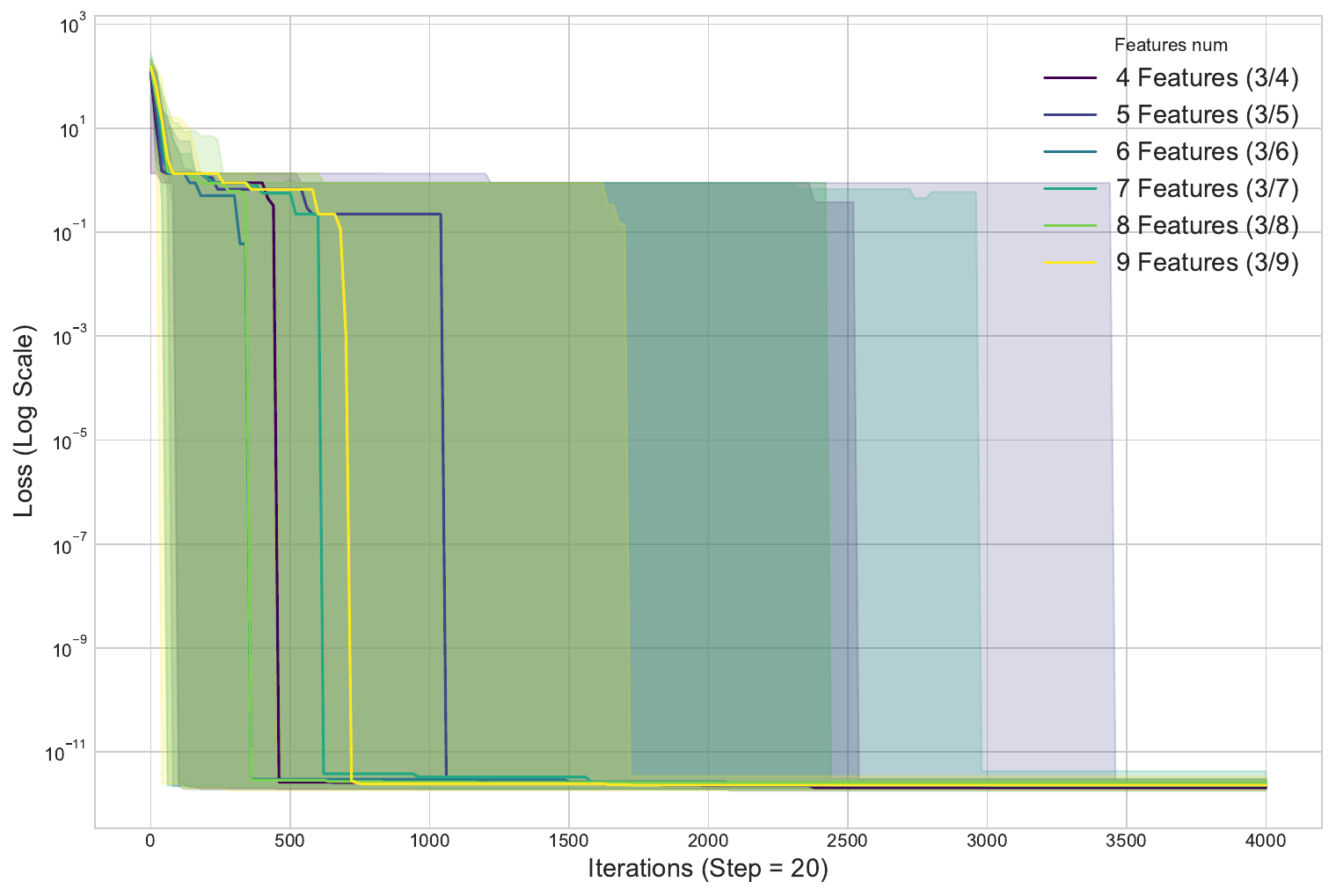}
        \caption{Recover-30}
        \label{fig:loss11}
    \end{subfigure}
    \hfill
    \begin{subfigure}[b]{0.28\textwidth}
        \includegraphics[width=\textwidth]{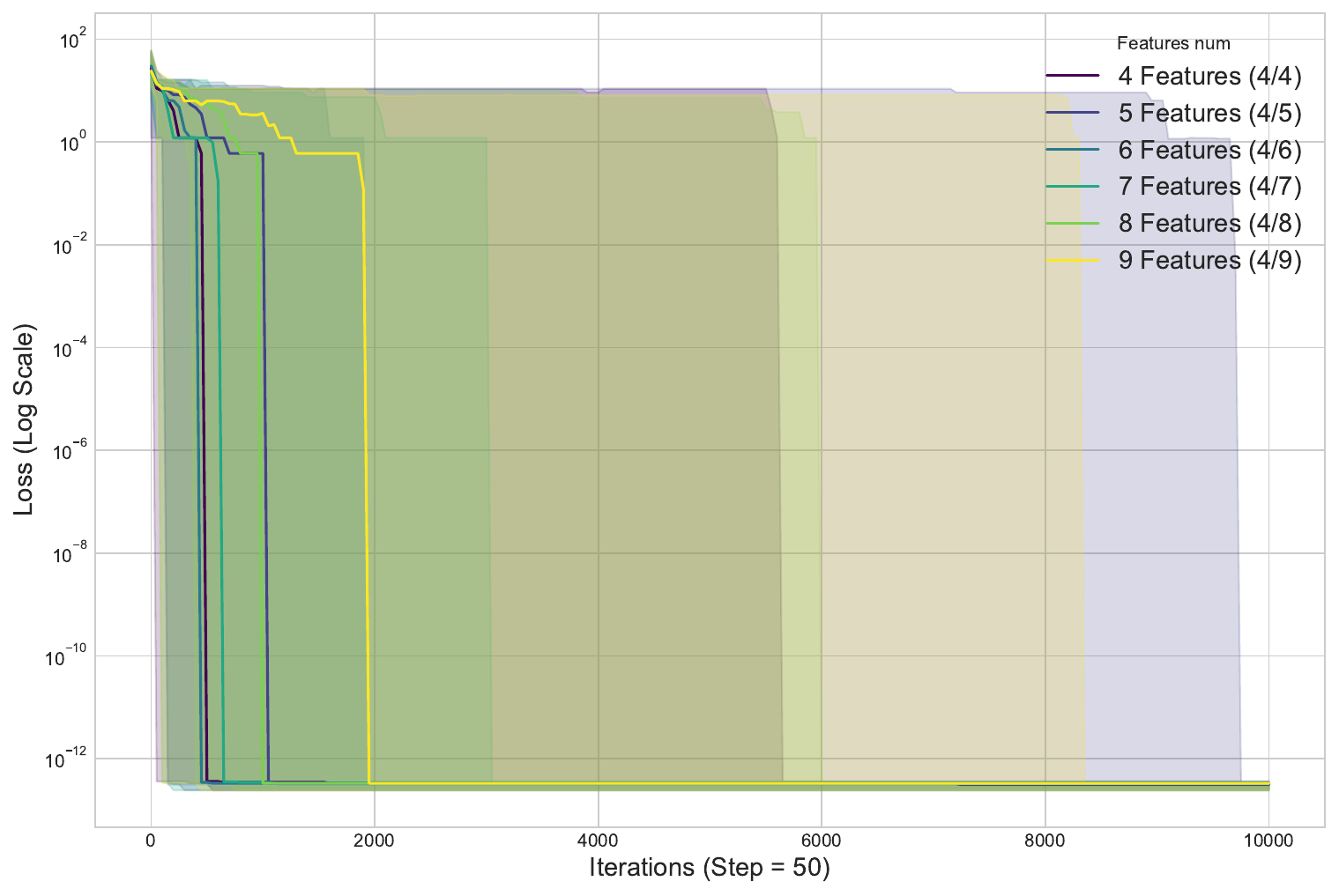}
        \caption{Recover-34}
        \label{fig:loss12}
    \end{subfigure}

    \caption{\textbf{Loss trajectories of FePySR under varying numbers of selected top-n features.} The x-axis represents the training epoch, with the value in parentheses indicating the recording interval. The y-axis represents the training loss on a logarithmic scale. Solid lines denote the median loss across 20 independent runs, and shaded regions denote the corresponding min-max range.}
    \label{loss_convergence}
\end{figure}

\newpage

\begin{figure}[H]
    \centering
    \captionsetup{font=small,labelfont=bf}
    \captionsetup[subfigure]{labelformat=empty}
    \begin{subfigure}[b]{0.28\textwidth}
        \includegraphics[width=\textwidth]{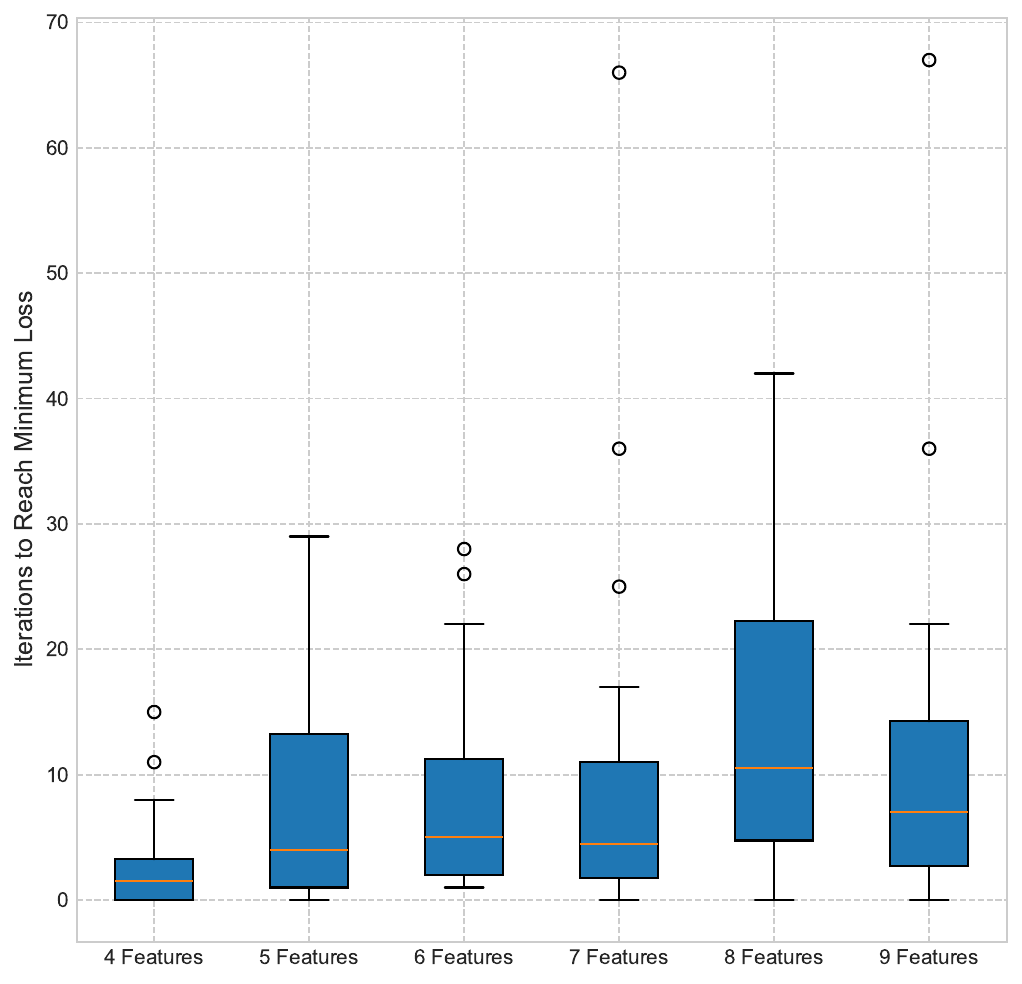}
        \caption{Recover-1}
        \label{fig:loss1}
    \end{subfigure}
    \hfill
    \begin{subfigure}[b]{0.28\textwidth}
        \includegraphics[width=\textwidth]{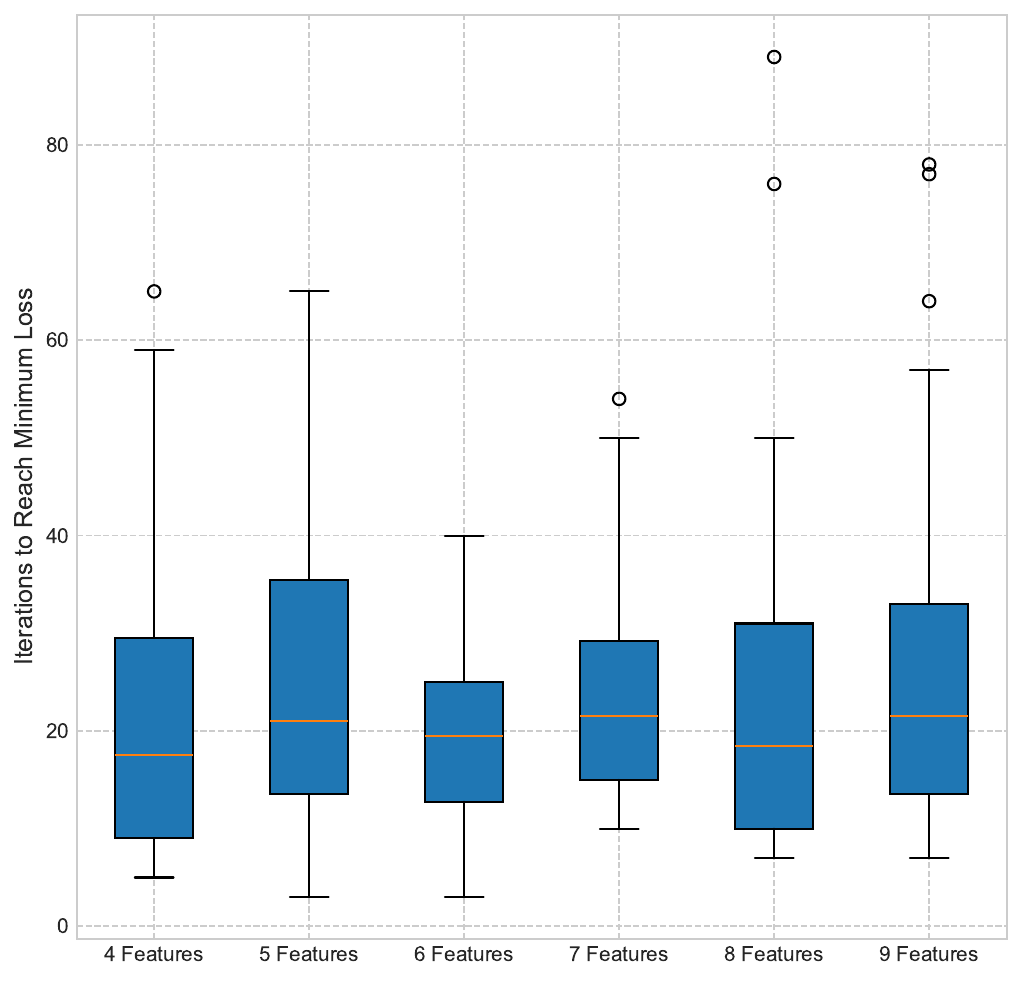}
        \caption{Recover-4}
        \label{fig:loss2}
    \end{subfigure}
    \hfill
    \begin{subfigure}[b]{0.28\textwidth}
        \includegraphics[width=\textwidth]{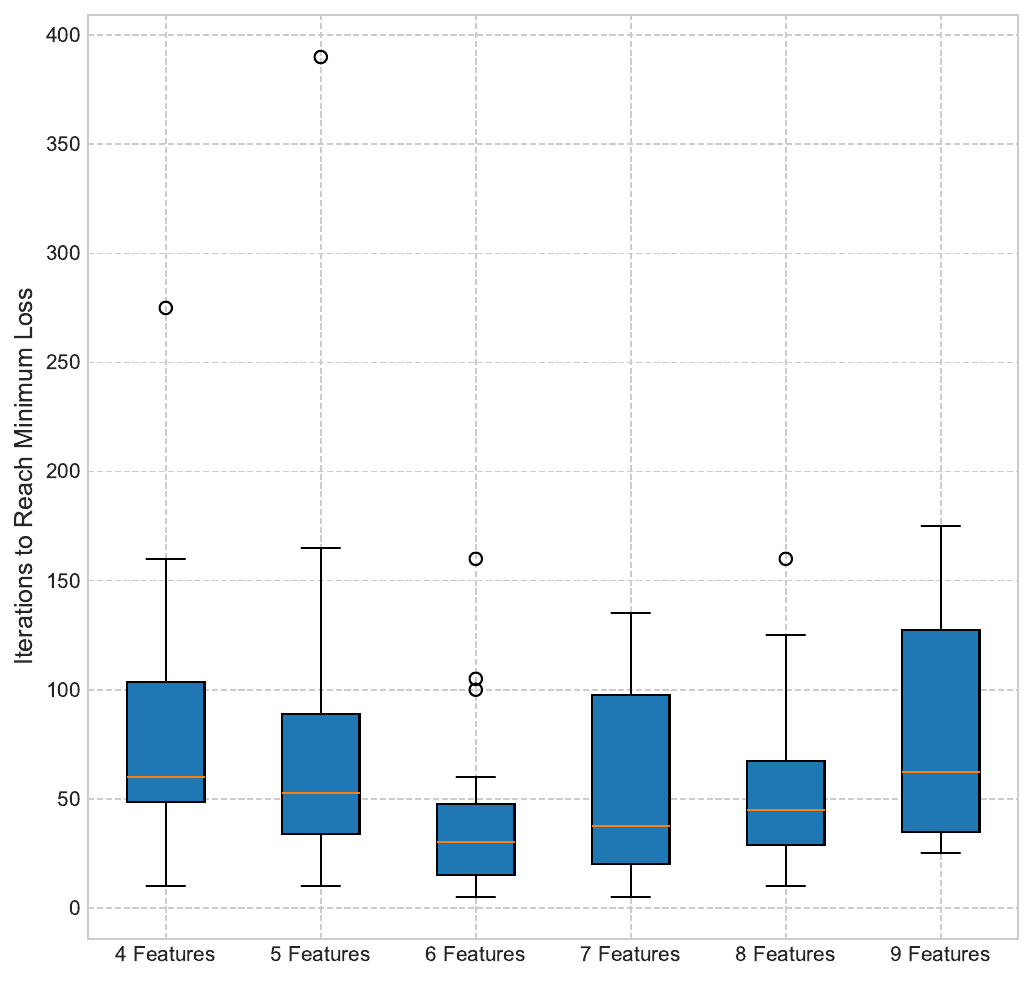}
        \caption{Recover-5}
        \label{fig:loss3}
    \end{subfigure}

    \vspace{1em} 

    \begin{subfigure}[b]{0.28\textwidth}
        \includegraphics[width=\textwidth]{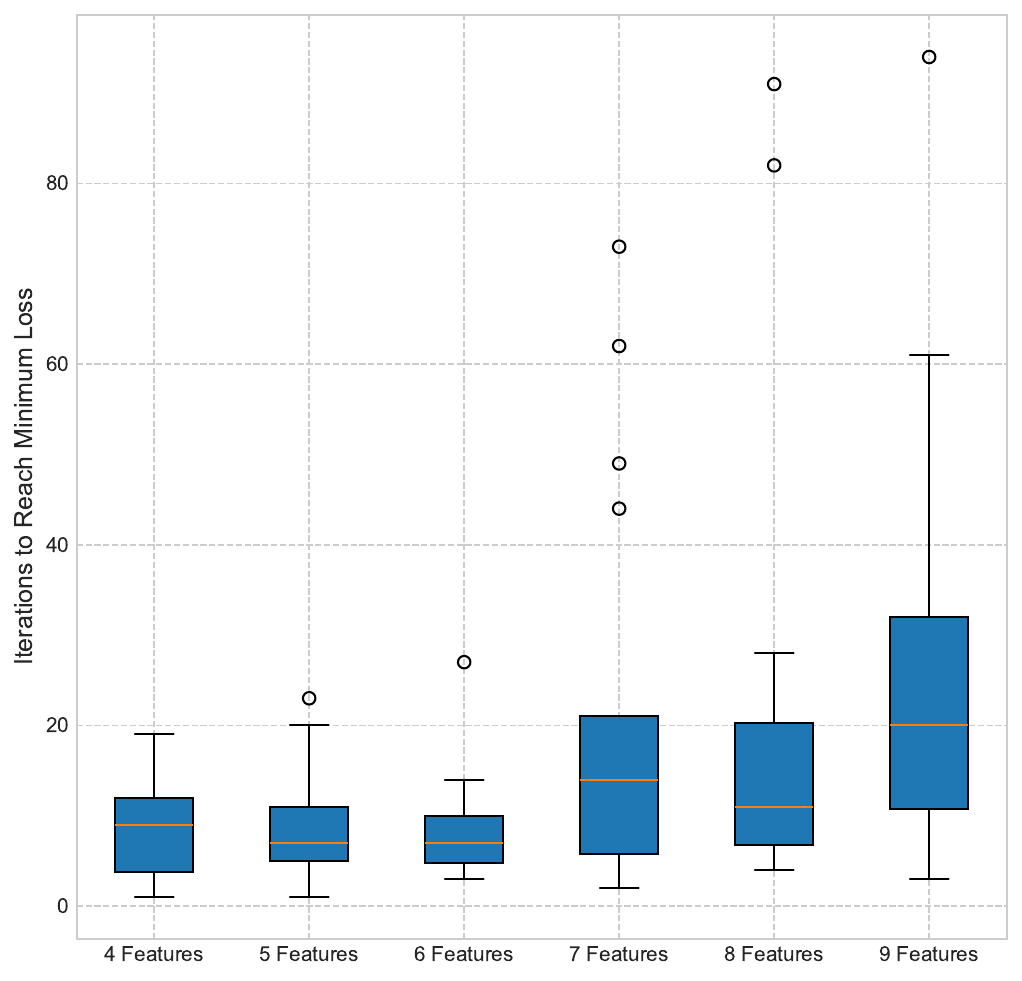}
        \caption{Recover-7}
        \label{fig:loss4}
    \end{subfigure}
    \hfill
    \begin{subfigure}[b]{0.28\textwidth}
        \includegraphics[width=\textwidth]{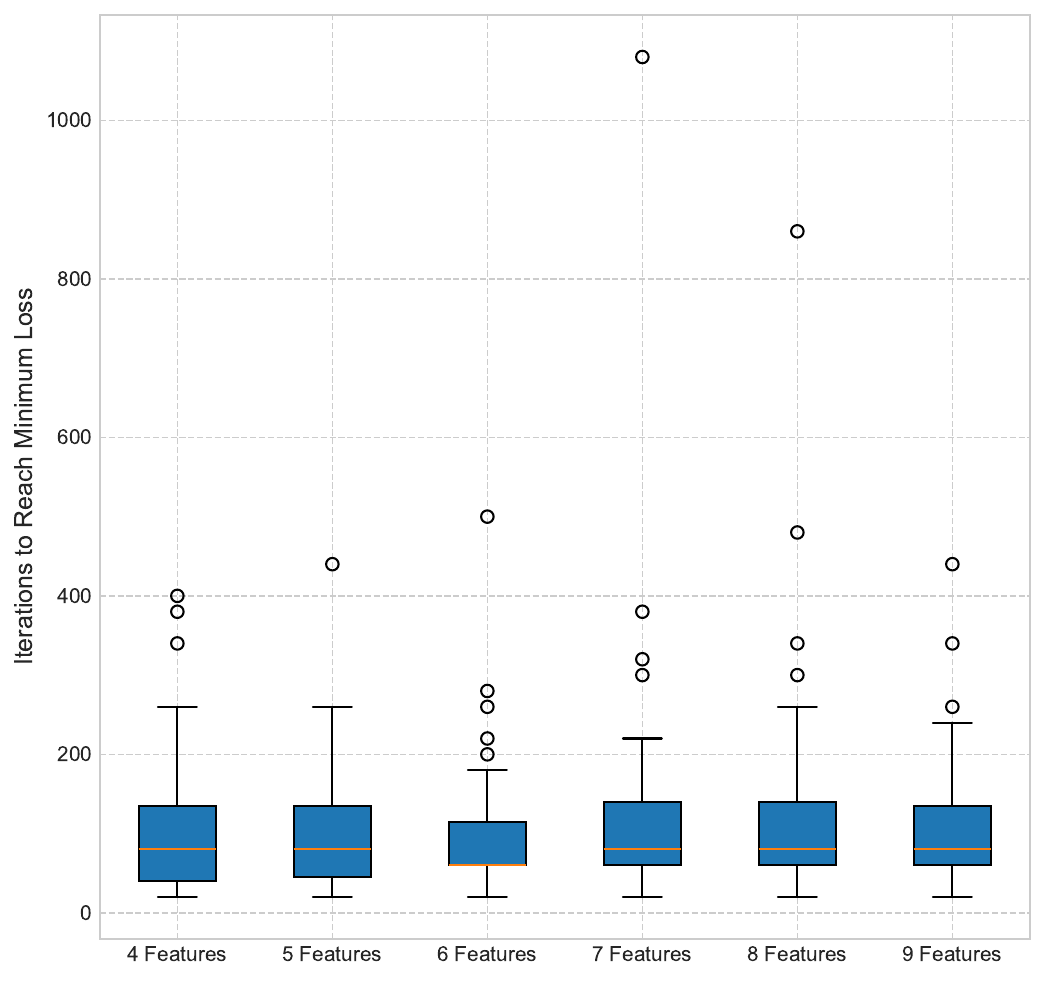}
        \caption{Recover-10}
        \label{fig:loss5}
    \end{subfigure}
    \hfill
    \begin{subfigure}[b]{0.28\textwidth}
        \includegraphics[width=\textwidth]{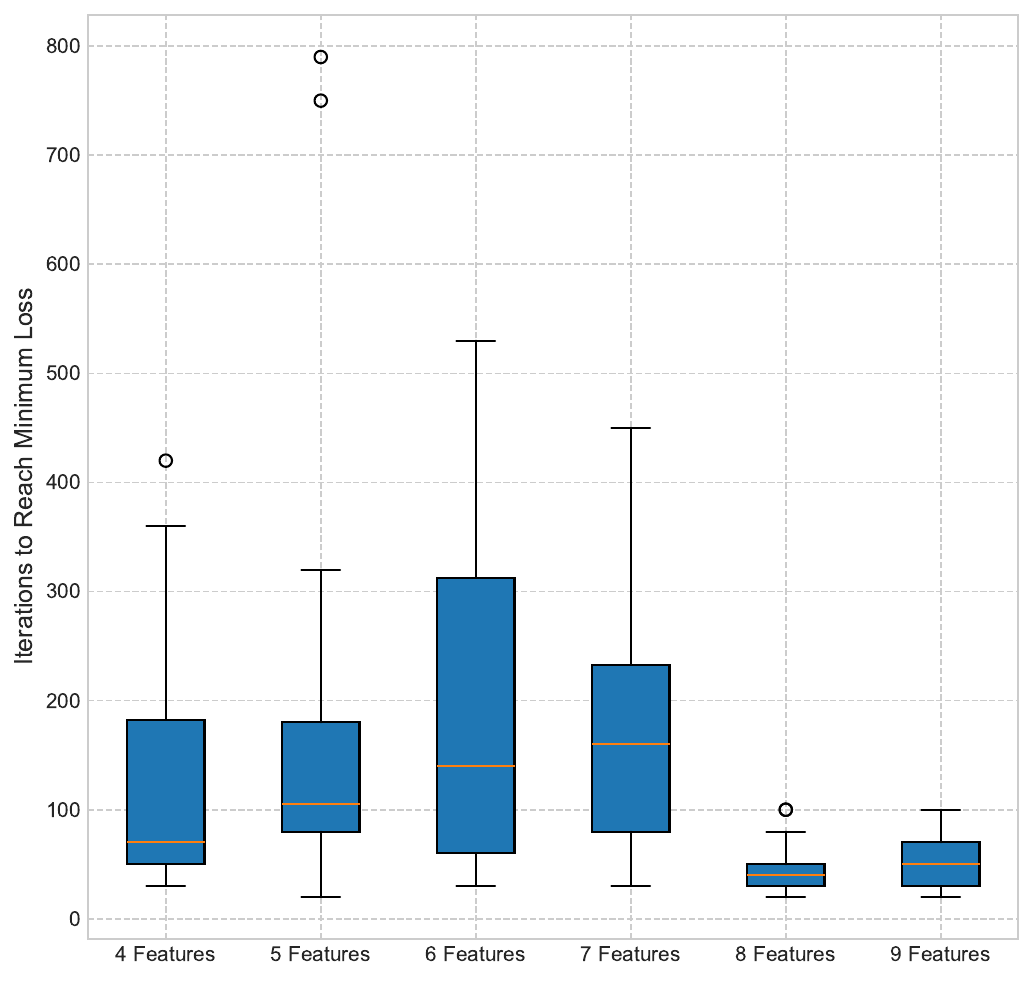}
        \caption{Recover-12}
        \label{fig:loss6}
    \end{subfigure}

    \vspace{1em} 

    \begin{subfigure}[b]{0.28\textwidth}
        \includegraphics[width=\textwidth]{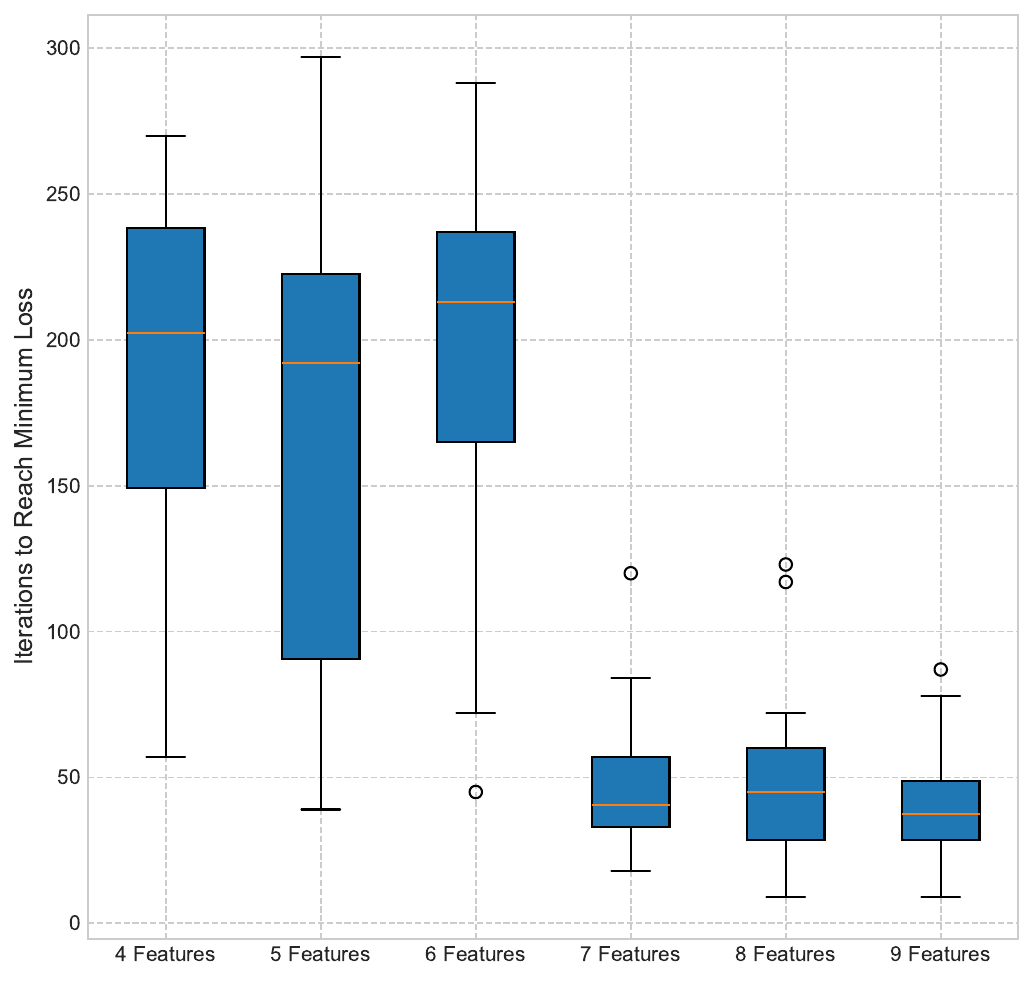}
        \caption{Recover-14}
        \label{fig:loss7}
    \end{subfigure}
    \hfill
    \begin{subfigure}[b]{0.28\textwidth}
        \includegraphics[width=\textwidth]{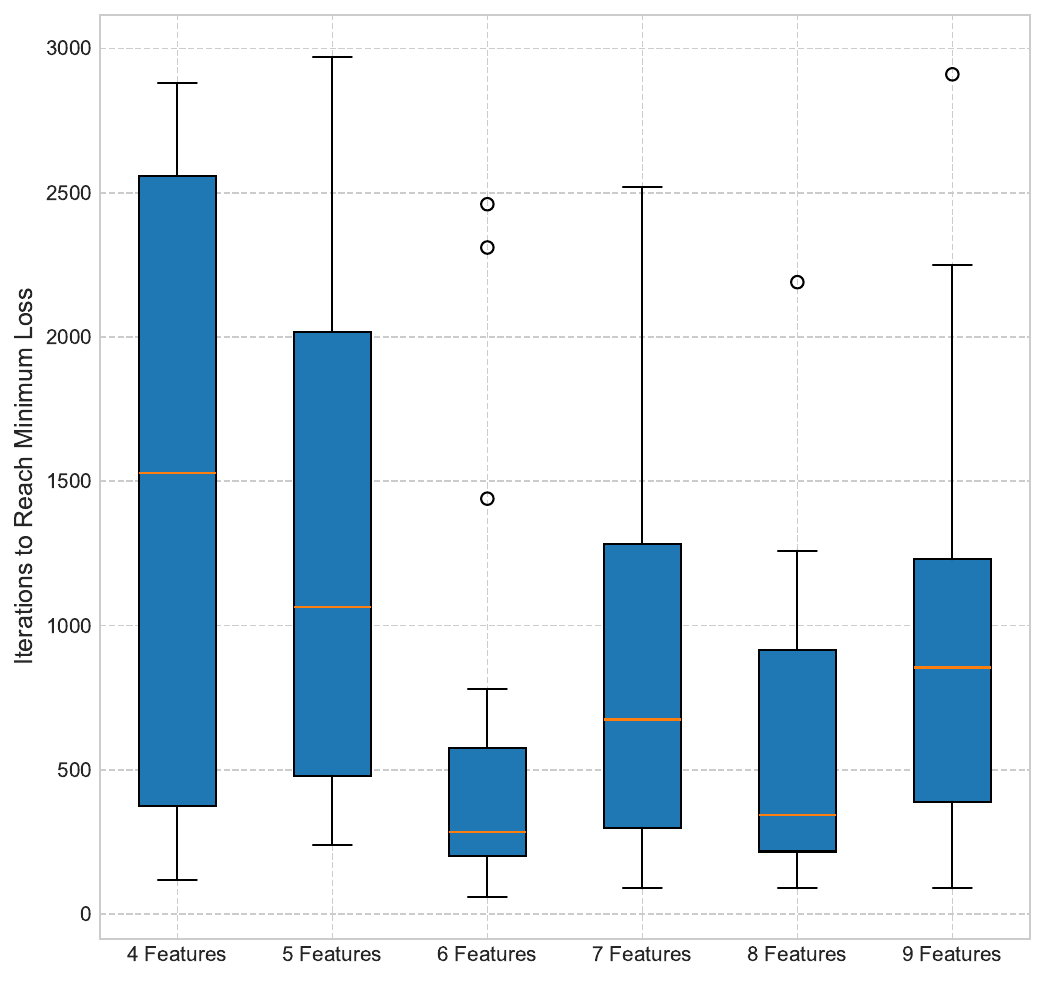}
        \caption{Recover-15}
        \label{fig:loss8}
    \end{subfigure}
    \hfill
    \begin{subfigure}[b]{0.28\textwidth}
        \includegraphics[width=\textwidth]{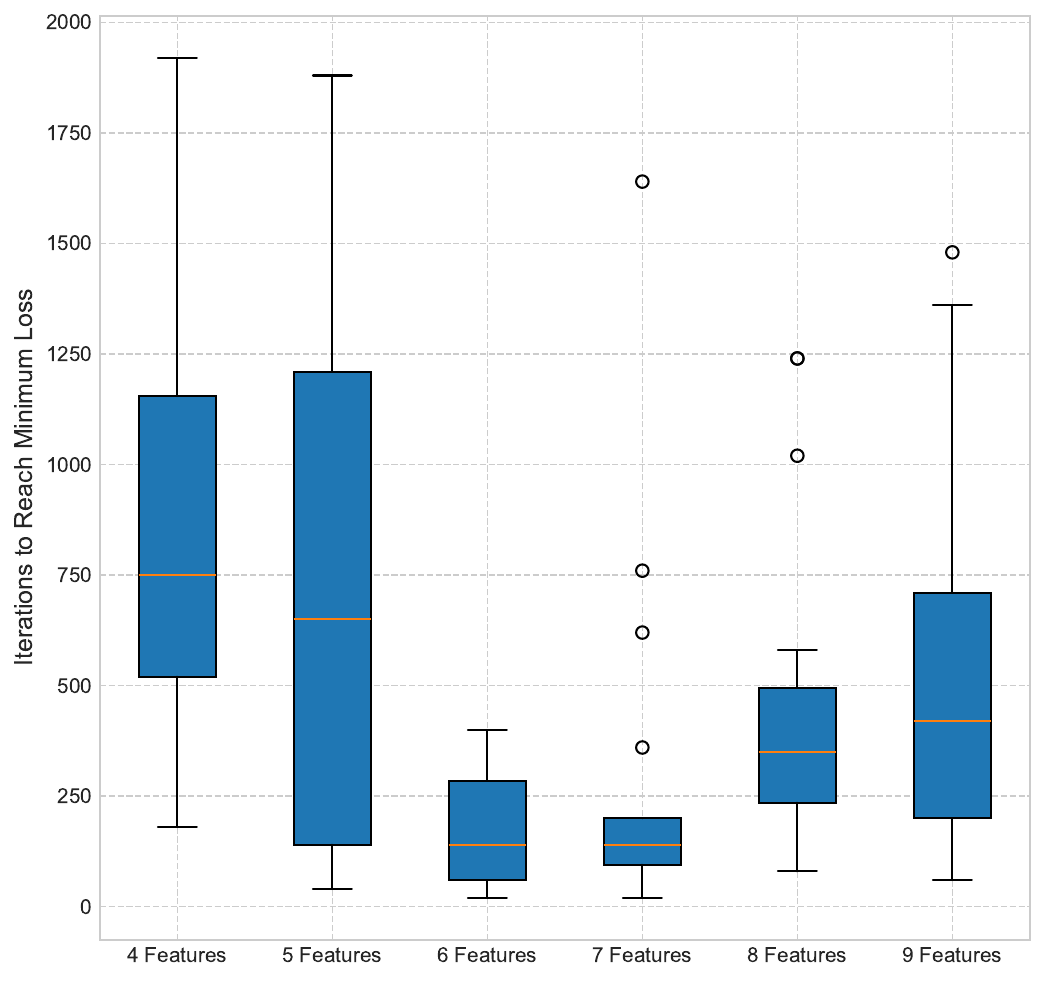}
        \caption{Recover-21}
        \label{fig:loss9}
    \end{subfigure}

    \vspace{1em} 

    \begin{subfigure}[b]{0.28\textwidth}
        \includegraphics[width=\textwidth]{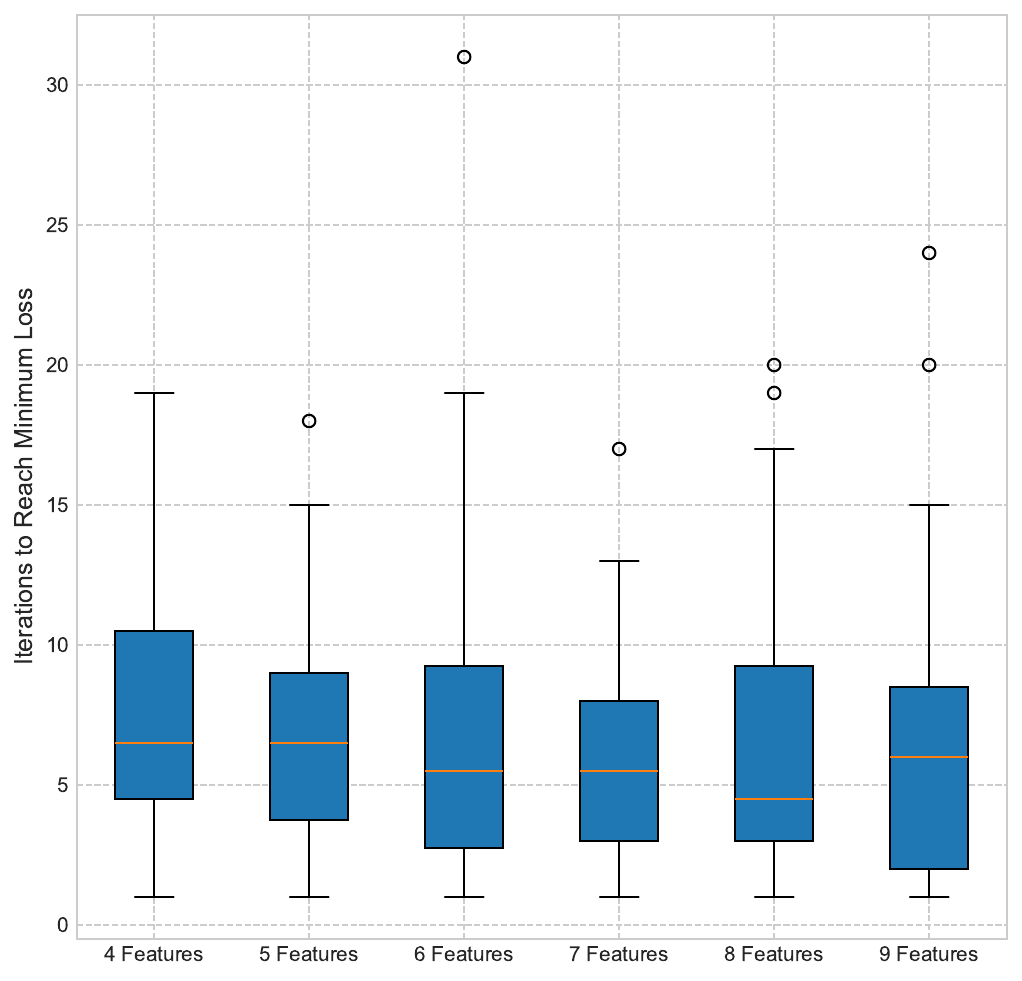}
        \caption{Recover-22}
        \label{fig:loss10}
    \end{subfigure}
    \hfill
    \begin{subfigure}[b]{0.28\textwidth}
        \includegraphics[width=\textwidth]{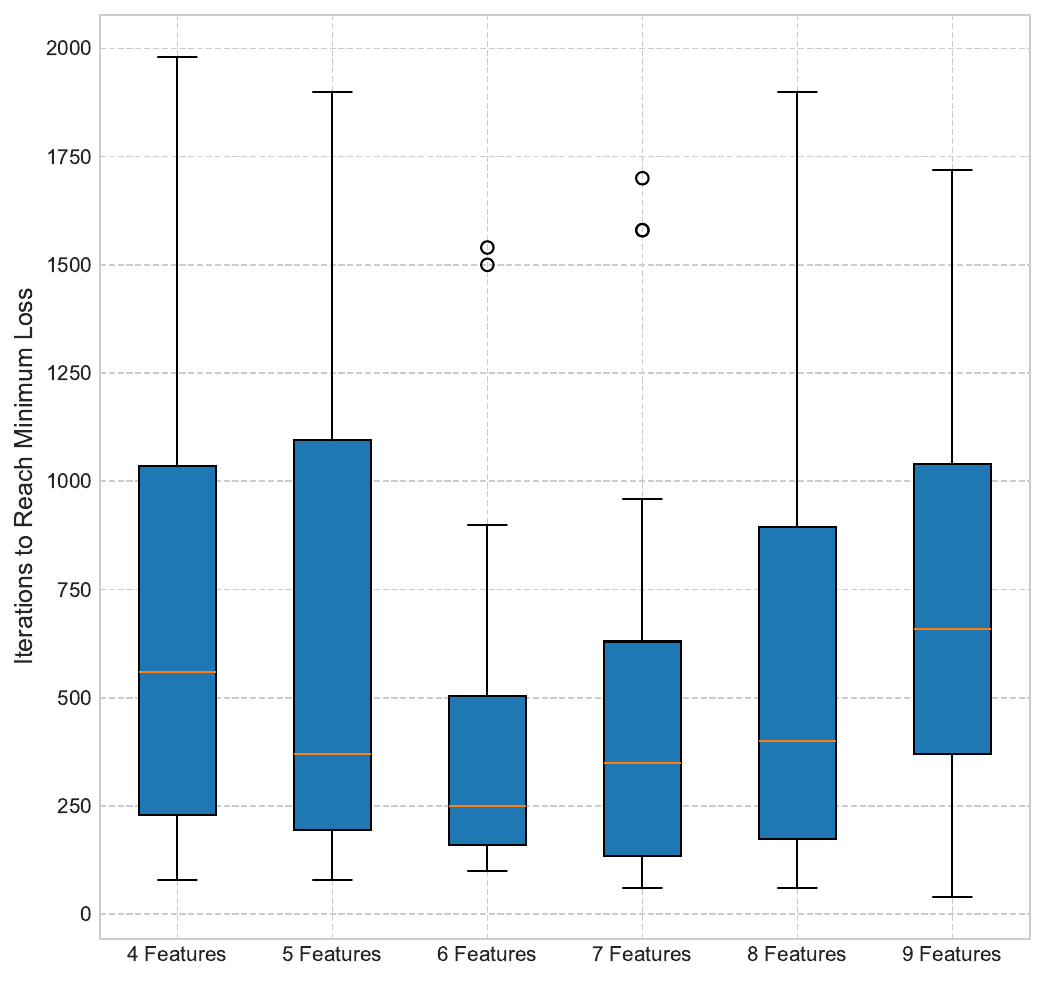}
        \caption{Recover-30}
        \label{fig:loss11}
    \end{subfigure}
    \hfill
    \begin{subfigure}[b]{0.28\textwidth}
        \includegraphics[width=\textwidth]{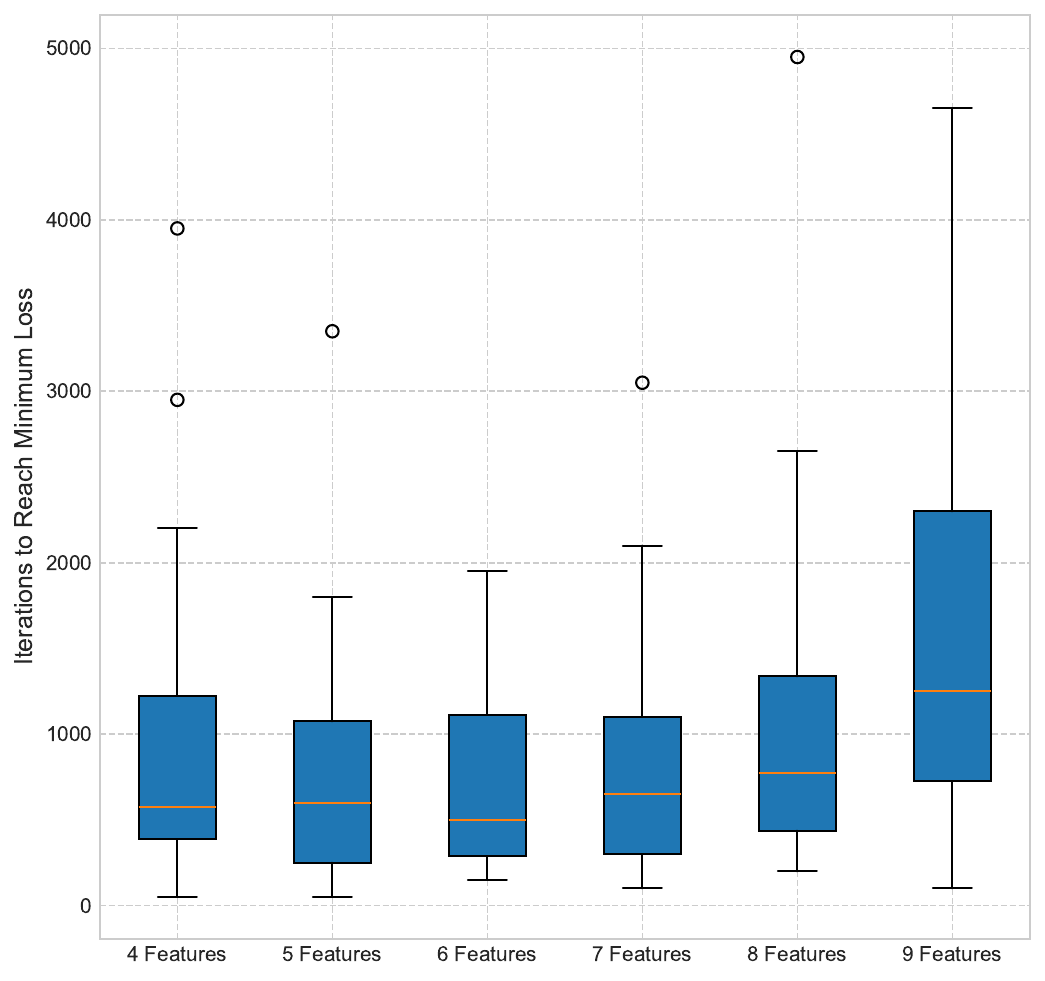}
        \caption{Recover-34}
        \label{fig:loss12}
    \end{subfigure}
    \caption{\textbf{Training epochs required to achieve the minimum MSE under varying numbers of selected top-n features.} The x-axis represents the number of selected top-n features. The y-axis represents the number of epochs required for the model to converge to the minimum MSE. Boxes represent the interquartile range (IQR; 25th, 50th, and 75th percentiles), orange lines denote the median, and circles indicate outliers.}
    \label{box}
\end{figure}

\newpage

\subsubsection{FePySR's Noise Sensitivity Evaluation}
\label{appendixB2.3}
\captionsetup{font=small,labelfont=bf} 
\begin{table}[H]
  \centering
  \caption{\textbf{EFR raw data for drawing Figure \ref{Combined_Noise_Sensitivity} in the main text.}}
  \label{tab:EFR}
  \begin{tabular}{cccccccc}
    \toprule  
    EFR& 0 & 0.02 & 0.04 & 0.06 & 0.08 & 0.1 & 0.2 \\
    \midrule  
    Recover-1 & 0.17029 & 0.17035 & 0.17120 & 0.16248 & 0.16638 & 0.17004 & 0.16962 \\
    Recover-4 & 0.17725 & 0.18445 & 0.18512 & 0.18341 & 0.18292 & 0.17804 & 0.18488 \\
    Recover-5 & 0.30945 & 0.31732 & 0.30811 & 0.31384 & 0.30969 & 0.30450 & 0.31000 \\
    Recover-7 & 0.11263 & 0.11576 & 0.12341 & 0.12533 & 0.12040 & 0.11886 & 0.11662 \\
    Recover-10 & 0.15344 & 0.15548 & 0.15291 & 0.12760 & 0.14880 & 0.12952 & 0.15780 \\
    Recover-12 & 0.17208 & 0.17444 & 0.17668 & 0.17554 & 0.18140 & 0.17912 & 0.17228 \\
    Recover-14 & 0.12223 & 0.12183 & 0.12150 & 0.11804 & 0.11983 & 0.11979 & 0.12439 \\
    Recover-15 & 0.14897 & 0.14762 & 0.14811 & 0.14897 & 0.14636 & 0.14368 & 0.14368 \\
    Recover-21 & 0.10754 & 0.11373 & 0.11422 & 0.10152 & 0.12569 & 0.10636 & 0.09583 \\
    Recover-22 & 0.10848 & 0.12801 & 0.11877 & 0.12854 & 0.12533 & 0.12667 & 0.11812 \\
    Recover-30 & 0.15169 & 0.13078 & 0.12394 & 0.12821 & 0.14913 & 0.12459 & 0.15120 \\
    Recover-34 & 0.13147 & 0.13460 & 0.13680 & 0.13322 & 0.13452 & 0.13241 & 0.12988 \\
    \bottomrule  
  \end{tabular}
\end{table}

\begin{table}[H]
  \centering
  \captionsetup{font=small,labelfont=bf} 
  \caption{\textbf{DCG-1 raw data for drawing Figure \ref{Combined_Noise_Sensitivity} in the main text.}}
  \label{tab:DCG-1}
  \begin{tabular}{cccccccc}
    \toprule  
    DCG-1& 0 & 0.02 & 0.04 & 0.06 & 0.08 & 0.1 & 0.2 \\
    \midrule  
    Recover-1 & 2.00081 & 1.53157 & 1.53157 & 1.46835 & 1.63159 & 1.53157 & 1.53157 \\
    Recover-4 & 2.16405 & 2.16405 & 2.16405 & 2.00081 & 2.16405 & 2.06403 & 2.06403 \\
    Recover-5 & 2.73382 & 2.73382 & 2.73382 & 2.76682 & 2.73382 & 2.73382 & 2.73382 \\
    Recover-7 & 1.85658 & 2.25292 & 2.14549 & 2.11249 & 2.11249 & 2.25292 & 2.25292 \\
    Recover-10 & 2.76997 & 2.76997 & 2.76997 & 2.16405 & 2.59834 & 2.35294 & 2.78723 \\
    Recover-12 & 3.36617 & 3.50858 & 3.32808 & 3.34534 & 3.42939 & 3.32808 & 3.39195 \\
    Recover-14 & 2.67795 & 2.64495 & 2.72216 & 2.64495 & 2.64495 & 2.64495 & 2.61917 \\
    Recover-15 & 3.29928 & 2.73382 & 3.24050 & 2.67060 & 2.73382 & 3.13604 & 3.23606 \\
    Recover-21 & 1.63159 & 1.63159 & 2.16405 & 1.63159 & 2.58108 & 2.35294 & 1.63159 \\
    Recover-22 & 1.53157 & 1.94860 & 1.94860 & 1.94860 & 2.49832 & 2.48106 & 2.48106 \\
    Recover-30 & 3.94644 & 3.20306 & 3.45517 & 3.42495 & 4.00522 & 3.34774 & 3.16811 \\
    Recover-34 & 3.15885 & 3.15885 & 3.34774 & 3.15885 & 3.15885 & 3.13307 & 3.13307 \\
    \bottomrule  
  \end{tabular}
\end{table}

\begin{table}[H]
  \centering
  \captionsetup{font=small,labelfont=bf} 
  \caption{\textbf{DCG-2 raw data for drawing Figure \ref{Combined_Noise_Sensitivity} in the main text.}}
  \label{tab:DCG-2}
  \begin{tabular}{cccccccc}
    \toprule  
    DCG-2    & 0       & 0.02    & 0.04    & 0.06    & 0.08    & 0.1     & 0.2     \\
    \midrule  
    Recover-1        & 3027.308 & 2174.147 & 2183.135 & 1995.925 & 2224.608 & 2182.597 & 2173.914 \\
    Recover-4        & 3335.511 & 3582.933 & 3530.275 & 3421.523 & 3531.718 & 3311.416 & 3463.963 \\
    Recover-5        & 3564.715 & 3701.384 & 3558.558 & 3640.664 & 3603.126 & 3570.567 & 3600.931 \\
    Recover-7        & 1722.497 & 2170.659 & 2227.845 & 2284.345 & 2163.386 & 2225.215 & 2168.296 \\
    Recover-10        & 3972.309 & 4019.649 & 4010.947 & 3485.551 & 3605.849 & 3832.284 & 4088.089 \\
    Recover-12        & 3837.552 & 4044.047 & 4013.557 & 4007.846 & 4140.607 & 4033.505 & 3852.622 \\
    Recover-14        & 2935.854 & 2932.037 & 2982.963 & 2848.361 & 2923.622 & 2881.062 & 3051.589 \\
    Recover-15        & 3145.091 & 2519.847 & 3090.538 & 2478.495 & 2520.645 & 2901.561 & 2977.224 \\
    Recover-21        & 2166.814 & 2285.715 & 3075.105 & 2043.055 & 3016.609 & 3081.675 & 1932.055 \\
    Recover-22       & 2132.882 & 2267.268 & 2115.370 & 2307.804 & 3036.837 & 3115.033 & 2897.672 \\
    Recover-30       & 3218.510 & 2797.675 & 2742.017 & 2902.841 & 3198.497 & 2786.817 & 2473.161 \\
    Recover-34       & 2779.991 & 2869.259 & 3107.165 & 2836.879 & 2863.059 & 2889.813 & 2810.401 \\
    \bottomrule  
  \end{tabular}
\end{table}

\begin{table}[H]
\centering
\caption{\textbf{Recovery rates of FePySR across 12 recoverable equations under varying noise levels $\alpha$.} For the first five equations, 25 tests are conducted per noise level. For the remaining seven equations, 10 tests are conducted per noise level.}
\label{noise_results_extended}
\small
\begin{tabular}{lccccccc}
\toprule
\textbf{Identifier} & \multicolumn{7}{c}{\textbf{Noise Level ($\alpha$)}} \\ \cmidrule(l){2-8} 
 & \textbf{0} & \textbf{0.02} & \textbf{0.04} & \textbf{0.06} & \textbf{0.08} & \textbf{0.1} & \textbf{0.2} \\ \midrule
Recover-1  & 100\% & 0\% & 0\% & 0\% & 0\% & 0\% & 0\% \\
Recover-4  & 100\% & 0\% & 0\% & 0\% & 0\% & 0\% & 0\% \\
Recover-5  & 100\% & 0\% & 0\% & 0\% & 0\% & 0\% & 0\% \\
Recover-7  & 100\% & 0\% & 0\% & 0\% & 0\% & 0\% & 0\% \\
Recover-10  & 100\% & 0\% & 0\% & 0\% & 0\% & 0\% & 0\% \\
Recover-12  & 100\% & 0\% & 0\% & 0\% & 0\% & 0\% & 0\% \\
Recover-14  & 94\% & 0\% & 0\% & 0\% & 0\% & 0\% & 0\% \\
Recover-15  & 39\% & 0\% & 0\% & 0\% & 0\% & 0\% & 0\% \\
Recover-21  & 100\% & 0\% & 0\% & 0\% & 0\% & 0\% & 0\% \\
Recover-22 &  84\% & 0\% & 0\% & 0\% & 0\% & 0\% & 0\% \\
Recover-30 & 55\% & 0\% & 0\% & 0\% & 0\% & 0\% & 0\% \\
Recover-34 & 62\% & 0\% & 0\% & 0\% & 0\% & 0\% & 0\% \\ \bottomrule
\end{tabular}
\end{table}

\newpage

\subsubsection{Model Performance on Unrecovered Equations}
\label{appendixB2.4}
{\renewcommand{\arraystretch}{1.5}\setlength{\extrarowheight}{1.5pt}
\begin{longtable}{clccc}
\captionsetup{font=small,labelfont=bf}
\caption{\textbf{Comparison of FePySR and PySR on unrecoverable equations generated by LLMs.} The minimum MSE and error ratios (FePySR / PySR) are reported in 50 tests. Among these runs, FePySR recovers the target expression once for Unrecover-7 and Unrecover-21; for Unrecover-15, FePySR and PySR succeed in seven and five runs, respectively. Given the low recovery rates, these equations are classified as unrecoverable. The reported minimum MSE values exclude the runs in which SR is successful.}
\label{more_complex_data} \\

\toprule
\textbf{Identifier} & \textbf{Equation} & \textbf{FePySR} & \textbf{PySR} & \textbf{rate} \\
\midrule
\endfirsthead

\multicolumn{5}{c}{\tablename\ \thetable{}} \\
\toprule
\textbf{Identifier} & \textbf{Equation} & \textbf{FePySR} & \textbf{PySR} & \textbf{rate} \\
\midrule
\endhead

\bottomrule
\endlastfoot

Unrecover-1 & $ \dfrac{\ln(x^2 + y^2)}{2 + \sin^2(xy)}$ & $3.53 \!\!\times\!\! 10^{-3}$ & $5.21 \!\!\times\!\! 10^{-3}$ & $0.678$ \\
Unrecover-2 & $\dfrac{|x|^{1.5} + |y|^{2.5} - xy}{\sqrt{1 + x^2 + 0.5y^2}}$ & $1.61 \!\!\times\!\! 10^{-3}$ & $1.56 \!\!\times\!\! 10^{-3}$ & $1.032$ \\
Unrecover-3 & $\dfrac{\cos(6 \arctan(y/x))}{1 + 0.2(x^2 + y^2)}$ & $2.83 \!\!\times\!\! 10^{-2}$ & $1.32 \!\!\times\!\! 10^{-2}$ & $2.144$ \\
Unrecover-4 & $ \dfrac{x^2 - y^2}{\sin^2(\pi x) - \cos^2(\pi y)} $ & $ 2.99\!\!\times\!\! 10^{+2} $ & $4.01 \!\!\times\!\! 10^{+2} $ & $0.746$ \\
Unrecover-5 & $ \dfrac{x^2 + y^3 - z}{|yz| + 1} + \sin(xz) $ & $ 1.20\!\!\times\!\! 10^{-1} $&$2.29 \!\!\times\!\! 10^{-1}$ & $0.524$ \\
Unrecover-6 & $\dfrac{\tanh(x^2 - y^2) \cos(y^3)}{x^4 + y^4 + \sin^2(xy) + 1}$ & $3.24\!\!\times\!\! 10^{-4} $ &$1.39\!\!\times\!\! 10^{-3} $ & $0.233$ \\
Unrecover-7 & $ \sqrt{|\sqrt{|x|} + \sqrt{|yz|}|} - x^2 $ &$1.14\!\!\times\!\!10^{-5}$ &$ 3.72\!\!\times\!\! 10^{-3}$ & $0.003$ \\
Unrecover-8 & $ \sin(x^2 e^{-|y|}) \cos(y^2 e^{-|x|}) $ & $1.61 \!\!\times\!\! 10^{-4} $ & $2.64 \!\!\times\!\! 10^{-4} $ & $0.610$ \\
Unrecover-9 & $e^{-0.1(x^2 + y^2)} \cos(x \sin(y))$ & $6.31 \!\!\times\!\! 10^{-9}$ & $2.26 \!\!\times\!\! 10^{-4}$ & $2.8\! \!\times\!\!\! 10^{\!-\!5}$ \\
Unrecover-10 & $\dfrac{x^2 \tanh(y) - y^2 \tanh(x)}{1 + \cosh(xy)}$ & $3.58 \!\!\times\!\! 10^{-5}$ & $1.62 \!\!\times\!\! 10^{-4}$ & $0.221$ \\
Unrecover-11 & $ \dfrac{\sin(x \ln(2 + y^2 \cos^2(y)))}{1 + 0.2e^{-|x - y^2|}} $ & $5.64  \!\!\times\!\! 10^{-5} $ & $1.17 \!\!\times\!\! 10^{-4} $ & $0.482$ \\
Unrecover-12 & $\ln(1 + \left| \sin(x^2 + \cos(y^2)) \right|)$ & $1.83 \!\!\times\!\! 10^{-4}$ & $2.00 \!\!\times\!\! 10^{-4}$ & $0.915$ \\
Unrecover-13 & $ \ln\left(\left|\dfrac{x^2 + y^2}{z + 1}\right|\right) + \sqrt{|x - z|} $ & $ 2.20\!\!\times\!\! 10^{-1} $& $ 2.19\!\!\times\!\! 10^{-1} $ & $1.005$ \\
Unrecover-14 & $ \sqrt{\ln(1 + |xy|)} + \sqrt{|z - xy|} $ &$7.05 \!\!\times\!\! 10^{-3}$ &$1.21\!\!\times\!\! 10^{-2}$ & $0.582$ \\
Unrecover-15 & $ \sin(x^2 \cos(3y) + y^2 \sin(3x)) $ &$8.73 \!\!\times\!\! 10^{-3} $ & $1.79 \!\!\times\!\! 10^{-2} $ & $0.488$ \\
Unrecover-16 & $ \dfrac{\ln(1 + |\sin(x)|) \tanh(y^2 - x)}{\ln(1 + |\cos(y)|) \cosh(x^2 - y)} $ &$3.32 \!\!\times\!\! 10^{-2}$ &$3.42 \!\!\times\!\! 10^{-2} $ & $0.971$ \\
Unrecover-17 & $ \ln(1 + |\sin(xy)|) \cdot \sqrt{|z^2 + y^2|} $ & $5.07\!\!\times\!\! 10^{-5} $ &$2.16\!\!\times\!\! 10^{-4} $ & $0.235$ \\
Unrecover-18 & $\tanh(x^2 - y) + e^{-|xy|} \sin(y) - x$ & $6.33 \!\!\times\!\! 10^{-3}$ & $6.47 \!\!\times\!\! 10^{-3}$ & $0.978$ \\
Unrecover-19 & $\dfrac{xy}{\ln(1 + |x| + |y|) + 0.1x^2 + 0.1y^2}$ & $1.71 \!\!\times\!\! 10^{-4}$ & $2.88 \!\!\times\!\! 10^{-4}$ & $0.594$ \\
Unrecover-20 & $ \sinh(x^2 - y) \cosh(yz) - \tanh(z^2) $ &$4.98\!\!\times\!\! 10^{-1}$ & $4.58 \!\!\times\!\! 10^{+0}$ & $0.109$ \\
Unrecover-21 & $\sin(x^2 y) \cos(y^2) - \cos(x^2) \sin(xy)$ & $9.23 \!\!\times\!\! 10^{-4}$ & $8.60 \!\!\times\!\! 10^{-3}$ & $0.107$ \\
Unrecover-22 & $ \sin(x^2 + \cos(y)) \cdot \operatorname{sgn}(\tan(z) + x) $ & $4.42 \!\!\times\!\! 10^{-2}$& $ 3.41\!\!\times\!\! 10^{-2}$ & $1.296$ \\
Unrecover-23 & $ \sin(\cos(x) + \tan(yz)) - \arctan(xz) $ & $3.46\!\!\times\!\! 10^{-2} $ & $4.12\!\!\times\!\! 10^{-2} $ & $0.840$ \\
Unrecover-24 & $ \tanh(50 (\sin(xy) - 0.5)) e^{-0.1(x^2+y^2)} $ &$9.49 \!\!\times\!\! 10^{-4} $ & $1.47\!\!\times\!\! 10^{-3} $ & $0.646$ \\
Unrecover-25 & $ |xy + 10^{-5}|^{1.5 + \exp(-0.2 (\sin(x) - \cos(y))^2)} $ &$6.66 \!\!\times\!\! 10^{-4} $ &$4.77 \!\!\times\!\! 10^{-3} $ & $0.139$ \\
Unrecover-26 & $ \begin{cases} \sin(xz) & \text{if } x > y \\ \cos(y - z) & \text{otherwise} \end{cases} + \operatorname{sgn}(zx) $ &$ 9.90\!\!\times\!\! 10^{-2}$ &$ 1.75\!\!\times\!\! 10^{-1}$ & $0.566$ \\
Unrecover-27 & $ \arctan\left(\dfrac{x^3 - 3xy^2}{3x^2y - y^3}\right) \ln(1 + x^2+y^2) $ & $4.29 \!\!\times\!\! 10^{-1} $ & $3.46 \!\!\times\!\! 10^{-1} $ & $1.240$ \\
Unrecover-28 & $ \arctan(\sinh(x) + \cosh(yz)) - \tanh(xy) $ &$6.15 \!\!\times\!\! 10^{-3}$ & $ 6.23\!\!\times\!\! 10^{-3}$ & $0.987$ \\
Unrecover-29 & $\makecell[l]{\left(0.1x^3y^2\! -\! 0.05x^4\! +\! 0.02y^5\! -\! xy\! +\! x\! -\! y\right)\!\!\times\!\! \\ \cos(2\pi\sqrt{x^2+ y^2}) + \sin(5x)\cos(5y)}$ & $4.99\!\!\times\!\! 10^{-1} $ &$3.71\!\!\times\!\! 10^{-1} $ & $1.345$ \\
Unrecover-30 & $ \ln\left(1 + \left|\tan\left(\dfrac{x}{y^2+1}\right)\right| + \left|\tan\left(\dfrac{y}{x^2+1}\right)\right|\right) $ & $2.28 \!\!\times\!\! 10^{-2} $ & $4.47 \!\!\times\!\! 10^{-2} $ & $0.510$ \\
Unrecover-31 & $\sin\left(e^{{x}/{10}}\right) \cos(xy) e^{\!-\sqrt{x^2\! +\! y^2}/{20}}\! +\! \ln(1 \!+\! x^2 y^2)$ & $1.28\!\!\times\!\! 10^{-4} $ &$2.33\!\!\times\!\! 10^{-4} $ & $0.549$ \\
Unrecover-32 & $J_0\left(\sqrt{x^4 + y^4}\right) + Y_1(x^2 + y^2 + 1) e^{-{(x^2 + y^2)}/{100}}$ & $3.31\!\!\times\!\! 10^{-3} $ &$4.02\!\!\times\!\! 10^{-3} $ & $0.823$ \\
Unrecover-33 & $\text{erf}(\sin(x) - \cos(y)) e^{-{(x - y)^2}/{10}} + \dfrac{2xy}{x^2 + y^2 + 0.1}$ & $2.90\!\!\times\!\! 10^{-3} $ &$1.41\!\!\times\!\! 10^{-2} $ & $0.206$ \\
Unrecover-34 & $ \left(1 - e^{-{1}/{\sqrt{x^2 + y^2}}}\right) \cos\left(5 \arctan\left(\dfrac{\sin(xy)}{\cos(x+y)}\right)\right) $ & $ 1.68\!\!\times\!\! 10^{-2} $ & $ 4.02\!\!\times\!\! 10^{-2} $ & $0.418$ \\
Unrecover-35 & $\dfrac{e^{-x^2}}{1 \!+\! \left( y \! -\! \sin(5x) e^{\!-\!x^2/10}\right)^2}\! +\! \dfrac{e{- y^2}}{1 \!+\! \left(x \!-\! \cos(5y) e^{\!-\!y^2/10}\right)^2}$ & $1.92\!\!\times\!\! 10^{-2} $ &$2.26\!\!\times\!\! 10^{-2} $ & $0.850$ \\
Unrecover-36 & $\ln\left(|\sin(x)\! +\! \cos(y)| \!+\! \sqrt{x^2 \!+\! y^2 \!+\! 1}\right) \sqrt[3]{|xy^2 \!-\! yx^2| \!+\! 1}$ & $4.37\!\!\times\!\! 10^{-3} $ &$9.16\!\!\times\!\! 10^{-3} $ & $0.477$ \\
Unrecover-37 & $ \sin(xy) + 0.3\sin(4x\cos(4y)) + 0.1\cos(12y\sin(12x)) $ &$4.73 \!\!\times\!\! 10^{-3} $ & $4.74 \!\!\times\!\! 10^{-3} $ & $0.998$ \\
Unrecover-38 & $ \left(\begin{cases} \sin(x) & \text{if } x+y > z \\ \cos(y) & \text{otherwise} \end{cases}\right) + \left(\begin{cases} z^2 & \text{if } z < 0 \\ -z^2 & \text{otherwise} \end{cases}\right) $ &$5.79 \!\!\times\!\! 10^{-2}$ &$5.80 \!\!\times\!\! 10^{-2}$ & $0.998$ \\
Unrecover-39 & $ \ln(|\sin(x)\! -\! \cos(y)|\!\! +\!\! 1)\!\! +\!\! \dfrac{\cos(7\!\arctan(y/x))}{\sqrt{x^2+y^2}}\!\! -\! \tanh(xy) $ &$4.74 \!\!\times\!\! 10^{-1} $ & $ 4.71\!\!\times\!\! 10^{-1} $ & $1.006$ \\

\end{longtable}
}
\restoregeometry
\end{document}